\documentclass[11pt,a4paper,dvips]{article}

\usepackage{jheppub}

\usepackage{amsmath,amssymb,bbbold}
\usepackage{graphicx}

\newcommand{\f}[2]{\frac{#1}{#2}}
\newcommand{\la}{\langle}
\newcommand{\ra}{\rangle}
\newcommand{\tr}{{\rm tr}}
\newcommand{\barq}{\protect{\bar{q}}}
\newcommand{\barQ}{\protect{\bar{Q}}}
\newcommand{\W}{{\cal W}}
\newcommand{\N}{{\cal N}}
\newcommand{\A}{{\cal A}}
\newcommand{\sla}[1]{{\raisebox{0.8pt}{\makebox[0pt][l]{\hspace{-2.5pt}\bf{
          /}}}{#1}}} 
\newcommand{\slatwo}[1]{{\raisebox{0.8pt}{\makebox[0pt][l]{\hspace{-4.2pt}\bf{
          /}}}{#1}}}

\newcommand{\slax}{{\raisebox{0.1pt}{\makebox[0pt][l]{\hspace{-4.2pt}\bf{
          /}}}{x}}} 
\newcommand{\slan}{{\raisebox{0.1pt}{\makebox[0pt][l]{\hspace{-4.2pt}\bf{
          /}}}{n}}} 
\newcommand{\slap}{{\raisebox{-0.14pt}{\makebox[0pt][l]{\hspace{-4.2pt}\bf{
          /}}}{p}}} 

\newcommand{\vphi}{\varphi}
\newcommand{\tvphi}{\tilde\vphi}

\newcommand{\acosh}{{\rm arccosh}\,}
\newcommand{\asinh}{{\rm arcsinh}\,}
\newcommand{\atanh}{{\rm arctanh}}
\newcommand{\acos}{{\rm arccos}\,}
\newcommand{\asin}{{\rm arcsin}}
\newcommand{\atan}{{\rm arctan}}

\newcommand{\de}{\partial}

\newcommand{\one}{{\mathbbb    1}}

\renewcommand{\Im}{{\rm Im}\,}

\title{Reggeon exchange from gauge/gravity duality}

\author[a]{Matteo Giordano}
\author[b]{and Robi Peschanski}
\affiliation[a]{Departamento de F\'isica Te\'orica, Universidad de
  Zaragoza, \\ 
  Calle Pedro Cerbuna 12, E--50009 Zaragoza, Spain}

\affiliation[b]{Institut de Physique Th\'eorique  CEA-Saclay
  \\ F-91191 Gif-sur-Yvette Cedex, France}
  \emailAdd{giordano@unizar.es}
  \emailAdd{robi.peschanski@cea.fr}

\abstract{  We perform the analysis of quark--antiquark Reggeon exchange in
  meson--meson scattering, in the framework of the gauge/gravity
  correspondence in a confining background. On the gauge theory side,
  Reggeon exchange is described as quark--anti\-quark exchange
  in the $t$ channel between fast projectiles.
  The corresponding amplitude is represented in terms of Wilson loops
  running along the trajectories of the constituent quarks and
  antiquarks. 
  The paths of the exchanged fermions are integrated over,  while the 
  ``spectator'' fermions are dealt with in an eikonal approximation.
  On the gravity side, we follow a previously proposed approach, and
  we evaluate the Wilson--loop expectation value by making use of
  gauge/gravity duality for a generic confining gauge theory. The
  amplitude is obtained in a saddle--point approximation through the
  determination near the confining horizon of a Euclidean ``minimal
  surface with floating boundaries'', i.e., by fixing the
  trajectories of the exchanged quark and antiquark by means of a
  minimisation procedure, which involves both area and length
  terms. After discussing, as a warm--up exercise, a simpler problem  
  on a {\it plane} involving a soap film with floating boundaries, we
  solve the variational problem relevant to Reggeon exchange, in which
  the basic geometry is that of a {\it helicoid}. A compact expression
  for the Reggeon--exchange amplitude, including the effects of a
  small fermion mass, is then obtained through analytic continuation  
  from Euclidean to Minkowski space--time. We find in particular a
  {\it linear}  Regge  trajectory, corresponding to a Regge--pole
  singularity  supplemented by a logarithmic cut induced by the
  non--zero quark mass. 
  The analytic continuation leads also to companion contributions,
  corresponding to the convolution of the same Reggeon--exchange
  amplitude with multiple elastic rescattering interactions between
  the colliding mesons. 
}

\keywords{Gauge-gravity correspondence, Nonperturbative Effects}

\arxivnumber{1105.6013}

\begin{document}

\maketitle

\section{Introduction and summary of results}

The main difficulty in the study of hadronic {\it soft} scattering 
at high energy\footnote{A {\it soft} two--body high--energy scattering
  process is characterised by a large center--of--mass energy squared
  $s$ and a small momentum transfer $t$ with respect to the typical
  hadronic scale, i.e., $\sqrt{s}\gg 1\, {\rm GeV} \gtrsim
  \sqrt{-t}$.} in the framework of Quantum Field Theory
is due to the fact that it involves the nonperturbative,
strong--coupling regime of the microscopic theory 
underlying strong interactions, namely Quantum Chromodynamics (QCD).
In recent years, a remarkable achievement of the general gauge/string
theories relationship (see~\cite{Polyakov1} and older relevant
references therein), 
namely the so--called {\it gauge/gravity duality}, has provided a new
possible practical tool to deal with strong--coupling physics in QCD,
and this has raised the hope to obtain new insights in this difficult
and long--standing problem. Gauge/gravity duality, whose first precise 
realisation has been provided by the AdS/CFT
correspondence~\cite{adscft1,adscft2,adscft3}, 
relates a gauge field theory at strong coupling with a dual gravity
theory in the weak coupling regime, and has been the subject of
intense research work over the last decade. 
The AdS/CFT correspondence, which is valid, strictly speaking, for the
conformal $\N=4$ SYM gauge field theory, appears to be physically
useful in the study of the high--temperature
quark--gluon plasma in QCD, where the confinement property is less relevant
\cite{QGP1,QGP2} (for a recent review see~\cite{QGP3}). The extension
of the gauge/gravity duality to non 
conformal confining theories is motivated by the possibility to obtain
a better understanding of those nonperturbative properties of strong 
interactions which are sensitive to the confinement scale.  

Although not yet completed, specifically for QCD, this
program has shown a few properties which the gravity theory dual to QCD
should have, in order to reproduce the main features of strong
interactions. 
In particular, the presence of a confinement
scale in the gauge theory translates into a characteristic scale in
the gravity theory, associated for example to the horizon
of a black hole~\cite{AdSBH}, or to the position of a hard
wall~\cite{Polchinski}, or to the scale associated to a soft
wall~\cite{Karch}. Our aim is to formulate results which 
may be valid in a generic confining case, independently of a specific 
realisation of the duality. Such an opportunity is provided by the
high--energy limit of {\it soft} two--body scattering
amplitudes~\cite{JP1,JP2,Janik,Jani,GP}, for which relatively general
properties may be obtained where the confinement scale plays a major
role. 

As it is well known, from the phenomenological point of view,
{\it soft} high--energy hadron--hadron scattering processes can be
described, in the language of Regge theory, in terms of the exchange
of ``families'' of states between the interacting hadrons. These
``families'' correspond to the singularities in the
complex--angular--momentum plane of the amplitude in the crossed
channel (see e.g.~\cite{Collins}). The leading contribution at high
energy comes from the so--called {\it Pomeron}, which carries the
quantum numbers of the vacuum, while subleading contributions are
usually called {\it Reggeons}, and correspond to various non--vacuum
quantum--number exchanges. One of the aims of the theoretical study of
{\it soft} high--energy scattering is to obtain an explanation of these
phenomenological concepts from the underlying microscopic field
theory.   

While a lot of work has been done in recent years regarding
the
Pomeron~\cite{Nacht,DFK,Meggiolaro96,Nachtr,BN,Dosch,Meggiolaro01,
LLCM1,LLCM2,ILM,Latt,Latt2}, especially in the context of the gauge/gravity
duality~\cite{JP1,JP2,Janik,
GP,Brow0,Brow1,Brow2,Brow3,Brow4,Corn1,Corn2,Corn3,Corn4,Corn5,
Tali1,Tali2,Muel,LP,Khar},   
the problem of Reggeon exchange has received less
attention.\footnote{We mention however Ref.~\cite{Yoshi}, where a
  unified treatment of the  signature--odd partner of the Pomeron,
  the so--called ``Odderon'', and of the signature--odd Reggeons is
  proposed, and Ref.~\cite{Makeenko}, where the Regge behaviour of
  scattering amplitudes in QCD is obtained in an effective string
  approach.}  
In this paper we shall focus on the approach to $q\bar{q}$--Reggeon
exchange in meson--meson scattering proposed
in~\cite{Jani}. This approach elaborates on the formalism
of~\cite{Nacht,DFK,Meggiolaro96,Nachtr,BN,Dosch}, valid for {\it soft}
high--energy processes, and on previous work on the use of
gauge/gravity duality for scattering amplitudes~\cite{JP1,JP2,Janik}. 
In particular, it assumes gauge/gravity duality for a
generic confining theory, and exploits the path--integral
representation for the fermion
propagator~\cite{Brandt,Brandt2,Polyakov,Korchemsky,Korchemsky2} to
provide an expression for the Reggeon--exchange scattering amplitude
in the energy regime under investigation.

In this approach, the Reggeon--exchange 
amplitude is put into a relation with the expectation
value of certain Euclidean Wilson loops, describing the exchange of a
(Reggeised) quark--antiquark pair between the interacting hadrons. 
More precisely, the loop contours are made up of a fixed part,
corresponding to the eikonal trajectories of the ``spectator''
fermions, and a ``floating'' part, corresponding to the trajectories
of the exchanged fermions. The Reggeon--exchange scattering amplitude
is obtained by summing up the contributions of these loops, through a
path--integration over the trajectories of the exchanged fermions. In
turn, these Wilson loops are related via the gauge/gravity duality to
minimal surfaces in a curved confining metric, having the loop contour
as boundary (Plateau problem), which correspond to the exchange of an
open string between the interacting hadrons. Finally, the physical 
amplitude in Minkowski space--time is recovered by means of
analytic
continuation~\cite{Meggiolaro97,Meggiolaro98,Meggiolaro02,Meggiolaro05,
  crossing,Meggiolaro07,EMduality}. 

The determination of the relevant minimal surfaces is in general a
difficult problem, and depends on the specific choice for the
confining background. 
The key idea is to observe that an approximate solution to this
problem can be found by solving a simpler Plateau problem, 
namely by finding a minimal surface in a flat Euclidean
4--dimensional space near the confinement scale (e.g., near a confining
horizon) in the bulk. Since the precise form of the metric does not
enter the simplified problem, the corresponding solution is expected
to be a valid approximation independently of the specific realisation
of the duality.
To leading order, the amplitude is then evaluated through a saddle--point
approximation of the Euclidean path--integral, which fixes the shape
of the floating boundary. 

The quantity to be minimised is the Euclidean
``effective action'' (see Section \ref{saddle}), which consists of a
linear combination of the area of the surface and of the length of its
floating boundaries, and which encodes the contribution to the
(Euclidean) scattering amplitude of a given shape of the floating
boundary.
For the sake of brevity, we will sometimes refer to our
variational problem as the {\it minimal surface problem with floating
boundaries}, which we will define precisely in Section \ref{saddle}.

It is worth noticing already at this stage an important practical
aspect of this approach. For large ``spectator''--quark mass (and thus
large meson mass) and small exchanged--quark mass, 
it is argued that the relevant contributions to the
path integral should come from those configurations in which the
floating part of the boundary lies on a specific surface, namely the
minimal surface having as boundaries two infinite straight
lines. Such a surface is a {\it helicoid}, which has been already
encountered in the study of {\it soft} quark--quark scattering at high
energy~\cite{JP2}. 

In~\cite{Jani}, this problem was solved using a null quark
mass approximation, which simplifies the minimisation procedure by
reducing the effective action to the area term only, and 
leads to a complex solution (thus not in Euclidean space) for the
saddle--point equation.\footnote{
Quadratic fluctuations of the string around the corresponding minimal
surface were also computed, leading to a constant shift in the
Reggeon ``intercept'', i.e., the exponent of $s$ at $t=0$ in the
high--energy behaviour of the amplitude~\cite{Janik}.} 
The physical amplitude was finally 
obtained by means of analytic continuation of this complex solution
onto a physically admissible one in Minkowski space--time.  
Interestingly enough, the resulting amplitude obtained in~\cite{Jani} 
was of Regge--pole type, with a linear Reggeon trajectory. 
Our aim in this paper is to revisit this method in a more general
minimisation setting, including the length terms by considering a
non--zero quark mass. In this way we find real Euclidean solutions, from
which we can obtain the physical scattering amplitude in Minkowski
space--time by means of a suitable analytic continuation.

In this paper, we investigate in detail and solve the Euler--Lagrange
equations corresponding to the above--mentioned {\it minimal surface
  problem with floating boundaries}, and we discuss the properties of
the resulting Reggeon--exchange amplitudes. We summarise here the main
results.  
\begin{itemize}
\item It is shown how a {\it real} solution to the minimal surface problem
  with floating boundaries is obtained in Euclidean space, provided
  the quark mass is non--zero. The minimal Euclidean ``effective
  action'' corresponding to the solution reads
  \begin{equation}
    \label{eq:eff_noprof_intro}
    S_{{\rm eff,\,E}} = \f{b^2}{2\pi\alpha_{\rm eff}'\theta}\,f(\tvphi)+
    \f{4mb}{\theta}\left(B(\vphi_0,\tilde\vphi) - 
      \sinh\tilde\vphi\right)\, ,
  \end{equation}
  where $1/2\pi\alpha_{\rm eff}'$ is the string tension, 
  $b$ is the impact--parameter distance and $\theta$ the 
  angle between the Euclidean trajectories of the two incoming
  particles, and $\vphi_0$ and $\tilde\vphi$ are 
  geometric parameters of the solution of the minimisation problem for
  the boundaries (see further Section \ref{Reggeon}). The shape of the
  boundaries enters the effective action through the functions
  $f(\tvphi)$ and $B(\vphi_0,\tilde\vphi)$, which are obtained in an
  implicit form, and can be easily evaluated numerically; an
  explicit analytic expression is obtained in two specific
  regimes. 
\item The saddle--point equation in Euclidean space admits a real
  solution only in a finite region for the impact parameter $b$,
  namely $b\le b_c = 4\pi\alpha_{\rm eff}' m$, and this limitation
  carries over to Minkowski space after analytic continuation. In
  order to investigate the region $b>b_c$, 
  and also in order to take the massless--quark limit, we show how an
  analytic continuation of the result in Minkowski space to the region
  $b>b_c$ can be performed, giving rise to a sensible scattering
  amplitude. Our main result for the Minkowskian effective action
  then reads  
  \begin{equation}
    \label{eq:act_large_b_intro}
    S_{{\rm eff,\,M}}(s,b)\sim \f{b^2}{2\pi\alpha_{\rm
        eff}'\chi}\acos\f{b_c}{b}  -
    \f{2bm}{\chi}\sqrt{1-\left(\f{b_c}{b}\right)^2} +
    2\pi^2\alpha_{\rm eff}'m^2  \,  ,
  \end{equation}
  where $\chi\sim \log s$ is substituted to $\theta$ by the analytic
  continuation $\theta\to -i\chi$, see Section \ref{Mink}.
 
\item Expanding Eq.~\eqref{eq:act_large_b_intro} for small  quark
  mass, one finds a Gaussian--like Reggeon--exchange amplitude in
  impact--pa\-ra\-me\-ter space (up to prefactors),  
  \begin{equation}
    \label{eq:small_m_intro}
    a(\vec{b},\chi)\equiv 
    \f{i}{2s} \int \f{d^2\vec{q}}{(2\pi)^2}\, e^{-i\vec q \cdot \vec b}\
    \A_{\cal R}(s,t)  
    \propto \exp{\left\{ 
        -\f{b^2}{4\alpha_{\rm eff}'\chi} + \f{4bm}{\chi}  
        - 2\pi^2\alpha_{\rm eff}' m^2\right\}}\, ,
  \end{equation} 
  where $\A_{\cal R}(s\!\sim\!e^\chi,t\!=\!-\vec{q}^{\,2})$ is the
  Reggeon--exchange scattering amplitude in momentum space. The result
  Eq.~\eqref{eq:small_m_intro} leads to a 
  {\it linear} Reggeon trajectory $\alpha_{\cal R}(t)=\alpha_0
  +\alpha'_{\cal R}t,$ with slope  $\alpha_{\cal R}'\equiv \alpha_{\rm 
    eff}'$ equal to the inverse string tension.\footnote{Note that this
    equality is non trivial, since the string tension may be
    independently obtained by evaluating the confining $Q\!-\!\barQ$
    potential.}  
  As discussed in Section \ref{mass_effects}, one is able to discuss
  modifications of the Regge singularity due to a small but non--zero
  quark mass.  We find the same Regge--pole singularity obtained at
  $m=0$ in~\cite{Jani}, plus a logarithmic singularity due to the
  non--zero quark mass.  Although the nature of the Reggeon
  singularity is changed, the Reggeon trajectory remains linear after
  the inclusion of (small) quark--mass effects.   
  Their main physical consequence is that the slope of the amplitude
  $\de\A_{\cal R}/\de t|_{t=0}$ is increased, and its shrinkage with
  energy is strengthened.   

\item The analytic continuation of the Euclidean action
  \eqref{eq:eff_noprof_intro} leads to other contributions in
  Minkowski space--time through the Riemann multi--sheet structure of
  the inverse cosine function in Eq.~\eqref{eq:act_large_b_intro}. 
  They take the form of a multiple convolution in momentum space 
  of the Reggeon--exchange amplitude \eqref{eq:small_m_intro} with an
  integer number of copies of the following amplitude, 
  \begin{equation}
    \label{eq:elastic_intro}
    \A_{el}(s,t) = -4i\pi s\int_0^\infty db\, b\,J_0(qb)%\,
    \exp\left\{-\f{b^2}{\alpha_{\rm eff}'\chi}\right\} \equiv -2i\pi
    s\,\alpha_{\rm eff}'\chi \exp\left\{\f{\alpha_{\rm eff}'
        t}{4}\chi\right\}\, ,  
  \end{equation}
  which happens to be a Regge--pole amplitude with intercept one. The
  properties of this amplitude allow one to interpret it as an
  elastic interaction between the incident mesons. 
  Moreover, this interaction has the same features of the one obtained
  some time ago for dipole--dipole elastic scattering amplitudes
  \cite{JP2}. The multiple convolution can be phenomenologically
  interpreted as the effect of the ``rescattering corrections'' to the 
  ``bare'' $q\barq$--Reggeon exchange. 
\end{itemize}

As we have already remarked, our results for confining gauge theories  
are expected to be quite general, and independent of the 
precise realisation of the gauge/gravity duality (assuming it exists),
since they rely only on general features of the dual geometry,
essentially the (effective) cut--off provided by the confinement scale
in the bulk. It is worth mentioning that in recent years the
holographic approach has been applied also to the issue of scattering
amplitudes in the context of ${\cal N}=4$ SYM theory, in particular
using the AdS/CFT correspondence and minimal surfaces to investigate
gluon--gluon elastic scattering at high energy~\cite{Alday}. 
It appears that in this case the resulting Regge
trajectory is {\it logarithmic} rather than {\it linear}
\cite{Naculich:2007ub,us}, which is a striking difference between the
predictions for conformal and confining gauge theories.

The plan of the paper is the following. In Section \ref{AdSBH} we
recall the relevant features of the gauge/gravity duality used in the
evaluation of Wilson--loop expectation values in a confining
theory. In Section \ref{saddle} we review the approach of~\cite{Jani}
to Reggeon exchange, discussing in some detail the approximations
involved and the Euler--Lagrange equations for the relevant minimal
surface with floating boundaries in the presence of a non--zero quark
mass. In Section \ref{planar} we solve explicitly the equations in a
planar case, which happens to be related to a classical problem
involving a soap film with floating boundaries. In Section
\ref{Reggeon} we solve the Euler--Lagrange equations related to 
the Reggeon--exchange problem, where the basic geometry is that of a
helicoid. We discuss in particular the issue of smoothness conditions,
and we obtain an exact solution in implicit form for the general
case. We then investigate analytically two limits of the solution, in
which we are able to write it down explicitly and to uncover the
dependence on the relevant variables. We also compare them with some
numerical results for the exact solution.   
In Section \ref{Mink} we perform the analytic continuation
into Minkowski space--time, and discuss the properties of the
resulting Reggeon amplitude, in particular regarding the dependence on
the energy and on the impact parameter.
In Section \ref{mass_effects} we discuss the effect of a non--zero
quark mass on the nature of the Reggeon singularity and on the Reggeon
trajectory. We also discuss the other companion contributions to the
amplitude coming from the multi--sheet structure of the Minkowskian
effective action. 
Finally, in Section \ref{concl} we draw our 
conclusions and show some prospects for the future. A few technical
details are given in the Appendices.

\section{Wilson loops and gauge/gravity duality for confining
  theories} 
\label{AdSBH}

In this Section we recall the relevant aspects of the gauge/gravity
duality which will be used in the following. We begin with the now
standard AdS/CFT correspondence~\cite{adscft1,adscft2,adscft3}, 
which relates type IIB string theory in $AdS_5\times S^5$ in the
weak--coupling, supergravity limit, to four--dimensional ${\cal N}=4$
SYM theory, which is a conformal and non confining field theory, in
the limit of large number of colours $N_c$ and strong 't Hooft
coupling $\lambda = g_{\rm   YM}^2 N_c$, where $g_{\rm YM}$ is the
coupling constant in the gauge theory. Expectation values in the field
theory can be obtained from the dual gravity theory with the
appropriate prescription; for further convenience, we focus on the
problem of the vacuum expectation value of Wilson loops. Going over to
Euclidean signature, the prescription for a Wilson loop running along
the path ${\cal C}$ is given by the following area
law~\cite{Wilson,Wilson2,GO,DGO}, 
\begin{equation}
  \label{eq:area_law}
  \la \W[{\cal C}] \ra \sim {\cal F}[{\cal C}]\ 
  e^{-\f{1}{2\pi\alpha'}A_{\rm min}[{\cal C}]}\, .
\end{equation}
Here $A_{\rm min}[{\cal C}]$ is the area of a minimal surface in the
Euclidean version of the $AdS_5$ metric, which is obtained from the
original metric 
\begin{equation}
  \label{eq:ads_metric}
ds_{AdS}^2 = \f{dz^2}{z^2} +
  \f{\eta_{\mu\nu}dx^\mu dx^\nu}{z^2} \, ,  
\end{equation}
where $\eta_{\mu\nu}$ is the four--dimensional Minkowski metric
($\mu,\nu=0,\ldots, 3$), by replacing $\eta_{\mu\nu}\to\delta_{\mu\nu}$.
The minimal surface has as boundary the contour ${\cal C}$ at
$z=0$, i.e., on the four--dimensional boundary of $AdS_5$; moreover,
${\cal F}$ is a prefactor due to quantum fluctuations around the
minimal surface, and $1/2\pi\alpha'$ is the string tension. 

In order to extend the duality to the confining case,
one has to properly modify the background metric in the dual gravity
theory, taking into account that the theory is no more conformal. 
Although the precise realisation of the duality (assuming it exists)
is not known yet, a common feature of various attempts to describe a
confining theory in terms of a gravity dual is the presence of a
characteristic scale $R_0$ in the metric, which separates the small
and large $z$ regions. With the appropriate choice of coordinates,
while for small $z$ the metric diverges as some inverse power of $z$,
for $z$  of the order of $R_0$ it turns out to be effectively
flat. The interpretation in the dual confining field theory is that
the scale $R_0$ provides the confinement scale. For example, in the
case of the AdS/BH metric of~\cite{AdSBH}, such a scale is provided
by the position in the fifth dimension of the black--hole horizon. The  
relevant part of the metric reads
\begin{equation}
  \label{eq:BH}
  ds_{AdS/BH}^2 = \f{16}{9} \f{1}{f(z)}\f{dz^2}{z^2} +
  \f{\eta_{\mu\nu}dx^\mu dx^\nu}{z^2} + \ldots\, ,
\end{equation}
where $f(z)=z^{2/3}(1-(z/R_0)^4)$. The
near--horizon geometry is effectively flat,
\begin{equation}
  \label{eq:BH_nh}
  ds_{\rm hor}^2 \simeq \f{1}{R_0^2}\eta_{\mu\nu}dx^\mu dx^\nu\, .
\end{equation}
The prescription used to calculate a Wilson loop
expectation value (in Euclidean signature) is the same as above, but
substituting the AdS metric with an appropriate confining
background~\cite{Sonn0,Sonn1,Sonn2}. 
Also, one has to replace $1/2\pi\alpha'$ with an effective 
string tension $1/2\pi\alpha'_{\rm eff}$, which depends on the particular
background metric: in the case of the AdS/BH metric \eqref{eq:BH}, for
example, it is given by $1/2\pi\alpha'_{\rm  eff}=\sqrt{2g_{\rm
    YM}^2N_c}/2\pi R_0^2$. Although it is not possible to determine
its explicit expression in the general case, $\alpha'_{\rm eff}$ can
be determined phenomenologically by comparison with the heavy
quark--antiquark confining potential $V_{Q\barQ}(R) = (1/2\pi\alpha'_{\rm eff})R$.

\begin{figure}[t]
   \centering
\includegraphics[width=0.7\textwidth]{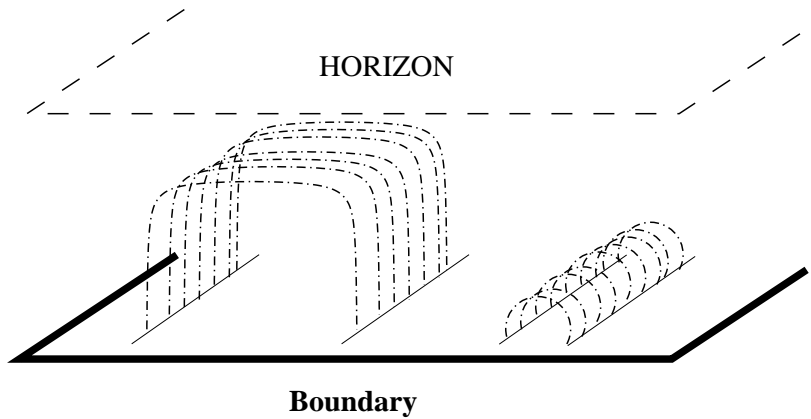}
  \caption{{\it Minimal surface in the confining black--hole
      geometry.} For simplicity, the Wilson lines are
    drawn here with vanishing angle of tilt $\theta=0$. For large
    enough impact parameter (left), the minimal surface rises as a
    vertical wall from the boundary, and is almost flat
    near the horizon. At small impact parameter (right) the surface is
    more similar to a non confining case. The picture is taken
    from~\cite{JP2}.}   
  \label{config}
\end{figure}

The analytic  solution of the Plateau problem\footnote{An analytic
  solution is required for performing the continuation from Euclidean
  to Minkowski space.} is a highly non trivial task 
already in flat Euclidean space, and it is even harder in a non--flat
metric such as \eqref{eq:BH}: some approximations are then necessary
in order to 
obtain an analytic expression. A reasonable and manageable 
scheme is obtained by means of a near--horizon approximation, taking
into account the above--mentioned features which the dual gravity
theory is expected to have~\cite{JP2,Jani} (see
also~\cite{Sonn1,Sonn2}). The small--$z$ behaviour suggests that, in
order to minimise the area, it is convenient for the surface to rise
almost vertically from the boundary, without appreciable motion in the
other directions, at least when the typical size $b$ of the Wilson
loop is not too small,\footnote{This
  approximation is expected to be valid when $b$ is greater than
  $R_0$, which should correspond to the distance at which the
  interquark potential becomes linear.} see Fig.~\ref{config}
(left). On the other hand, the geometry of the surface is different
for smaller values of $b$, see Fig.~\ref{config} (right).  
The presence of a horizon puts an upper bound on this vertical rise; 
moreover, when $z\sim R_0$, the surface lives effectively in flat
space. As a result, the minimal surface is expected to be constituted
by two parts: an almost vertical wall rising from the boundary up to
the horizon, and transporting there the boundary conditions, and a
solution of the Plateau problem in flat space.

A schematic representation of this geometrical configuration in the
bulk for parallel Wilson lines, relevant to the determination of the 
confining potential, is displayed in Fig.~\ref{config} (left). 
The area of the vertical wall is divergent, but in the expression for
scattering amplitudes it is usually cancelled by appropriate
normalisation factors.\footnote{This is the case, for example, for the
dipole--dipole scattering amplitude expressed in terms of Wilson--loop
correlation functions~\cite{JP2}: here the relevant size is the
distance $b$ between the two loops, which play the role of
disconnected boundary for the minimal surface, and the normalisation
factor is the product of the Wilson--loop expectation values. } 
The problem is then reduced to a calculation in  
four--dimensional Euclidean space: this is reminiscent 
of the old ``QCD string'' approach (see~\cite{Frampton} for a
comprehensive review), although in
this case it should be the result of an approximation to a 
higher--dimensional critical string theory, and thus it should not
suffer the problems of the old approach. 

Due to the generality of the geometrical picture leading to the
considered flat--space approximation near the confinement scale,
analogous to a near--horizon approximation, one expects to get results 
valid for a gauge/gravity duality for a generic confining gauge field
theory, and hopefully for QCD assuming the existence of its yet
unknown gravity dual. We shall now turn to the determination of a
Reggeon--exchange amplitude in this context.

\section{Reggeon--exchange amplitude}
\label{saddle}

In this Section we recall the method of Ref.~\cite{Jani} for the
determination of the Reggeon--exchange contribution to the
meson--meson scattering amplitude in the {\it soft} high--energy regime, 
developing on a few points which are relevant for our analysis
and more briefly discussed in that work. 

The starting point is to adopt a description of the interacting
hadrons in terms of their constituent partons. 
Such an approach to {\it soft} high energy 
hadron--hadron scattering has been introduced in~\cite{Nacht}, where it
was used, together with an LSZ reduction scheme and an eikonal
approximation for the propagators, in order to derive
approximate nonperturbative formulas for the scattering amplitudes.
The basic idea is that the leading Pomeron--exchange contribution to the
elastic amplitude comes from processes which are elastic and {\it
  soft} at the level of 
the constituent partons, justifying an eikonal--like approach. In a
space--time picture of these processes, 
the partons travel along their classical, straight--line trajectories,
exchanging only {\it soft} gluons which leave these trajectories practically 
unperturbed. This approach to the Pomeron--exchange amplitude has been
investigated and extended in a number of
papers~\cite{DFK,Meggiolaro96,Nachtr,BN,Dosch,Meggiolaro01,LLCM1,LLCM2,ILM,JP1, 
JP2,Janik,Latt,Latt2,GP}. 

In particular, in the case of meson--meson scattering, one can
describe the mesons, in a first approximation, in terms of a
wave packet of transverse colourless quark--antiquark  
dipoles~\cite{DFK,Nachtr,BN,LLCM1}. The mesonic scattering amplitude
is reconstructed, after folding with the appropriate wave functions,
from the scattering amplitude of such dipoles. Since here we are
interested only in the Reggeon trajectory, which, being a universal
quantity,  should not depend on the details of the meson wave
function, we can focus on the dipole--dipole amplitude, which is
expected to encode the relevant features of the process. Stated
differently, invoking the universality of Reggeon exchange, one can
consider mesons whose wave function is strongly peaked around some
average value $|\vec{R}|$ of the dipole size.

\begin{figure}[t]
  \centering
  \includegraphics[width=0.7\textwidth]{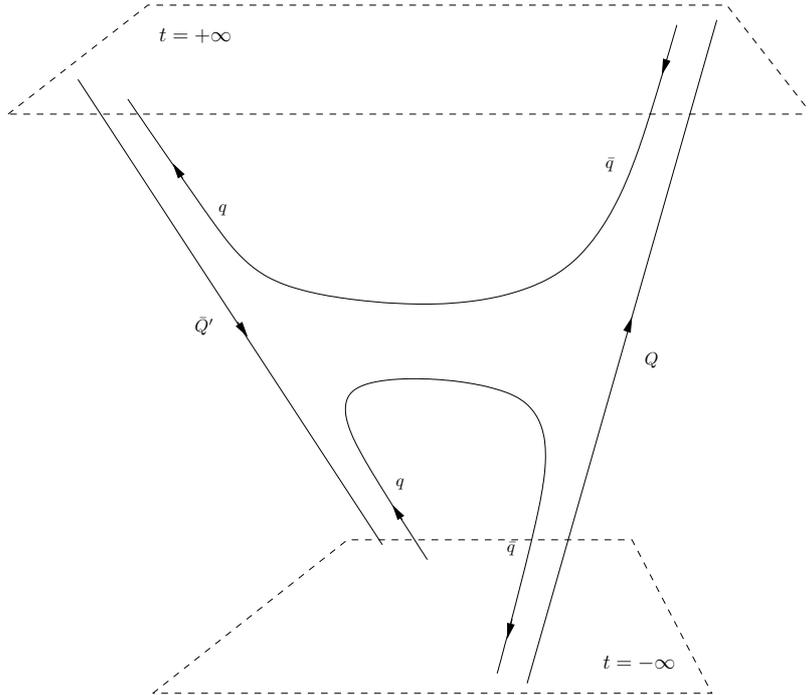}
  \caption{{\it Space--time picture of the
    Reggeon--exchange process.} $Q, \barQ'$ are heavy and fast quark
  and antiquark, which follow straight--line  trajectories in
  the eikonal approximation. $q,\barq $ are the exchanged light quark
  and antiquark, describing the Reggeon exchange between the incident $Q
  \barq$ and  $\barQ'q$ mesons (see text).} 
  \label{fig:1}
\end{figure}

Using this simplified description for the mesons, Reggeon exchange is
identified as an inelastic process at the partonic level, involving the
exchange of a quark--antiquark pair between the colliding dipoles. 
More precisely, the corresponding space--time picture 
is the following (see Fig.~\ref{fig:1}). Before and after the 
interaction time (which may be long for a {\it soft} interaction), the 
partons inside the high--energy mesons travel approximately along
their classical, straight--line ({\it eikonal\/}) trajectories. During
the interaction 
time, a pair of valence partons is exchanged in the $t$ channel 
between the mesons, and thus their trajectories bend, connecting the
incoming and outgoing eikonal trajectories; the other partons exchange
only {\it soft} gluons, and their straight--line trajectories are left
practically undisturbed. The softness of the process requires that the
exchanged fermions carry a small fraction of longitudinal momentum of
the mesons.\footnote{
  A rigorous quantitative
  formulation of this statement is still lacking and requires a
  more detailed study of Reggeon exchange from first principles
  \cite{Giordano_wp1}.} 

In order to avoid inessential complications, we consider the
scattering of two heavy--light mesons $M_{1,2}$ of large mass
$m_{1,2}$, i.e., $M_1 = Q\barq$ and $M_2=\barQ' q$, where $Q$ and
$\barQ'$ are heavy and of different flavours, while $q$ and $\barq$
are light and of the same flavour. In this way the total scattering
amplitude amounts to a single type\footnote{For physical mesonic
  amplitudes, different Regge trajectories are introduced depending on
  the exchanged quantum numbers and the quark flavours. In the present
  case, we consider only the simplest case with only light--quark
  exchange. For completion however, as we will discuss in Section
  \ref{mass_effects}, the Reggeon exchange is not isolated. It is
  expected to be accompanied by  contributions corresponding to the
  so--called ``rescattering corrections'', which we will also obtain 
  from holography.} of Reggeon--exchange process, namely the 
one in which $q$ and $\barq$ are exchanged in the $t$ channel, plus
the Pomeron--exchange component, where there are no exchanged
fermions. Moreover, the choice of heavy mesons is made in order for
the typical size of the dipoles to be small, since in this case
$|\vec{R}_{1,2}|\sim m_{1,2}^{-1}\ll \Lambda_{\rm QCD}^{-1}$; the
reasons for this choice will be explained later on.  

\subsection{Impact--parameter amplitude}

At this point, let us describe in some detail the expression for the
Reggeon--exchange contribution $\A_{\cal R}(s,t)$ to the scattering
amplitude proposed in~\cite{Jani}. To this extent, let us introduce
the impact--parameter amplitude $a(\vec{b},\chi)$, 
\begin{equation}
  \label{eq:ampli}
  \A_{\cal R}(s,t) =  -i 2s \int d^2b\ e^{i\vec{q}\cdot\vec{b}}
  a(\vec{b},\chi)\, ,
\end{equation}
where $\chi$ is the hyperbolic angle between the classical
trajectories of the colliding mesons, related to the center--of--mass
energy squared $s$ through $\chi\simeq \log s/(m_1 m_2)$ (for
${s\to\infty}$), with $m_{1,2}$ the masses of the mesons, and $t= 
-\vec{q}\,{}^2$. Here we do not write explicitly the dependence on
the orientation of the dipoles. According to the space--time picture
of the process given above, the eikonal approximation can no
longer be used to describe the propagation of the light quarks, and
different techniques are required. 
Working in Euclidean space, the authors of~\cite{Jani} exploit the
path--integral representation for the fermion propagator in an external
non--Abelian gauge
field~\cite{Brandt,Brandt2,Polyakov,Korchemsky,Korchemsky2}, in order
to write down a Euclidean ``amplitude'' 
$\tilde{a}(\vec{b},\theta,T)$ in terms of a path--integral over the
trajectories of the light quarks. Here $\theta$ is the angle between
the Euclidean trajectories of the mesons, and $T$ is an IR
cutoff, which will be explained shortly. 
The physical Minkowskian amplitude $a(\vec{b},\chi)$ in
Eq.\eqref{eq:ampli} is finally recovered by means of the analytic
continuation $\theta\to -i\chi$, $T\to
iT$~\cite{Meggiolaro97,Meggiolaro98,Meggiolaro02,Meggiolaro05,crossing,
  Meggiolaro07,EMduality},
contracting with the appropriate Dirac spinors for the quarks and
antiquarks, and removing the IR cutoff by taking the limit
$T\to\infty$:  
\begin{equation}
  \label{eq:mink_ampl}
  \begin{aligned}
      a(\vec{b},\chi) =& \lim_{T\to\infty} 
      [\tilde{a}(\vec{b},-i\chi,iT)]_{\alpha'\beta'\gamma'\delta';
        \alpha\beta\gamma\delta}\,  
      \bar{u}^{(s_Q')}_{\alpha'}(p_Q') u^{(s_Q)}_{\alpha}(p_Q)
      {v}^{(t_{\barQ'}')}_{\beta'}(p_{\barQ'}') 
      \bar{v}^{(t_{\barQ'})}_{\beta}(p_{\barQ'}) \\
      &
      \phantom{\lim_{T\to\infty}a_{\alpha'\beta'\gamma'\delta';\alpha\beta\gamma\delta}
        (\vec{b},-i\chi,iT)}      \times
      \bar{v}^{(t_\barq)}_{\gamma}(p_\barq) u^{(s_q)}_{\delta}(p_q)
      {v}^{(t_\barq')}_{\gamma'}(p_\barq') \bar{u}^{(s_q')}_{\delta'}(p_q')
    \, .
  \end{aligned}
\end{equation}
Even in the simplified setting that we are considering here, 
in order to reconstruct the mesonic amplitudes one has still to 
average over the orientation of the dipoles; moreover, one should also
contract the spin indices with the appropriate wave functions. 
As we have already said, we are interested here only in the Reggeon
trajectory, and so the detailed dependence on spin should not be
relevant, and it will not be discussed in this work. As for the
dependence on the orientation, the choice of large meson masses, or
equivalently of small dipole sizes, will make it negligible in a 
first approximation, as it will be discussed further on. 

 \begin{figure}[t]
   \centering
  \includegraphics[width=.5\textwidth]{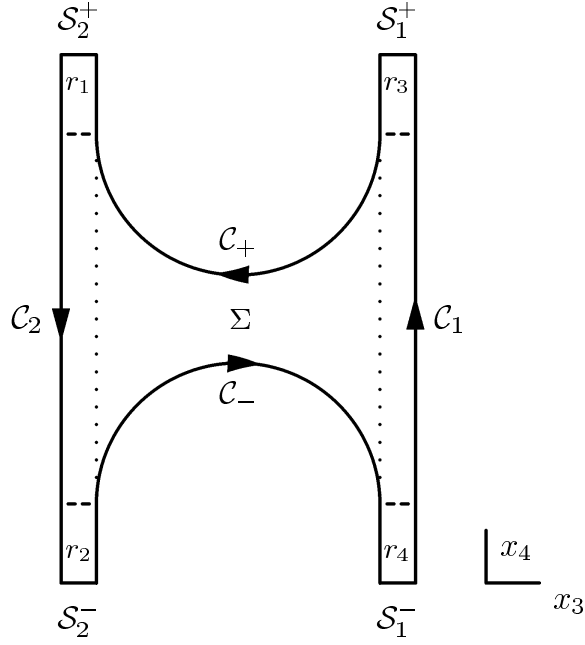}
  \caption{{\it Schematic representation of the Wilson loop contour
    relevant to Reggeon exchange.} The ``tilted'' contour (solid line)
  is projected on the $(x_4-x_3)$ plane for simplicity. The dashed
  lines delimit the various regions of the corresponding minimal
  surface, to be discussed below. The dotted lines correspond to the 
  ``virtual'' eikonal trajectories of the light quarks, which together
  with ${\cal C}_{1,2}$ describe the free propagation of the mesons. 
}   
\label{fig:1bis}
\end{figure}

All in all, the Euclidean ``amplitude'' $\tilde{a}(\vec{b},\theta,T)$, which
should encode the features of the Reggeon trajectory, can be written
symbolically as 
\begin{equation}
  \label{eq:sym_PI}
  \tilde{a}(\vec{b},\theta,T) = {\cal Z}^{-1}\int {\cal D}{\cal C}_+
  {\cal D}{\cal C}_-
  \, \la \W[{\cal C}]\ra \ e^{-m_0L[{\cal C}]}\ {\cal I}[{\cal C}]\, ,  
\end{equation}
where the different terms $\la \W[{\cal C}]\ra,L[{\cal C}],{\cal
  I}[{\cal C}],{\cal Z}$ are defined as follows.

$\bullet$ 
$\la\W[{\cal C}]\ra$ is the expectation value of the
Euclidean Wilson loop running along the path ${\cal C}$ (see
Fig.~\ref{fig:1bis}), composed essentially of the Euclidean
trajectories of the partons,  
\begin{equation}
  \label{eq:path}
  \begin{aligned}
    {\cal C} &= {\cal C}_1 \circ {\cal S}_1^- \circ {\cal C}_+
    \circ {\cal S}_2^-
    \circ{\cal C}_2 \circ {\cal S}_2^+ \circ{\cal C}_- \circ {\cal
      S}_1^+     \, ,\\
    \W[{\cal C}] &= \tr \left\{W[{\cal C}_1]W[{\cal S}_1^-]W[{\cal
      C}_-]W[{\cal S}_2^-]W[{\cal C}_2]W[{\cal S}_2^+] W[{\cal
      C}_+]W[{\cal S}_1^+] \right\}\, ,  
  \end{aligned}
\end{equation}
where $W[{\cal C}_i]$ is the Wilson line along the path ${\cal
  C}_i$. 

More precisely, ${\cal C}_{1}$ and ${\cal C}_{2}$ are the
straight--line paths corresponding to the heavy partons $Q$ and
$\barQ'$, respectively, which are
fixed,\footnote{Here and in the following we denote with $\vec{v}$ a
  two--dimensional vector. The components of the Euclidean vectors are
chosen to be $x=(x_4,x_1,\vec{x})$ with $\vec{x}=(x_2,x_3)$.}  
\begin{equation}
  \label{eq:path_straight}
  \begin{aligned}
    {\cal C}_{1} :&\,\, X_1(\nu) = u_1\nu + \f{b}{2} + \f{R_1}{2}\, ,
    && \nu\in[-T,T]\, , \\
    {\cal C}_{2} :&\,\, X_2(\nu) = -u_2\nu - \f{b}{2} - \f{R_2}{2}\, ,
    &&\nu\in[-T,T]\, , \\
    &\,\, u_1=(\cos\f{\theta}{2},\sin\f{\theta}{2},\vec{0})\, , &&
    u_2=(\cos\f{\theta}{2},-\sin\f{\theta}{2},\vec{0})\, , \\
    &\,\, b=(0,0,\vec{b})\, , && R_i=(0,0,\vec{R}_i) \, , 
  \end{aligned}
\end{equation}
while ${\cal C}_{+,-}$ are the curved paths corresponding to the  
exchanged light partons, which have to be integrated over,
\begin{equation}
  \label{eq:path_curved}
  \begin{aligned}
     {\cal C}_+:&\,\, X_+(\nu)\, , \, \,   \dot{X}_+^2(\nu)=1\, , 
     && \nu\in [0,L_+]\, ,\\ 
     &\,\, X_+(0) = u_1 T + \f{b}{2} - \f{R_1}{2}\, ,  && X_+(L_+) =
     u_2 T -\f{b}{2}  +    \f{R_2}{2}\, ,    \\
     {\cal C}_- :&\,\, X_-(\nu)\, , \, \,   \dot{X}_-^2(\nu)=1\, ,
     &&\nu\in [0,L_-]\, ,\\ 
     &\,\, X_-(0) = -u_2 T -\f{b}{2} +  \f{R_2}{2}\, ,  && X_-(L_-) = -u_1
     T + \f{b}{2} -  \f{R_1}{2}\, ,
  \end{aligned}
\end{equation}
and ${\cal S}_{1,2}^\pm$ are straight--line paths in the transverse
plane (see Fig.~\ref{fig:1bis}), connecting the four pieces above,
which are introduced in order to make the whole expression
gauge--invariant. The path--integration over the exchanged--quark
trajectories ${\cal C}_\pm$ is denoted simbolically by $\int{\cal DC}_\pm$.

In the expressions
above, the condition $\dot{X}_i^2=1$ makes of $\nu$ the natural
parameter along the curve: this condition comes from the integration
over momenta in the path integral for the Euclidean fermion
propagator~\cite{Korchemsky,Korchemsky2}. As we have 
already said, the sizes $|\vec{R}_i|$ of the dipoles are of the order
of the (small) inverse mass of the mesons, $|\vec{R}_i|\sim m_i^{-1}$. 

$\bullet$ $L[\cal C]$ is the length of the path traveled by the
light quarks,  
\begin{equation}
  \label{eq:length}
  L[{\cal C}] \equiv L[{\cal C}_+]+ L[{\cal C}_-] =
L_+ + L_-\, ,
\end{equation}
and $m_0$ is the (bare) mass of the light quark. As we will see below in more
detail, the length--term factor $e^{-m_0L[{\cal C}]}$ in
Eq.~\eqref{eq:sym_PI} plays an important stabilisation role in the
minimisation procedure related to the saddle--point approximation of
the path--integral.

$\bullet$ ${\cal I}[{\cal C}]\equiv \otimes_{i=\{ 1,2,+,- \}}\,{\cal
  I}_i[{\cal C}_i]$  is the product of the spin
factors~\cite{Korchemsky,Korchemsky2} corresponding to the various
fermionic trajectories, and it comes from the integration over momenta
in the path--integral representation for the fermion propagator. Its
subfactors are given by the path--ordered products
\begin{equation}
  \label{eq:spin_fact}
  {\cal I}_i[{\cal C}_i] = \prod_{X_i(\nu)\in \,{\cal C}_i}
  \f{1+\sla{\dot{X}_i}(\nu)}{2} \, ,
\end{equation}
where we have used the notation $\slax\equiv x_\mu\gamma_{E\mu}$, 
with $\gamma_{E\mu}$ the Euclidean Dirac matrices (see Appendix
\ref{app:A}), and where we 
understand that terms corresponding to larger values of the parameter
$\nu$ appear on the left. 

$\bullet$ ${\cal Z}$ is a normalisation constant, which was 
implicitly assumed in~\cite{Jani}, whose role is to make the amplitude
IR--finite. In principle, one should be able to determine it from first
principles; at the present stage, we adopt a more pragmatic approach,
fixing it ``by hand'' in order to remove infrared divergences. 

Before proceeding, a comment is in order. Although there are
reasonable arguments for the validity of
Eq.~\eqref{eq:sym_PI} as a nonperturbative, approximate expression 
for the Reggeon--exchange amplitude~\cite{Jani}, a direct derivation
of it from first principles is not yet known, contrary to the
Pomeron--exchange case. Also, the analytic continuation 
used to obtain the Minkowskian amplitude has been proved to be the
correct one in the case of the Pomeron--exchange amplitude: there is
not yet an explicit proof that it is the 
correct one also in the case of Reggeon exchange, although it seems
quite plausible. These two issues are currently 
under investigation,\footnote{
 A first analysis indicates that the basic formula
is essentially correct, apart from minor modifications which do not 
change the results on the intercept and the slope of the Reggeon
trajectory; moreover, the analytic continuation required to
obtain the physical amplitude turns out to be essentially the same as in
the Pomeron--exchange case. A detailed study of these
issues is delayed to a future publication~\cite{Giordano_wp1}.} and
since the approach described in this Section appears to be basically
correct, we will use it as the starting point for our analysis.

\subsection{Gauge/gravity correspondence and minimal surfaces}

The following step is the application of the gauge/gravity
correspondence, which, as discussed in Section \ref{AdSBH}, allows to
write the Wilson--loop expectation value as
\begin{equation}
  \label{eq:wloop_conf}
   \la \W[{\cal C}]\ra = {\cal F}[{\cal C}]\ 
   e^{-\f{1}{2\pi\alpha'_{\rm eff}}A_{\rm  min}[{\cal C}]}\, ,
\end{equation}
where $A_{\rm  min}[{\cal C}]$ is the area of the minimal surface
having the contour ${\cal C}$ as boundary, and ${\cal F}[{\cal C}]$
contains the contributions of fluctuations around this surface. In
this work we stick to the ``quenched'' approximation,\footnote{
Expression \eqref{eq:wloop_conf} does not contain the effect of
dynamical fermions, 
which are subleading at large $N_c$. A way to include such effects,
going beyond the ``quenched'' approximation, has been suggested
in~\cite{Armoni}, making use of the world--line formalism to express
the fermion--matrix determinant as a sum of Wilson loops over all
possible contours.} while loop corrections will be considered in a future
work~\cite{Giordano_wp2}. 
 
At this point, one should in principle solve the Plateau problem in a
curved background for a general boundary, and then integrate over all
possible boundaries: this is a formidable task, which is currently out
of reach. In order to simplify the problem, it is useful to recall the
physical picture of the process, already discussed above, and sketched
in Fig.~\ref{fig:1}. Before and after the interaction, the partons
travel along their eikonal, straight--line trajectories, and during
the time of interaction the light quarks are exchanged between the two
mesons. Translating this picture to Euclidean space, we then expect
that the main contributions to the path integral come from those paths
${\cal C}_\pm$ which away from the central (interaction) region are
straight lines, coinciding with the eikonal trajectories of the light
quarks. As a consequence, the relevant minimal surfaces are essentially
made up of a central strip (corresponding to region $\Sigma$ in
Fig.~\ref{fig:1bis}), bounded by the curved part of the light--quark
trajectories, which corresponds to the exchanged Reggeon, and four
rectangles (regions $r_{1,2,3,4}$ in Fig.~\ref{fig:1bis}),
corresponding to the free propagation of mesons before and after the
interaction.   

In the case that we are considering,
namely small dipole sizes corresponding to heavy mesons, the part of
the minimal surface corresponding to these rectangles is determined by
the near--boundary behaviour of the metric, reaching a maximal value
$z_{\rm max}\sim {\cal O}(|\vec{R}_{1,2}|)$ in the $z$--direction, and
thus not feeling the confinement scale, see Fig.~\ref{config}
(right). On the other hand, for large enough $b$, in the central
region we can use the approximation scheme 
discussed in Section \ref{AdSBH}. In this region the minimal surface
is expected to be made up of an almost vertical wall of area $A_{\rm
  wall}$, extending from the boundary of AdS up to the region where
the metric is effectively flat (e.g., the black--hole horizon of
Ref.~\cite{AdSBH}), and a minimal surface living in the effectively
flat metric, bounded by the light--quark trajectories transported from
the boundary of AdS to the effectively flat region, see
Fig.~\ref{config} (left).

Within this configuration, the geometry of the flat part of the
Reggeon strip is governed by the (almost) infinite straight lines
corresponding to the eikonal trajectories of the heavy quarks,
transported to the effectively flat region. This suggests that the
relevant contributions come from configurations in which the floating
boundaries lie on the corresponding {\it helicoid}.  
Indeed, the helicoid has been recognised as the minimal surface
associated with {\it soft} elastic quark--quark (and also quark--antiquark)
scattering at high energy~\cite{JP2}.  This assumption is expected to
be sensible only for small quark mass (more precisely for small {\it
  constituent} quark mass, see footnote
\ref{foot:constmass}), as we will discuss further on. We
then recover the same 
basic geometry already found in the treatment of Pomeron exchange, the
difference being the presence of partially floating, instead of fully
fixed boundaries.  

Notice that since we are considering the case $|\vec{R}_{1,2}|\ll b$,
we can neglect the size of the dipole in the interaction region, so
that the eikonal trajectories of the light and heavy quarks coincide
at the given level of approximation.  
Therefore, to first order the flat part of the ``strip'' $\Sigma$
takes the form  
\begin{equation}
  \label{eq:helico}
  \begin{aligned}
&X^{\rm hel}(\bar{\tau},\sigma) = \left(
\cos\left(\f{\theta\sigma}{b}\right)\bar{\tau}\,,\,
\sin\left(\f{\theta\sigma}{b}\right)\bar{\tau}\,,\,      
\f{\vec{b}}{b}\, \sigma \right) 
\, , 
\\
& \sigma\in [-b/2,b/2] 
\,,\quad
\bar{\tau}\in[-\tau^-(\sigma),\tau^+(\sigma)]\,,
\quad\tau^\pm(\sigma)\ge 0 \,. 
  \end{aligned}
\end{equation}
The path--integral is then reduced to the integration over
the curved part of the light--quark trajectories, constrained now to
lie on the helicoid, i.e., over the ``profiles'' $\tau^\pm(\sigma)$
which constitute the boundary of the relevant piece of helicoid; the
remaining parts of the paths ${\cal C}_\pm$ lie on the eikonal
light--quark trajectories. Notice that for any choice of
$\tau(\sigma)$ in Eq.~\eqref{eq:helico}, the resulting surface is
automatically a minimal surface in flat space, i.e., a surface with
zero mean curvature. 

The remaining part of the minimal surface is made up of the vertical
wall and of the four rectangles. In turn, the vertical wall is made of
four pieces, corresponding to the paths ${\cal C}_\pm$  and to those
pieces of the paths ${\cal C}_{1,2}$ bounding the interaction region
(i.e., between the dashed lines in Fig.~\ref{fig:1bis}). The
rectangles are deformed in the region where they connect to the
interaction region (near the dashed lines in Fig.~\ref{fig:1bis}),
where the surface rises steeply to the 
effectively flat region; nevertheless, the area of these regions is
proportional to $|\vec{R}_{1,2}|$, and can be neglected.

In this approximation, the dependence of the minimal surface on the 
orientation of the dipoles is trivial, as anticipated. Moreover, in
this case the spin factor simplifies considerably, and it can be
explicitly evaluated.\footnote{The first calculation of
  Ref.~\cite{Jani} has been redone with a different 
  result, see Appendix \ref{app:A}.} The details of the
calculation are given in Appendix \ref{app:A}, here we quote only the
final result, 
\begin{equation}
  \label{eq:spin_fact_eval_0}
{\cal I}[{\cal C}_i] = {\cal N}_i \  P(\dot{X}_i(\nu_f))\, {\cal U}_i\,
    P(\dot{X}_i(\nu_i)) \,, 
\end{equation}
where the various quantities are defined as follows,
\begin{equation}
  \label{eq:spin_fact_eval_1}
  \begin{aligned}
      P(n) &= \f{1+\slan}{2}\,,  \qquad 
    {\cal N}_i = \left(\f{1 + \dot{X}_i(\nu_f)\cdot
        \dot{X}_i(\nu_i)}{2}\right)^{-\f{1}{2}}\,,\\
{\cal U}_i &=  {\rm diag}\left(e^{-\f{i}{2}\Phi({\cal
          C}_{\vec{u}_i})}, e^{\f{i}{2}\Phi({\cal
          C}_{\vec{u}_i})},e^{\f{i}{2}\Phi({\cal
          C}_{\vec{u}_i})},e^{-\f{i}{2}\Phi({\cal C}_{\vec{u}_i})}\right)\,.
  \end{aligned}
\end{equation}
Here ${X}_i(\nu_{i,f})$ are the initial and final points of ${\cal
  C}_i$, and the (real) phases $\Phi({\cal C}_{\vec{u}_i})$, which depend on
the shape of the path, are given in Appendix~\ref{app:A}.

Since the paths $X_{1,2}$ are fixed straight lines, and moreover, for
the relevant paths, $X_\pm$ lie on the eikonal trajectories of the
light quarks near the initial and final points, it is possible to
factor out of the path integral the quantities ${\cal
  N}\equiv\prod_{i=\{1,2,+,-\}}\,{\cal N}_i$ and
$\Omega^{(i,f)}\equiv\otimes_{i=\{1,2,+,-\}}\,P(\dot{X}_i(\nu_{i,f}))$. 
Denoting in short
${\cal U}=\otimes_{i=\{1,2,+,-\}}\,{\cal U}_i$, we have
\begin{multline}
  \label{eq:sym_PI_2}
  \tilde{a}(\vec{b},\theta,T) \sim 
  \, {\cal Z}^{-1} {\cal N} \Omega^{(f)}\bigg\{\int {\cal
    D}\tau^+ {\cal D}\tau^- 
  \, {\cal F}[\tau^+,\tau^-] e^{-\f{1}{2\pi\alpha'_{\rm
        eff}}\{A_{\rm min}^{\rm hel}[\tau^+,\tau^-]+
    A_{\rm rect} + A_{\rm  wall}\}} \\ \times e^{-m_0\{ 
L^{\rm hel}[\tau^-] + L^{\rm hel}[\tau^+] + 4T -L_0[\tau^-]
-L_0[\tau^+]\}}\  {\cal U}[\tau^+,\tau^-] \bigg\}\ \Omega^{(i)}
\, ,
\end{multline}
where the appropriate contraction of indices among $\Omega^{(f)}$,
${\cal U}$ and $\Omega^{(i)}$ is understood, and we have made explicit
the dependence of ${\cal U}$ on $\tau^\pm$. 
The area $A_{\rm min}^{\rm hel}$ of the helicoidal ``Reggeon strip'',
and the length $L^{\rm hel}$ of the boundaries, can now be written
explicitly as functionals of $\tau^\pm(\sigma)$, 
\begin{equation}
  \label{eq:functional}
  \begin{aligned}
      A_{\rm min}^{\rm hel}[\tau^+,\tau^-] &=
      \int_{-\f{b}{2}}^{+\f{b}{2}}d\sigma\,\int_{-\tau^-(\sigma)}^{+\tau^+(\sigma)}dx\,
      \sqrt{1+(px)^2}\, 
      ,\\ 
      L^{\rm hel}[\tau^\pm(\sigma)] &=
      \int_{-\f{b}{2}}^{+\f{b}{2}}d\sigma\,
      \sqrt{1+(p\tau^\pm(\sigma))^2+(\dot{\tau}^\pm(\sigma))^2}\, .
  \end{aligned}
\end{equation}
where we have used the notation
\begin{equation}
  p = {\theta}/{b}\, . 
\end{equation}
In \eqref{eq:sym_PI_2}, the contributions $A_{\rm rect}$ 
and $A_{\rm  wall}$ correspond to the four rectangles $r_{1-4}$ (see 
Fig.~\ref{fig:1bis})  and to the vertical wall, respectively. 
Moreover, $2T -L_0[\tau^\pm]$ is the length of the straight--line part
of the light--quark trajectories, with $L_0$ depending only on the
endpoints,  
\begin{equation}
  \label{eq:subtr}
  L_0[\tau^\pm] = \tau^\pm\left({b}/{2}\right)+\tau^\pm\left(-{b}/{2}\right)\, .
\end{equation}
The expression Eq.~\eqref{eq:sym_PI_2} is almost the final answer, but
we still have to deal with infrared problems. 
It is immediate to see that there are two possible
sources of infrared singularities, 
which should be removed by the normalisation constant ${\cal Z}$. The first
one comes from the area of the rectangles, and can be removed by including
in ${\cal Z}$ the quantity
\begin{equation}
  \label{eq:norm}
  {\cal Z}_1 = \la \W_1 \ra \la \W_2 \ra \, ,
\end{equation}
where $\W_{1,2}$ are the rectangular Wilson loops describing the free
propagation of the mesons (see Fig.~\ref{fig:1bis}). Such a term plays
the role of renormalisation constant for the dipoles in a LSZ approach
to dipole--dipole scattering~\cite{DFK}, and in the gauge/gravity
duality approach it is given by
\begin{equation}
  \label{eq:norm_2}
 {\cal Z}_1 \simeq e^{-\f{1}{2\pi\alpha'_{\rm eff}} A_{\rm rect}'}\, ,
\end{equation}
where $A_{\rm rect}'$ is the sum of the areas of the minimal surfaces
corresponding to the Wilson loops $\W_{1,2}$. 
For almost vertical walls, the difference $A_{\rm wall}+A_{\rm
  rect}-A_{\rm rect}'$ is approximately of the form
\begin{equation}
\label{eq:walls}
\begin{aligned}
\f{1}{2\pi\alpha'_{\rm eff}}(A_{\rm wall}+A_{\rm rect}-A_{\rm rect}')
& \simeq  \delta m \left(L^{\rm hel}[\tau^-] + L^{\rm 
  hel}[\tau^+]-L_0[\tau^-]-L_0[\tau^+]\right) \\ & \phantom{\simeq} + \delta c
\left(L_0[\tau^-]+L_0[\tau^+]\right) \, ,
\end{aligned}
\end{equation}
and so it is independent of $T$. The UV divergencies coming from the
part of the surface near the boundary $z=0$ are contained in $\delta
m$ which, as we will see in a moment, amounts simply to a renormalisation of
the (bare) mass parameter $m_0$. The second term originates from the
incomplete cancellation between the area of the rectangles in the
central region (i.e., the region between dashed and dotted lines in
Fig.~\ref{fig:1bis}). The quantity $\delta c$ is a UV--finite
quantity, which at the present stage we are not able to compute 
explicitly. However, this term would not affect the variational
problem: indeed, it depends on the light--quark trajectories only
through $L_0[\tau^\pm]$, which, as we will see in the next subsection,
does not enter the minimisation procedure. Therefore, the contribution
of this term to the effective action (in the saddle--point
approximation considered in this paper) could be determined if $\delta
c$ were known. For this reason, we will discuss the possible role of
this term in footnotes \ref{foot:c1}, \ref{foot:c2} and \ref{foot:c3},
dropping it from the main derivation. 

The second source of IR singularities is the length of
the straight--line part of the light--quark trajectories, which for
the dominant paths is expected to be of order
$4T+ {\cal O}(1)$ for large $T$, so that after analytic continuation
$T\to iT$ we would get an infinite phase $\sim e^{i4m_0T}$. 
This phase corresponds to the self--interaction of the quarks, which
plays no role in the scattering process, and has therefore to be
removed. We then insert a second factor
\begin{equation}
  \label{eq:norm_3}
  {\cal Z}_2 = e^{-4m_0T}\, ,
\end{equation}
which accomplishes this task already at the Euclidean level. 
The normalisation constant is then taken to be ${\cal Z}={\cal
  Z}_1{\cal Z}_2$.  
All in all, we obtain for the Euclidean amplitude
\begin{multline}
  \label{eq:sym_PI_3}
  \tilde{a}(\vec{b},\theta,T) \sim {\cal N}\Omega^{(f)} \bigg\{
\int {\cal D}\tau^+ {\cal D}\tau^-
  \,{\cal F}[\tau^+,\tau^-] e^{-\f{1}{2\pi\alpha'_{\rm
        eff}}A_{\rm  min}^{\rm hel}[\tau^+,\tau^-]} \\ \times e^{-m\{
    L^{\rm hel}[\tau^-] + L^{\rm hel}[\tau^+]-L_0[\tau^-]
    -L_0[\tau^+]\}}\,{\cal U}[\tau^+,\tau^-]\bigg\} \Omega^{(i)} \, ,  
\end{multline}
where we have reabsorbed the contribution of the first term in
Eq.~\eqref{eq:walls} in a renormalisation of $m_0$, namely $m\equiv
m_0+\delta m$.  
As we show in Appendix \ref{app:B}, the bispinors are eigenvectors with 
eigenvalue 1 of the (analytic continuation of the) projectors
$P(\dot{X}_i(\nu_{i,f}))$ acting on them, and so we can replace
$\Omega^{(i,f)}$ in Eq.~\eqref{eq:sym_PI_3} with the identity. On the
other hand, the phase factors $e^{\pm\f{i}{2}\Phi({\cal
    C}_{\vec{u}_i})}$  contained in  ${\cal U}$ (see
Eq.~\eqref{eq:spin_fact_eval_1}) do not cancel, and their effect has
to be properly taken into account.

\subsection{Saddle--point approximation}

As anticipated, the final step is a saddle--point approximation of
\eqref{eq:sym_PI_3}: exploiting the symmetry of the
configuration in order to restrict to the case
$\tau^+(\sigma)=\tau^-(\sigma)\equiv \tau(\sigma)$,  one has to solve
the Euler--Lagrange equations $\delta S_{{\rm eff,\,E}}[\tau_{\rm
  s.p.}(\sigma)] = 0$, to find the profile $\tau_{\rm
  s.p.}(\sigma)$ which minimises the ``effective action'',
\begin{equation}
  \label{eq:variational}
  \begin{aligned}
    S_{{\rm eff,\,E}}[\tau(\sigma)] & \equiv \f{1}{2\pi\alpha'_{\rm
        eff}}A_{\rm  min}^{\rm hel}[\tau(\sigma)] + 2m(L^{\rm
      hel}[\tau(\sigma)]-L_0[\tau(\sigma)]) 
    \, , 
  \end{aligned}
\end{equation}
where, with a slight abuse of notation, we 
have avoided the repetition of the argument in the area
functional. In the general
case, the variational problem defined in Eq.~\eqref{eq:variational} is
aimed at the determination of an ``optimal'' boundary, involving in
the minimisation procedure both the area of the resulting surface and
the length of the boundary. This is what we have called ``minimal
surface problem with floating boundaries'' in the
Introduction.\footnote{Although somewhat similar, this problem must not
be confused with the ``minimal surface problem with partially free
boundary'', known in the mathematical literature (see
e.g.~\cite{min_surf}). In that case, part of the boundary is not
competely fixed, but only constrained to lie on a given surface, as in
our problem; however, only the area of the surface enters the
minimisation procedure, and not the length of the boundary.}

Substituting the solution $\tau_{\rm s.p.}(\sigma)$ in
Eq.~\eqref{eq:sym_PI_3} we obtain for the Euclidean amplitude  
\begin{equation}
  \tilde{a}(\vec{b},\theta,T) \sim  {\cal N}\,{\cal U}[\tau_{\rm s.p.}(\sigma)]
{\cal F}[\tau_{\rm s.p.}(\sigma)]e^{-\f{1}{2\pi\alpha'_{\rm
      eff}}A_{\rm 
      min}^{\rm hel}[\tau_{\rm s.p.}(\sigma)]} e^{-2m\{L^{\rm
      hel}[\tau_{\rm s.p.}(\sigma)]-L_0[\tau_{\rm s.p.}(\sigma)]\}} \, ,  
\end{equation}
where we have avoided the repetition of the argument also in the
fluctuation functional and in ${\cal U}$,  and we have dropped the
$\Omega^{(i,f)}$.   
The Reggeon trajectory is now encoded in the 
solution $\tau_{\rm s.p.}$ of the saddle--point equation: a detailed
study of this equation is the subject of the rest of this paper.
Since there is no possibility of confusion, in the following we will drop
the subscript ``s.p.'' to keep the notation simpler.

A few remarks are now in order. 
In~\cite{Jani} it was considered explicitly only the case of massless
light--fermions. It is immediate to see that in this case a real
solution of the Euler--Lagrange equations does not exist, and the
minimal value of $S_{{\rm eff,\,E}}$ for real $\tau(\sigma)$ is obtained for
$\tau(\sigma)\equiv 0$, 
corresponding to a strip of vanishing width connecting the central
points of the eikonal trajectories, for which $S_{{\rm eff,\,E}}=2mb$. 
The reason for this can be easily understood. The area
term in the effective action is of ``attractive'' nature for the
boundaries, since it tends to bend inwards the boundaries in order to
minimise the area in between them. On the contrary, the length term is
of ``repulsive'' nature, since it tends to minimise the curvature of
the boundaries, in order to minimise their lengths. The value of the
quark mass sets the rigidity of the boundaries, and if $m=0$ there is
nothing preventing the boundaries to collapse to a strip of vanishing
width. In order to start with  a non trivial real solution, before
analytic continuation, we thus cannot ignore the effect of the
length term.  

This qualitative argument applies also in a more general setting, with
the floating boundaries not constrained to lie on a specific
surface. In this setting, we expect that when the quark mass is
large the minimisation procedure is dominated by the length term, so
that the floating boundaries tend to become straight lines, and the
helicoid geometry is lost. As a consequence, the approximation
considered here is expected to be valid only for small quark mass.

The solution of the saddle--point equations at $m=0$, found
in~\cite{Jani}, is indeed a complex solution; more precisely, it is
a purely imaginary constant trajectory $\tau(\sigma)\equiv \pm
i/p$. Using this solution, and choosing the minus sign for physical
reasons, one obtains, after analytic continuation, a Gaussian
impact--parameter amplitude, which in turn yields a linear Reggeon
trajectory. However, one can immediately check that this solution
corresponds to a singular point of the area functional, which reads 
explicitly
\begin{equation}
  \label{eq:area_expl}
  A^{\rm hel}[\tau(\sigma)] = \f{1}{p}\int_{-\f{b}{2}}^{\f{b}{2}}
  d\sigma \left[p\tau(\sigma)\sqrt{1+\left(p\tau(\sigma)\right)^2} +
    \asinh p\tau(\sigma) 
\right]\, ,
\end{equation}
so that the applicability of the saddle--point method is not guaranteed.  
In order to investigate this problem more rigorously, it is convenient
to start from the case $m\neq 0$, where regular real solutions can be
found in Euclidean space: the limit $m\to 0$ will be considered only
after the analytic continuation into Minkowski space--time has been
performed.

In the next Sections we will study in details the variational problem
at hand, which involves the minimisation of a functional which
contains both an area and a length term. In particular, in the next
Section we will study a simpler case, where we can determine exactly
and explicitly the solution, in order to obtain a few insights on this
kind of problem. The case relevant to Reggeon exchange is discussed in
Section \ref{Reggeon}, where we provide an exact solution in implicit
form for the general case, and an approximate solution in explicit
form in two specific regimes.

\section{Warm--up exercise: soap film with floating boundaries}
\label{planar}

Before attacking the minimisation problem relevant to Reggeon
exchange in full generality, we want to discuss a simpler case, namely
the case in which the straight lines forming the fixed part of the
boundary are parallel, i.e., 
$\theta=0$. 
This configuration is of limited interest for our problem, since our
purpose is to obtain an analytic dependence on $\theta$; nevertheless,
the mathematical problem is similar, and moreover in this case the
variational equations can be solved explicitly, so that we can obtain
a few indications in the study of the more complicated ``tilted''
case $\theta\neq 0$. 
We consider then the minimisation of the functional
\begin{equation}
  H = \f 1{2\pi \hat \alpha'} A[{\cal C}_1,{\cal C}_2] + \hat m
  (L[{\cal C}_1]+L[{\cal 
    C}_2])\, , 
\end{equation}
where $A$ is the area of a surface bounded on two opposite sides by
two parallel straight 
lines of length $2T$ at a distance $R$, which are held fixed. 
On the other sides, the surface is bounded by two {\it a priori} free
lines following the paths ${\cal C}_{1,2}$, of length $L[{\cal  
  C}_{1,2}]$, which have to be determined by the minimisation
procedure.  

\begin{figure}[t]
  \centering
  \includegraphics[width=0.65\textwidth]{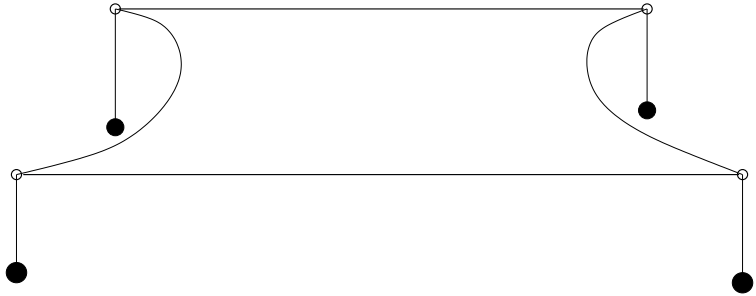}
  \caption{{\it Soap film with partially floating 
      boundary.} The straight lines
    correspond to rigid rods, the curved lines to flexible wires,
    attached to four equal masses (black balls).}
  \label{fig:0}
\end{figure}

For want of a physical interpretation, this functional 
corresponds to the energy of an ideal soap 
film of vanishing mass and of surface tension $1/{2\pi \hat \alpha'}$,
extending between two rigid rods (the straight lines) parallel to the
ground, and between two flexible (massless) wires (of length larger
than $2T$), each passing through two rings positioned at the endpoints
of the rods (see Fig.~\ref{fig:0}); moreover, two equal masses $M$ are
attached at the endpoints of each wire, with $Mg=\hat m$, and their
potential energy in the gravitational field contributes the length
term. 

\begin{figure}[t]
  \centering
  \includegraphics{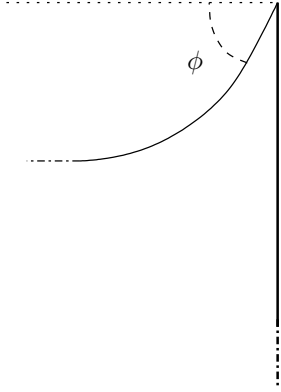}
  \caption{{\it The angle $\phi$ formed by the fixed and the floating
      boundaries at the soap film corner.} The angle is given by the
    tangent of $\tau(\sigma)$ at $\sigma= R/2$, cfr. 
    Eq.~\eqref{eq:str_brk_geo}.}  
  \label{fig:cusp}
\end{figure}

Given the symmetries of the problem, the solution will be a planar
surface, and the two floating boundaries will be one the reflection
of the other. The problem is thus effectively two--dimensional, and we
can parameterise the relevant surfaces in terms of a single function
$\tau(\sigma)$, i.e., 
\begin{equation}
  \label{eq:surface_plan}
  X^{\rm plan}[\tau(\sigma);\bar{\tau},\sigma] = (\bar{\tau},\sigma)\, , \quad  
  \sigma\in[-R/2,R/2]\,, \,\, 
  \bar{\tau}
  \in[-\tau(\sigma),\tau(\sigma)]\,, \,\, \tau(\sigma)>0\, .
\end{equation}
The expression of the functionals simplifies therefore to
\begin{equation}
\begin{aligned}
  L &= \int_{-\f{R}{2}}^{+\f{R}{2}}d\sigma
  \sqrt{1+(\dot{\tau}(\sigma))^2}\, , \qquad 
  A = 2\int_{-\f{R}{2}}^{+\f{R}{2}}d\sigma\ \tau(\sigma)\, .
\end{aligned}
\end{equation}
Notice that $\tau$ must satisfy $\tau(\sigma)=\tau(-\sigma)$ because
of the symmetries of the problem.
The Euler--Lagrange equation is easily derived, and reads
\begin{equation}
\label{var_1}
 \ddot{\tau} -  2R_c^{-1}\left(1+\dot{\tau}^2\right)^{\f{3}{2}}=0\, , 
\end{equation}
where the combined parameter 
\begin{equation}
\label{Rc}
R_c\equiv 4\pi \hat \alpha' \hat m \, ,
\end{equation}
will play an important  role as a critical value for $R$ in the
minimisation problem. Notice that for $R_c>0$ we have $\ddot{\tau}>0$.  
This equation reflects the general expectation on the nature of the
two terms contributing to the energy functional, discussed in the
previous Section. For large $R_c$ the first ``length'' term in
\eqref{var_1} dominates, so that the equation reduces to that of a
straight line; the second ``area'' term increases the curvature of the
free boundary, bending it inwards. This equation is solved in the
standard way by setting\footnote{The 
  equation could have been solved by direct integration, but the
  present approach generalises immediately to non--constant $R_c$,
  which will be relevant in the next Section.} 
\begin{equation}
  v(\tau) =
  \sqrt{1+\dot{\tau}^2}\, , \qquad v' \equiv \f{dv}{d\tau} =
  (\dot{\tau})^{-1}\f{dv}{d\sigma} = 
  \f{\ddot{\tau}}{v}\, . 
\end{equation}
The equation becomes then
\begin{equation}
  v'  = 2R_c^{-1} v^2 \, , 
\end{equation}
which is easily solved by
\begin{equation}
  v(\tau) = \f{v(\tau_0)}{1 - 
    2R_c^{-1}v(\tau_0)(\tau\!-\!\tau_0)} = \f{1}{1 -  2R_c^{-1}(\tau-\tau_0)}\, ,
\end{equation}
where $\tau_0=\tau(0)$, and we have taken into account that $v(\tau_0)
= \sqrt{1+(\dot{\tau}(0))^2} = 1$, since $\dot{\tau}(0)=0$. 
Notice that, since $1 \le v \le \infty$, 
we have to satisfy $0\le (\tau-\tau_0) \le R_c/2$.
\begin{figure}[t]
  \centering
  \includegraphics[width=0.7\textwidth]{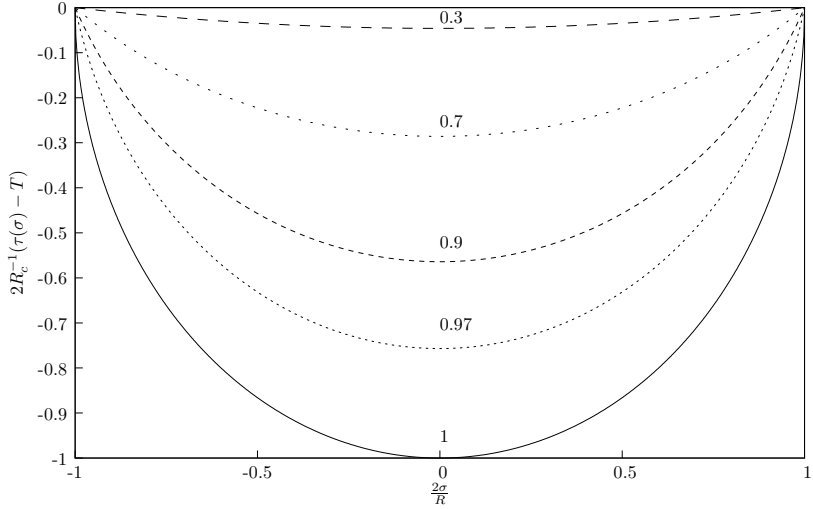}
  \caption{{\it Minimisation profile of the floating boundary.} Half of
    the floating boundary is represented after minimisation for
    various values of $R/R_c$. The opposite half of the floating
    boundary is obtained by reflection with respect to the horizontal
    axis.} 
  \label{fig:prof}
\end{figure}
Proceeding in the usual way, we write 
\begin{equation}
\label{eq:thetazero_sol}
  |\sigma| = \int_{\tau_0}^{\tau(\sigma)} \f{d\tau}{\sqrt{v^2-1}} =
   \sqrt{\left(\tau(\sigma)\!-\!\tau_0\right)
    \left[R_c-\left(\tau(\sigma)\!-\!\tau_0\right)\right]}
  \, ,  
\end{equation}
which can be inverted to give
\begin{equation}
  \tau(\sigma)-\tau_0 =
  \f{R_c}{2}\left[1-\sqrt{1-\left(\f{2\sigma}{R_c}\right)^2}\,\right]\, ,
 \end{equation}
where the minus sign for the square root has been chosen since the
left--hand side of the equation vanishes at $\sigma=0$. 
Finally, imposing the boundary condition $\tau(R/2)=T$, we obtain
the complete solution\footnote{
As we will discuss in detail in the next Section, this is not the
boundary condition that we impose in the $\theta\ne 0$ case, so this
result does not give the $\theta\to 0$ limit of the calculation
relevant to Reggeon exchange.} (see Fig.~\ref{fig:prof})
\begin{equation}
\label{eq:taumin}
  \tau_{min}(\sigma) = T -
  \f{R_c}{2}\left[\sqrt{1-\left(\f{2\sigma}{R_c}\right)^2}-
    \sqrt{1-\left(\f{R}{R_c}\right)^2}\/ \right]\, .
\end{equation} Notice that in order for $\tau(\sigma)$
to be real we need the following condition to be satisfied: 
\begin{equation}
\label{eq:string_brk}
  R \le R_c \, . 
\end{equation}
This justifies the notation chosen for $R_c\equiv 4\pi \hat \alpha'
\hat m$. 
The geometric meaning of this condition is clarified by computing
the derivative of $\tau(\sigma)$ at $\sigma= R/2$,
\begin{equation}
  \label{eq:str_brk_geo}
  \dot{\tau}(\sigma)\vert_{\sigma={R}/{2}} = 
  \f{{R}/{R_c}}{\sqrt{1-
      \left({R}/{R_c}\right)^2}}  =  \tan\phi\, ,
\end{equation}
where the angle $\phi$ is shown in Fig.~\ref{fig:cusp}. It is then
immediate to see that the condition Eq.~\eqref{eq:string_brk} simply
means that $\phi\le\pi/2$, and when the bound is reached the flexible
wire runs parallel to the rigid rod at the junction point.  The
physical interpretation of this condition will be discussed in a
moment. 

We turn now to the computation of the energy corresponding to 
the solution $\tau_{min}$. 
Making use of the properties of the minimal solution, it can be
expressed in a simple form, which is readily evaluated:
\begin{equation}
\label{eq:thetazero_energy}
  \begin{aligned}
    H_{min} &=  \f 2{\pi \hat \alpha'}\int_0^{\f{R}{2}} d\sigma \left\{ \tau_0 +
      \f{R_c}{2}\left(1+
        \f{\left(2R_c^{-1}\sigma\right)^2}{\sqrt{1-
            \left(2R_c^{-1}\sigma\right)^2}}\right)  
    \right\} = \\ 
    &= \f 1{2\pi \hat \alpha'}\left\{2 T R + \f{RR_c}{2}
    \sqrt{1-\left(\f{R}{R_c}\right)^2} +
    \f{R_c^2}{2}\asin \f{R}{R_c}\right\} 
    \, .
\end{aligned}
\end{equation}
In the limit $R_c\to\infty$, the weight of the masses attached to the
wires wins over the surface tension, so that the wires are kept straight,
and we recover the result for a soap film with a fixed rectangular
boundary, i.e., $\tau_{min}\to T$, and 
\begin{equation}
 H_{min} \simeq \f 1{2\pi \hat \alpha'}\{2T R +  RR_c\}\, .
\end{equation}
At fixed $R$ we cannot go to the limit $R_c\to 0$, since in this case
Eq.~\eqref{eq:string_brk} implies that $R_c$ must be bounded from
below. The meaning of Eq.~\eqref{eq:string_brk} 
is the following: if $R$ exceeds the critical value $R_c$ at fixed
$\hat{\alpha}$ and $\hat{m}$ (more precisely, at fixed
$\hat{\alpha}\hat{m}$), or equivalently  if $R_c$ becomes smaller than
$R$ (e.g. for too large surface tension or too small mass), 
the force due to the surface tension is stronger than the
gravitational force on the masses, and it makes the soap film
collapse. This is essentially a Gross--Ooguri transition~\cite{GO},
which we expect to find also in the case $\theta\neq 0$, and $R_c$
appears to be the corresponding critical value at which the transition
takes place. Notice that rewriting the energy in terms of the angle 
$\phi$ as 
\begin{equation}
  \label{eq:energy_2}
   H_{min} = \hat m\left\{4T\sin\phi +
     R_c\left(\phi +  \sin
       \phi\cos
       \phi\right)\right\}\, ,
\end{equation}
we easily see that the maximal value is reached for $\phi=\pi/2$,
i.e., right before the collapse, so
that the maximal energy that can be stored in this system is $\hat
m\left\{4T +\pi/2\ R_c\right\}$. 

As a final remark, we note in passing that identifying $\f 1{2\pi\hat
  \alpha'}$ with the string tension $\sigma$ and $\hat m$ with the
{\it constituent} quark mass $m$ (the functional $H$ having now the
dimensions of an action), the minimisation procedure reproduces the
static $Q-\barQ$ linear potential $V_{Q\barQ}(R)=\sigma R$ as
$V_{Q\barQ}(R)=H_{min}/2T$ for $T\to\infty$, and moreover the bound
$R\le R_c$ gives the well--known string--breaking condition $\sigma
R\le 2m$.    

\section{Variational problem for the  Reggeon--exchange amplitude
in Euclidean space}
\label{Reggeon}

In this Section we discuss the Euclidean variational problem relevant
to the Reggeon--exchange amplitude, i.e., for the ``tilted''
configuration of Fig.~\ref{fig:1}. As we have already said in Section 
\ref{saddle}, the scattering amplitude is reconstructed through
analytic continuation from the solution of the minimal surface problem
with floating boundaries, that involves the ``effective action''
functional Eq.~\eqref{eq:variational}, which we rewrite here for
convenience: 
\begin{equation}
  \label{eq:variational_2}
    S_{{\rm eff,\,E}}[\tau(\sigma)]  = \f{1}{2\pi\alpha'_{\rm
        eff}}A_{\rm  min}^{\rm hel}[\tau(\sigma)] + 2m(L^{\rm
      hel}[\tau(\sigma)]-L_0[\tau(\sigma)]) \, .
\end{equation}
Our aim is to find a smooth ``profile'' $\tau(\sigma)$, bounding a
piece of helicoid which connects two straight lines at a transverse
distance $b$, and forming an angle $\theta$ in the longitudinal
($x_4-x_1$) plane. In order to do so, it is convenient to pass to
dimensionless coordinates by making the change of variables 
\begin{equation}
  t(s)=p\tau(\sigma),\quad s=p\sigma,\quad px=y\, ,\quad {\rm with}\quad
  p=\theta/b\, . 
\end{equation}
Note  that $\dot{t}\equiv\f{dt}{ds}=\f{d\tau}{d\sigma}=\dot{\tau}.$ In
terms of these reduced variables, the expressions for the area and
length functionals Eq.~\eqref{eq:functional} read 
\begin{equation}
  \begin{aligned}
    A_{\rm min}^{\rm hel} &=  \f{1}{p^2}
    \int_{-\f{\theta}{2}}^{+\f{\theta}{2}}ds\,\int_{-t(s)}^{+t(s)}dy\,\sqrt{1+y^2}\,  ,
    \\ 
    L^{\rm hel} &=
    \f{1}{p}\int_{-\f{\theta}{2}}^{+\f{\theta}{2}}ds\,\sqrt{1+[t(s)]^2+
      [\dot{t}(s)]^2}\, 
    , 
  \end{aligned}
\end{equation}
and moreover
 \begin{equation}
   \label{eq:subtr_2}
   L_0 = \f{1}{p}\left[t\left(\textstyle\f{\theta}{2}\right) +
     t\left(\textstyle -\f{\theta}{2}\right) \right]\,,
 \end{equation}
for the subtraction term. This term will not enter the variational
equations, since  the value of $\tau(\pm\f{b}{2})$, and so that of
$t(\pm\f{\theta}{2})$,  is determined by requiring a smooth transition
to the eikonal straight--line paths: in other words, we perform the
variation of the effective action at $t(\pm\f{\theta}{2})$ 
fixed, we solve the equation and we subsequently determine the value
which makes the path smooth. In terms of our parameterisation, in
order for the part of the path on the helicoid to be smoothly 
connected with the incoming and outgoing straight lines, we need
that $\dot{t}(\pm\f{\theta}{2})=\pm\infty$. We will discuss this point in more
detail after solving the general equation.

\subsection{Exact solution in implicit form}

It is straightforward to obtain the Euler--Lagrange equation
corresponding to the minimisation of the functional, 
which reads explicitly
\begin{equation}
  \f{2m}{p}\f{1}{(1+t^2+\dot{t}^2)^{\f{3}{2}}}[(\ddot{t}-t)(1+t^2)-2t\dot{t}^2]
  - \f{1}{\pi\alpha_{\rm eff}'p^2}\sqrt{1+t^2}=0\,. 
\end{equation}
After setting
\begin{equation}
  \label{eq:change}
t(s)  = \f{\theta}{b}\tau(\sigma) \equiv \sinh\vphi(s)
\, , \qquad \displaystyle
\lambda\equiv\f{1}{2\pi\alpha_{\rm eff}'mp} = 
\f{b}{2\pi\alpha_{\rm eff}'m\theta}\, ,  
\end{equation} the equation takes the simpler form
\begin{equation}
\label{eq:eq_phi}
  \ddot{\vphi} - (1+\dot{\vphi}^2)\tanh\vphi -
  \lambda(1+\dot{\vphi}^2)^{\f{3}{2}}\cosh\vphi=0\, .
\end{equation}
In some loose sense, the variable $\vphi$ parameterises in a
scale--invariant way the development in ``time'' of the quark--exchange
process in Euclidean space.

As shown in the previous Section, in order to solve this equation one sets
\begin{equation}
  v(\vphi) = \sqrt{1+\dot{\vphi}^2} \longrightarrow vv'=\ddot{\vphi}\, ,
\end{equation}
where the prime denotes differentiation with respect to $\vphi$. The
equation becomes then 
\begin{equation}
\label{eq:diffv}
  v' - v\tanh\vphi  - \lambda v^2\cosh\vphi=0\, ,
\end{equation}
which has the general solution 
\begin{equation}
    v(\vphi) =
    \f{v_0\cosh\vphi}{\cosh\vphi_0+v_0\f{\lambda}{2}(f(\vphi_0)-f(\vphi))}\,
    , \qquad v_0 = v(\vphi_0) \,  , 
\end{equation}
where the function
\begin{equation}
    f(x) \equiv x + \sinh x\cosh x
    \label{function}
\end{equation}
plays an important role in the following.

The symmetries of the problem imply
$\vphi(-s)=\vphi(s)$,\footnote{Strictly speaking, this is true only if
  the solution is unique. Since we are dealing with a boundary value
  problem for a nonlinear differential equation, as we will explain
  shortly, we are not guaranteed {\it a priori} of the existence and
  unicity of the solution. Nevertheless, we have verified that the
  solution that we have found is actually unique.}  
and as a consequence $\dot{\vphi}(0)=0$; in turn, we have
$v(\vphi_0)=1$. Notice that since $\ddot{\vphi}> 0$ (unless
$\lambda=0$, in which case the area term is absent) we have
$\vphi(s)>\vphi_0$ for $s\ne 0$, as one expects for a
minimum. Moreover, the positivity of $v$ and the monotonicity of
$f(x)$ imply that $\vphi(s)$ must satisfy $0\le f(\vphi)-f(\vphi_0)\le 
(2/\lambda)\cosh\vphi_0$. 

It is immediate at this point to write down the general solution of our
variational equation, which reads
\begin{equation}
\label{eq:form_sol}
|s| = \int_{\vphi_0}^{\vphi(s)}dx\,\f{1}{\sqrt{v(x)^2-1}}\, .
\end{equation}
To fully determine the solution of the variational problem we still
have to impose the appropriate boundary conditions.
In order to do so, it is convenient to define $\tilde\vphi$
through the equation 
\begin{equation}
\label{eq:boundary_1}
  \cosh\vphi_0+\f{\lambda}{2}(f(\vphi_0)-f(\tilde\vphi))=0\, ,
\end{equation}
and so rewrite $v$ as
\begin{equation}
\label{eq:exact_v}
  v(\vphi) = \f{\cosh\vphi}{\f{\lambda}{2}(f(\tilde\vphi)-f(\vphi))}\, .
\end{equation}
The value $\tvphi$ is the maximal allowed value for $\vphi(s)$ which
respects the positivity of $v$, i.e., whatever is the boundary
condition that we choose, we still need the inequality $\vphi_0\le
\vphi(s)\le \tvphi$ to be satisfied. Since we look for a path on the
helicoid that at $\sigma=\pm b/2$, or equivalently at $s=\pm
\theta/2$, joins smoothly the eikonal incoming and outgoing
trajectories of the exchanged light fermions, the derivative
$\dot{t}=\dot{\vphi}\cosh\vphi$ has to diverge at $s=\pm
\theta/2$. As the function $v$, and thus $\dot{\vphi}$, have a
divergence at the point $\tilde\vphi$, the appropriate boundary
condition is then precisely
$\vphi(\pm\f{\theta}{2})=\tvphi$.\footnote{The other possibility would
  be $\vphi(\pm\f{\theta}{2})=\infty$, but since $\vphi(s)\le\tvphi$  
  this would again require
  $\infty=\vphi(\pm\f{\theta}{2})\le\tvphi=\infty$.}    
We will comment further on this point at the end of this
Section. Using Eq.~\eqref{eq:form_sol}, this  
boundary condition is expressed as  
\begin{equation}
\label{eq:boundary_2}
  \f{\theta}{2} =
  \int_{\vphi_0}^{\tilde\vphi}d\vphi\,\f{1}{\sqrt{v(\vphi)^2-1}} =
  \int_{\vphi_0}^{\tilde\vphi}d\vphi\,
  \f{\f{\lambda}{2}(f(\tilde\vphi)-f(\vphi))}
  {\sqrt{\cosh^2\vphi-[\f{\lambda}{2}(f(\tilde\vphi)-f(\vphi))]^2}}     
  \, , 
\end{equation}
and the mathematical problem is now completely specified. Equations
\eqref{eq:boundary_1} and \eqref{eq:boundary_2} form a coupled set of
equations, whose solution would give the explicit form of the profile
$\tau(\sigma)$. We have not yet been able to find an
analytic solution in the general case. Some approximate solutions will
be discussed in the next subsection; here we discuss some 
general properties of the result.

\begin{figure}[t]
   \centering
   \includegraphics[width=0.7\textwidth]{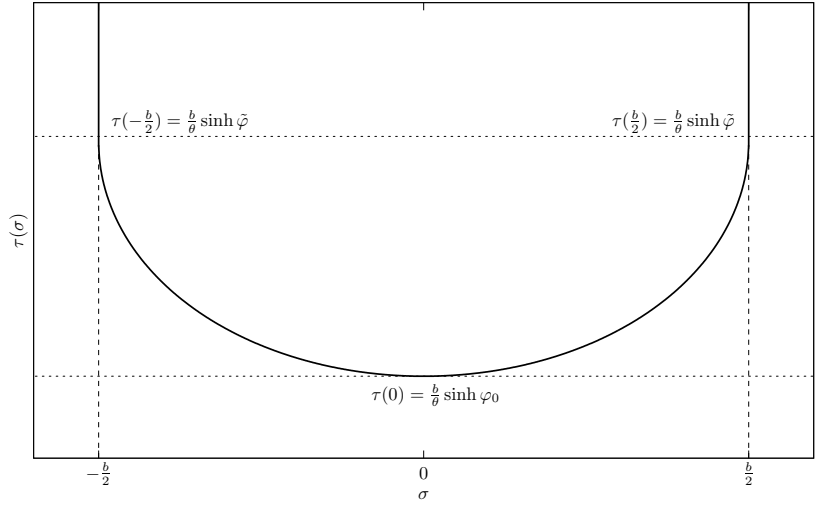}
   \caption{{\it Sketch of the minimisation profile $\tau(\sigma)$
       described by the 
       trajectories of the exchanged fermions on the helicoid.} 
     The solid line represents the trajectory of the exchanged fermion.
     The dashed (vertical) lines are the eikonal trajectories, plotted
     for reference. The dotted (horizontal) lines indicate the minimal
     and maximal values of $\tau(\sigma)$, i.e.,
     $\tau(0)=\f{b}{\theta}\sinh\vphi_0$ and
     $\tau(\pm\f{b}{2})=\f{b}{\theta}\sinh\tvphi$.} 
   \label{fig:sketch}
\end{figure}
As a first observation, we notice that the values $\vphi_0$ and
$\tvphi$ are related to the characteristic time scales of the
quark--exchange interaction in Euclidean space.  
Recall that in the dual string picture the interaction is described
by the exchange in the $t$--channel of an open string with helicoidal
world--sheet between the colliding mesons. The world--sheet
coordinates are $\sigma\in[-b/2,b/2]$ and
$\bar{\tau}\in[-\tau(\sigma),\tau(\sigma)]$ (see
Eq.~\eqref{eq:helico}), and they are related to the Euclidean time
$t_E=x_4$ by the relation $t_E=\bar{\tau}\cos(\theta\sigma/b)$. The
development of the interaction, seen as a process taking place in
Euclidean time, passes through the following three stages.
In the first stage, at time $t_{Ei}=-\tilde{\tau}\cos(\theta/2)$,
with $\tilde{\tau}=(b/\theta)\sinh\tvphi$, the strings corresponding
to the two scattering mesons in the initial state begin to expand; the
expansion continues until $t_{E-}=-\tilde{\tau}_0$, where 
$\tilde{\tau}_0=(b/\theta)\sinh\vphi_0$, when they join forming a
single, unstable string. The second stage corresponds to the existence
of this unstable string, which lasts until $t_{E+}=\tilde{\tau}_0$,
when it decays and splits in two. In the third stage, the decay
products shrink, returning to their initial size at $t_{Ef}=
\tilde{\tau}\cos(\theta/2)$, thus reconstituting the mesons in the
final state.

An important remark is that the solution does not depend on the length
variable $T$:\footnote{Strictly speaking, a solution exists only if $T\ge
  \tau(\pm b/2)$, but since we are interested in the limit
  $T\to\infty$ this restriction is irrelevant.} this guarantees
that our result will be free of IR divergencies. 
This is actually expected, since we are working with colourless 
objects, and it is in contrast with the divergencies arising in
quark--quark and gluon--gluon scattering~\cite{JP2,Alday}. 
The only way in which $T$ could 
have entered the solution is through the boundary conditions, since
the equations do not depend on it, but our choice for them is again 
independent of $T$. 
As a consequence, the relevant analytic
continuation from Euclidean to Minkowski space--time reduces simply
to $\theta\to -i\chi$.

It is easy to prove that a regular solution, for which $\tau(s)>0$, and
thus $\vphi(s)>0$, can exist only in a limited range
for the impact parameter. 
From Eq.~\eqref{eq:eq_phi}, using the fact that $\vphi(s)\ge\vphi_0$,
we derive the inequality 
\begin{equation}
  \label{eq:ineq_1}
  \f{\ddot{\vphi}}{(1+\dot{\vphi}^2)^{\f{3}{2}}} \ge \lambda\cosh\vphi
  \ge \lambda\cosh\vphi_0\, ,
\end{equation}
which, integrating between $0$ and $\theta/2$, and using
$\dot{\vphi}(0)=0$ and  $\dot{\vphi}(\theta/2)=\infty$, provides 
a bound on $b$,
\begin{equation}
\label{eq:b_bound}
  \f{\lambda\theta}{2} = \f{b}{4\pi\alpha_{\rm eff}' m} \le
  \f{1}{\cosh\vphi_0 } \le 1\, . 
\end{equation}
This defines a {\it critical} value 
\begin{equation}
  \label{eq:critical_b}
  b_c\equiv 4\pi\alpha_{\rm eff}' m\,,
\end{equation}
beyond which the Euclidean solution ceases to be a positive real
quantity.
The limitation imposed by this bound is analogous to the one found in
the case $\theta=0$, Eq.~\eqref{eq:string_brk}, i.e., for too large
$b$ the four--dimensional Euclidean ``soap film'' corresponding to the
string world--sheet collapses due to the attractive effect of the
string tension.\footnote{\label{foot:constmass}The interpretation of 
  Eq.~\eqref{eq:b_bound} as a string--breaking condition suggests
  that the mass parameter $m$ represents the {\it constituent} mass of
  the light quark. Together with the lower bound on $b$ discussed
  previously, this condition determines a window $R_0\lesssim b \le
  b_c$, where the flat--space approximation is expected to
  be valid.} Moreover, the fact that $b_c$ vanishes when $m=0$  
reflects the necessity of a ``repulsive'' boundary--length term to
compensate for the ``attractive'' area term in the minimisation
procedure.  

Another inequality can be obtained by multiplying by $\dot{\vphi}$
both sides of the first relation in Eq.~\eqref{eq:ineq_1}, and
integrating between $0$ and $\theta/2$, 
\begin{equation}
  \label{eq:ineq_bis_2}
  1\ge \lambda(\sinh\tvphi - \sinh\vphi_0) =
  \f{1}{2\pi\alpha_{\rm eff}'m}\big(\tau(\textstyle\f{b}{2}) - \tau(0)\big)\, .
\end{equation}
This result implies that the variation of $\tau$ along the boundary of
the helicoid is of order ${\cal O}(\alpha_{\rm eff}'m)$: as a consequence, in
the massless case the profile which minimises the effective action
should be a constant. However, no constant real solution of
Eq.~\eqref{eq:eq_phi} exists; moreover, even if it existed, it could
not be smooth at the junction with the eikonal incoming and 
outgoing trajectories. It is simple to derive from
\eqref{eq:ineq_bis_2} an inequality for $\Delta\equiv\tvphi-\vphi_0$, 
\begin{equation}
  \label{eq:ineq_bis_3}
  1 \ge 
  \lambda\cosh\vphi_0\sinh\Delta 
  \ge \lambda\Delta
  \, ,
\end{equation}
which will be useful in the next subsection. 

We can now exploit the formal solution, and the relations
Eq.~\eqref{eq:boundary_1} and \eqref{eq:boundary_2} satisfied by 
$\vphi_0$ and $\tvphi$, in order to rewrite the effective--action
functional in a rather compact way, namely\footnote{\label{foot:c1} We
  note in passing 
that the term neglected in Eq.~\eqref{eq:walls} would give to 
Eq.~\eqref{eq:eff_noprof} the contribution $\delta S_{{\rm eff,\,E}} =
\delta c (4 b/\theta) \sinh\tilde\vphi$.} 
\begin{equation}
\label{eq:eff_noprof}
  S_{{\rm eff,\,E}} = \f{b^2}{2\pi\alpha_{\rm eff}'\theta}\,f(\tvphi)+
\f{4mb}{\theta}\left(B(\vphi_0,\tilde\vphi) - 
\sinh\tilde\vphi\right)\, ,
\end{equation}
where
\begin{equation}
 B(\vphi_0,\tilde\vphi)=
 \int_{\vphi_0}^{\tvphi}d\vphi\,\sqrt{(\cosh\vphi)^2-
   \left[\f{\lambda}{2}(f(\tvphi)-f(\vphi))\right]^2}\, ,
\label{B}\end{equation}
and we recall that
\begin{equation}
f(\vphi)= \vphi+\sinh\vphi \cosh\vphi\, .
\label{f}\end{equation}
The last term in Eq.~\eqref{eq:eff_noprof} is simply the subtraction
term $2mL_0$, rewritten in 
terms of $\tvphi$. The other two terms are obtained by combining the
expressions for the area of the piece of helicoid and the length of
its curved boundaries. Since the area  of a portion of helicoid with a
constant profile $\tau(\sigma)\equiv \bar{\tau}$ can be expressed as 
$A(\theta,b,\bar{\tau})= (b^2/\theta)f(\asinh(\theta\bar{\tau}/b))$, one
recognises in the first term the area $\bar{A}$ of such a surface, with
$\bar{\tau}=(b/\theta)\sinh\tvphi$, times a factor $1/2\pi\alpha_{\rm
  eff}'$. Moreover, it is easily proved that
$B(\vphi_0,\tilde\vphi)-\sinh\tvphi\le 0$, so that the effective 
action is actually smaller than $\bar{A}/2\pi\alpha_{\rm eff}'$. 

As anticipated in the Introduction, the minimal effective action
Eq.~\eqref{eq:eff_noprof}, with the functions $B(\vphi_0,\tilde\vphi)$
defined by \eqref{B} and $f(\tvphi)$  by \eqref{f}, represents 
the main result of the variational problem discussed in the present
Section, and it encodes the properties of Reggeon exchange in a
compact analytic form. In order to be of 
practical use, it requires the explicit solution of the system of
equations \eqref{eq:boundary_1} and \eqref{eq:boundary_2}. Before
moving on to this issue, which is the subject of the next subsection, we
want to comment briefly on two points.

Although in this paper we have focussed only on positive real
solutions of \eqref{eq:boundary_1} and \eqref{eq:boundary_2}, this
system of equations admits also solutions for which $\vphi_0$ is
negative. As long as the equations for the boundary conditions give
$\tvphi\ge 0$, a profile can be formally defined, which at a certain
value $\pm\bar{s}$ vanishes, and is negative for $|s|\le\bar{s}$. This
simply means that the curves corresponding to the propagation of the
light fermions cross at a certain point; as a consequence, the profile
obtained by replacing the piece between the two crossing points with a
straight line would yield a smaller value for the effective action.
More precisely, if $\vphi(s)$ is a solution of the 
minimisation equations which vanishes at $\pm\bar{s}$, one substitutes
$\vphi(s)\to \vphi(s)\theta(s^2-\bar{s}^2)$, resulting in a surface
contracted to a vanishing strip in the central region. Of course, the value
which results for the effective action is no longer given by formula
\eqref{eq:eff_noprof}, which has to be modified taking into account
the different shape of the central region. The resulting profile is
continuous, but it has cusps at $\pm \bar{s}$, which require a careful
evaluation of the spin factor. Finally, when $\tvphi=0$, the surface
becomes just a thin strip of vanishing width connecting the eikonal
trajectories; the cusps are found at $\pm \theta/2$, and the cusp
angle is $\pi/2$. In the rest of this paper we will not discuss these 
configurations anymore, focusing only on smooth solutions with
$\tau(\sigma)\ge 0$.

We would also remark that the choice $\vphi(\pm\f{\theta}{2})=\tvphi$,
dictated by the smoothness condition, corresponds actually to the
minimal value of the effective action among the solutions of 
Eq.~\eqref{eq:eq_phi}. Indeed, one can consider the most general
choice $\vphi(\pm\f{\theta}{2})=\bar{\vphi}\le\tvphi$ (larger values
are not allowed, see the comment after Eq.~\eqref{eq:exact_v}), thus
introducing cusps at $s=\pm\f{\theta}{2}$. This
yields for the effective action
\begin{equation}
\label{eq:eff_gene}
  S_{{\rm eff,\,E}}(\bar{\vphi}) = \f{b^2}{2\pi\alpha_{\rm eff}'\theta}\,f(\tvphi)+
\f{4mb}{\theta}\left(\bar{B}(\vphi_0,\tilde\vphi,\bar{\vphi}) - 
\sinh\bar{\vphi}\right)\, ,
\end{equation}
where
\begin{equation}
\bar{B}(\vphi_0,\tilde\vphi,\bar{\vphi})=
 \int_{\vphi_0}^{\bar{\vphi}}d\vphi\,\sqrt{(\cosh\vphi)^2-
   \left[\f{\lambda}{2}(f(\tvphi)-f(\vphi))\right]^2}\, ,
\label{barB}\end{equation}
and $\tvphi$ is again defined by Eq.~\eqref{eq:boundary_1}. The
boundary condition Eq.~\eqref{eq:boundary_2}  becomes
\begin{equation}
\label{eq:boundary_gene}
  \f{\theta}{2} =
  \int_{\vphi_0}^{\bar\vphi}d\vphi\,
  \f{\f{\lambda}{2}(f(\tilde\vphi)-f(\vphi))}
  {\sqrt{\cosh^2\vphi-[\f{\lambda}{2}(f(\tilde\vphi)-f(\vphi))]^2}}     
  \, .
\end{equation}
Equations \eqref{eq:boundary_1} and \eqref{eq:boundary_gene} define
implicitly the dependence of $\vphi_0$ and $\tilde\vphi$ on the
boundary value $\bar\vphi$. In order to find the value of $\bar\vphi$
which minimises the effective action, one has to compute the
derivative
\begin{equation}
  \label{eq:derivative}
  \f{dS_{{\rm eff,\,E}}}{d\bar\vphi} = \f{d\vphi_0}{d\bar\vphi}\f{\de
    S_{{\rm eff,\,E}}}{\de\vphi_0} + \f{d\tvphi}{d\bar\vphi}\f{\de
    S_{{\rm eff,\,E}}}{\de\tvphi} + \f{\de S_{{\rm
        eff,\,E}}}{\de\bar\vphi}\,; 
\end{equation}
however, one easily sees that $\f{\de S_{{\rm eff,\,E}}}{\de\vphi_0}=0$ upon
use of  Eq.~\eqref{eq:boundary_1}, and moreover $\f{\de S_{{\rm
      eff,\,E}}}{\de\tvphi}=0$ upon use of
Eq.~\eqref{eq:boundary_gene}. One 
is thus left with
\begin{equation}
  \label{eq:derivative2}
   \f{dS_{{\rm eff,\,E}}}{d\bar\vphi} = \f{\de S_{{\rm eff,\,E}}}{\de\bar\vphi} = 
\f{4mb}{\theta}\left(\sqrt{\cosh^2\bar\vphi-[\textstyle
    \f{\lambda}{2}(f(\tilde\vphi)-     f(\bar\vphi))]^2} -
  \cosh\bar\vphi\right)\,,
\end{equation}
and so $\f{dS_{{\rm eff,\,E}}}{d\bar\vphi}< 0$ for $\bar\vphi<
\tvphi$. Therefore, $S_{{\rm eff,\,E}}$ is minimal for the maximal allowed
value of $\bar\vphi$, i.e., $\bar\vphi=\tvphi$.

\subsection{Explicit solutions: analytical and numerical results}

In order to perform correctly the analytic continuation, one should
obtain the exact dependence on $\theta$ by solving the equations
Eq.~\eqref{eq:boundary_1} and \eqref{eq:boundary_2} for
$\vphi_0$ and $\tvphi$, and inserting them in the formula for the
effective action. This is a hard problem, which we have not been able 
to solve in the general case, and so, in order to investigate the
analytic dependence on $\theta$, and on the impact parameter $b$, we
have to focus on some specific regimes where the relevant expressions
simplify, and the equations become manageable.

One possibility is to consider the case of large $\vphi_0$ (and
therefore also large $\tvphi$): in this case we can approximate
$f(x)\simeq e^{2x}/4$ in \eqref{f}, and solve explicitly equations
Eq.~\eqref{eq:boundary_1} and \eqref{eq:boundary_2}. From 
Eq.~\eqref{eq:ineq_bis_3}, we see that this approximation holds in the
region where $\lambda\sinh\Delta\ll 1$, which corresponds to small
$b$, as we will show. As we will see, this region is of limited
physical interest; moreover, it is not possible to calculate
higher--order corrections straightforwardly.   

Another possibility is to consider the case of large $\lambda$, which, 
according to the inequality Eq.~\eqref{eq:ineq_bis_3}, implies
$\Delta\ll 1$.  
Since $b$ is limited, this regime corresponds essentially to small
values of $\theta$.\footnote{Notice that there is a partial overlap 
  with the range of validity of the approximation discussed above.}  
In this case we can perform an expansion in powers of $\Delta$ of the
various quantities, and then solve explicitly the equations. This can
be done in a systematic way, but here we focus on the lowest order
approximation only, briefly commenting on higher--order
corrections. As we will see, this case turns out to be physically
relevant after analytic continuation to Minkowski space--time.

In order to obtain an overview of the
general features of the solution, and of the corresponding value of
the effective action, in a wider range of values of $b$ and
$\theta$, we have solved the equations numerically. Although we
cannot determine the analytic form of the solution for $\vphi_0$ and
$\tvphi$, and therefore that of the effective action, nevertheless the
numerical results can help in understanding better the various regimes
of the solution for the minimisation problem, and the range of
validity of our approximate analytic expressions. 
Moreover, the numerical investigation of the
solution reveals a few features which are not captured by the
available analytic results. Our numerical results are shown in
Figs.~\ref{fig:2}--\ref{fig:4}, and compared with the analytic
approximations.

\subsubsection{Case $\vphi_0\gg 1$}

We consider first the case of large $\vphi_0$. As we will see, this
corresponds to a region where $b/4\pi m\alpha_{\rm eff}'=b/b_c$ is
small. We begin by solving the equations \eqref{eq:boundary_1} and
\eqref{eq:boundary_2} for the boundary values $\vphi_0$ and $\tvphi$. 
Retaining only the leading terms in Eq.~\eqref{eq:boundary_1}, i.e.,
approximating $f(x) \simeq e^{2x}/4$ and $\cosh x \simeq e^x/2$, we find
\begin{equation}
\label{eq:smallb_1}
  1 \simeq \f{\lambda}{2}e^{\tilde\vphi}\sinh\Delta\, .
\end{equation}
It is easy to see that if this equation has a solution with large
$\tvphi$, then $\lambda\sinh\Delta$ must be small, as expected from 
\eqref{eq:ineq_bis_3}. Making the same approximation in
Eq.~\eqref{eq:boundary_2} we obtain 
\begin{equation}
  \f{\theta}{2} 
  \simeq \int_0^\Delta dx
  \f{\sinh x}{\sqrt{\sinh^2 \Delta-\sinh^2 x}} = \asin \tanh\Delta\, ,
\end{equation}
where we have used \eqref{eq:smallb_1} and the change of variables 
\begin{equation}
\label{eq:ch_var_1}
y =  \sqrt{1-\left(\f{\cosh x}{\cosh\Delta}\right)^2} \, .
\end{equation}
We can then write down the solution as
\begin{equation}
\label{eq:sol_smallb}  
\begin{aligned}
  \tilde\vphi &\simeq \log
  \left(\f{b_c}{b}\theta\cot\f{\theta}{2}\right)\,,\\ 
  \Delta &\simeq \atanh
  \sin\f{\theta}{2}=\f{1}{2}\log\left({\f{1+\sin\f{\theta}{2}}{1-
        \sin\f{\theta}{2}}}\right)\,   . 
\end{aligned}
\end{equation}
Explicitly, we have for $\vphi_0$
\begin{equation}
   \vphi_0 = \tilde\vphi - \Delta = \log\left(
   \f{b_c}{b}\theta\cot{\f{\theta}{2}}
   \sqrt{\f{1-\sin{\f{\theta}{2}}}{1+\sin\f{\theta}{2}}}\right) \, ,   
 \end{equation}
and the condition $\vphi_0\gg 1$ implies then
\begin{equation}
\label{eq:regime_1}
\lambda\sinh\Delta  = \f{b}{b_c}\f{1}{\theta}\,\tan\f{\theta}{2} \ll
 \sqrt{\f{1-\sin\f{\theta}{2}}{1+\sin\f{\theta}{2}}} < 1
\, . 
\end{equation}
Moreover, in order to have $0<\Delta <\infty$,
Eq.~\eqref{eq:sol_smallb} implies that the angle $\theta$ has to lie
in the range $0<\theta<\pi$: this implies that $b/b_c$ has to be much
smaller than a function of $\theta$ bounded by 
$2$, and thus small, as anticipated. 
We can now obtain the profile $\vphi(s)$ as
\begin{equation}
  \begin{aligned}
    s \simeq  \int_{\tilde\vphi-\vphi(s)}^\Delta dx &
  \f{\sinh x}{\sqrt{\sinh^2 \Delta-\sinh^2 x}} =
\\ &  \phantom{kkkkkkkkkk} 
    \asin\left(\tanh\Delta  
  \sqrt{1-\left(\f{\sinh(\tilde\vphi-\vphi(s))}{\sinh\Delta}\right)^2}\right)\,
, 
\end{aligned}
\end{equation}
which inverted gives 
\begin{equation}
\label{eq:profile_smallb}
  \sinh(\tilde\vphi-\vphi(s)) = \tan\f{\theta}{2}\sqrt{1-\left(\f{\sin
      s}{\sin \f{\theta}{2}}\right)^2}\, .
\end{equation}
In order to obtain the effective action we still need to evaluate the
integral $B(\vphi_0,\tilde\vphi)$, which in the given approximation
reads  
\begin{equation}
  B(\vphi_0,\tilde\vphi) \simeq \f{1}{\lambda}\int_0^\Delta
  \f{dx}{\sinh\Delta}e^{-x}\sqrt{1-\left(\f{\sinh x}{\sinh\Delta}\right)^2}\, . 
\end{equation}
Setting $\cos\phi = {\sinh x}/{\sinh\Delta}$ the integral is easily
evaluated, and gives
\begin{equation}
  B(\vphi_0,\tilde\vphi) \simeq 
 \f{\pi\alpha_{\rm eff}'m\theta}{b}\left[\f{\pi}{2} -
  \f{\theta}{2\sin^2\f{\theta}{2}} + \cot\f{\theta}{2} \right]\, . 
\end{equation}
In conclusion, we have for the effective action
\begin{equation}
\label{eq:eff_act_largephi}
  S_{{\rm eff,\,E}}|_{\vphi_0\gg 1} = 
  2\pi\alpha_{\rm eff}'m^2\left[\theta\cot^2\f{\theta}{2}
    + \pi - \f{\theta}{\sin^2\f{\theta}{2}}
    -2\cot\f{\theta}{2} \right] \, ,
\end{equation}
which at this level of approximation turns out to be independent of
$b$, and of order ${\cal O}(m^2)$. This is of course due to the fact
that we are neglecting important subleading contributions: indeed, a
logarithmic term $\propto \log \alpha_{\rm 
  eff}' m /b$ would appear if we naively included the contribution
coming from the $\tvphi$ term in $f(\tvphi)$ (see
Eq.~\eqref{eq:eff_noprof}). This term is of the same order of
contributions neglected in the approximation above, and thus it is not
consistent to include it; nevertheless, it shows how a non trivial
dependence on $b$ could appear at subleading order.

\subsubsection{Case $\lambda\gg 1$}

\begin{figure}[t]
  \centering
  \includegraphics[width=0.78\textwidth]{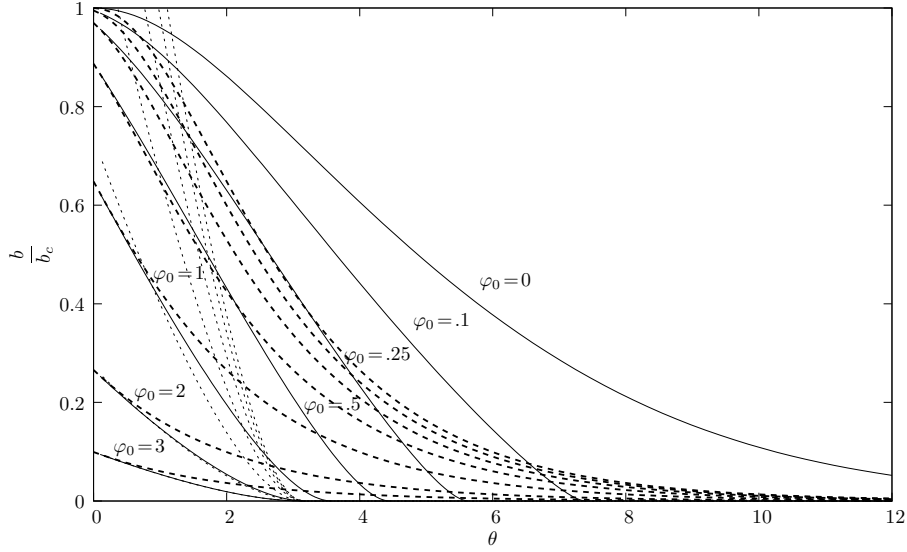}
  \caption{{\it Curves of constant $\vphi_0$ in  the $\theta-b$
      plane.} The solid lines represent the numerical results, the
    dotted lines represent the analytical result obtained in the case
    $\vphi_0\gg 1$, and the dashed lines represent the analytical
    result obtained in the case $\lambda\gg 1$.}
  \label{fig:2}
\end{figure}

We determine now the explicit form of the solution in the case of
large $\lambda$, which is expected to describe the
small--$\theta$ region, $\theta\ll b/(2\pi\alpha_{\rm eff}'m)=2b/b_c\le 2$.
According to Eq.~\eqref{eq:ineq_bis_3}, $\Delta=\tvphi-\vphi_0$ is of
order ${\cal O}(\lambda^{-1})$,\footnote{Actually, from
  Eq.~\eqref{eq:ineq_bis_3}, we can infer that $\Delta$ is of order
  ${\cal O}(\lambda^{-1-\epsilon})$, with $\epsilon\ge 0$. The actual
  value $\epsilon=0$ comes out of the calculation.} 
and so we can perform an expansion in powers of $\Delta$ of the
relevant quantities.  

We begin again by solving the equations \eqref{eq:boundary_1} and
\eqref{eq:boundary_2} for the boundary values. Expanding
Eq.~\eqref{eq:boundary_1} as 
\begin{multline}
  \cosh \tilde\vphi - \Delta\sinh  \tilde\vphi + \f{1}{2}\Delta^2\cosh
  \tilde\vphi + {\cal O}(\Delta^3) \\= \lambda\Delta\cosh
  \tilde\vphi\left[\cosh \tilde\vphi - \Delta\sinh  \tilde\vphi  +
    {\cal O}(\Delta^2) \right]\, ,
\end{multline}
we see that up to ${\cal O}(\Delta^2)$ we have
\begin{equation}
  (1-\lambda\Delta\cosh \tilde\vphi)(\cosh \tilde\vphi - \Delta\sinh
  \tilde\vphi)= {\cal O}(\Delta^2)\, .
\end{equation}
The term in the second pair of brackets is positive for $\Delta<1$, and
of order ${\cal O}(\Delta^0)$, and so we infer
\begin{equation}
  \cosh \tilde\vphi = \f{1}{\lambda\Delta} + {\cal O}(\Delta^2)\, .
\end{equation}
Expanding now Eq.~\eqref{eq:boundary_2} we obtain
\begin{equation}
  \begin{aligned}
    \f{\theta}{2} &= \Delta \int_0^1 dx \f{x(\cosh \tilde\vphi - \Delta\sinh
      \tilde\vphi)+{\cal O}(\Delta^2)}{\sqrt{1-x^2}(\cosh \tilde\vphi
      - \Delta\sinh 
      \tilde\vphi) +{\cal O}(\Delta^2)} \\ &= \Delta \int_0^1 dx
    \f{x}{\sqrt{1-x^2}} + {\cal O}(\Delta^3)  = \Delta + {\cal
      O}(\Delta^3)\, .
\end{aligned}
\end{equation}
Summarising, we have the solution\footnote{Due to the different
  boundary conditions, this solution is not expected to reduce to the
  one obtained in the previous Section in the limit $\theta\to 0$.} 
\begin{align}
\label{eq:sol_smallth}
  \cosh\tilde\vphi &= \f{4\pi\alpha_{\rm eff}' m}{b} = \f{b_c}{b}
\,,\\
  \Delta &= \f{\theta}{2}\, ,
\end{align}
and thus  
\begin{equation}
\vphi_0 =  \tilde\vphi  - \Delta=\acosh \f{b_c}{b}- \f{\theta}{2}\, .
\end{equation}
\begin{figure}[t]
  \centering
  \includegraphics[width=0.78\textwidth]{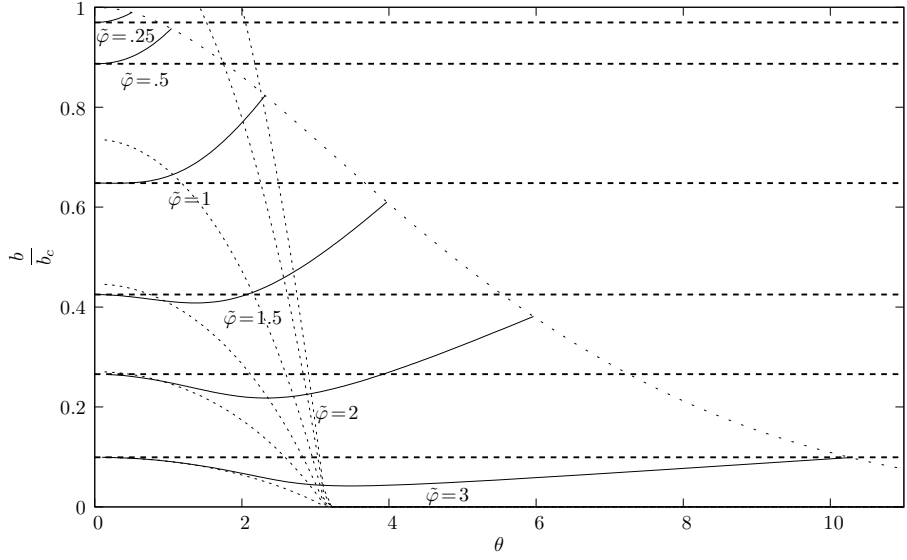}
  \caption{{\it Curves of constant $\tvphi$ in the $\theta-b$ plane.}
    The key is as in Fig.~\ref{fig:2}. The dotted line with sparse
    dots corresponds to the limiting case $\vphi_0=0$, i.e.,
    $\tau(0)=0$, and is plotted for reference.}  
  \label{fig:3bis}
\end{figure}
Since the left--hand side of Eq.~\eqref{eq:sol_smallth} is larger than one,
in order to have a real solution we must satisfy 
\begin{equation}
\f{b}{b_c} \le 1 \, ,
\end{equation}
which is exactly the general bound derived in the previous Section,
Eq.~\eqref{eq:b_bound}.  
A more restrictive requirement is expected from $\vphi_0\ge 0$, which
however yields 
\begin{equation}
  \label{eq:pos_phiz}
\f{b}{b_c}  \le \f{1}{\cosh\f{\theta}{2}} = 1 + {\cal O}(\theta^2)\, , 
\end{equation}
that at the given level of approximation is the same constraint found
above. Of course, higher--order corrections are expected to modify
this result. For completeness, we give also the explicit form of the
profile, which is obtained by integrating 
\begin{equation}
  s = \Delta \int_{\f{\tilde\vphi-\vphi(s)}{\Delta}}^1 dx
    \f{x}{\sqrt{1-x^2}} =
    \Delta\sqrt{1-\left(\f{\tilde\vphi-\vphi(s)}{\Delta}\right)^2}\, ;
\end{equation}
inverting this relation we obtain
\begin{equation}
\label{eq:profile_smallth}
  \tilde\vphi-\vphi(s) =
  \f{\theta}{2}\sqrt{1-\left(\f{2s}{\theta}\right)^2}\, .
\end{equation}
\begin{figure}[t]
  \centering
  \includegraphics[width=0.78\textwidth]{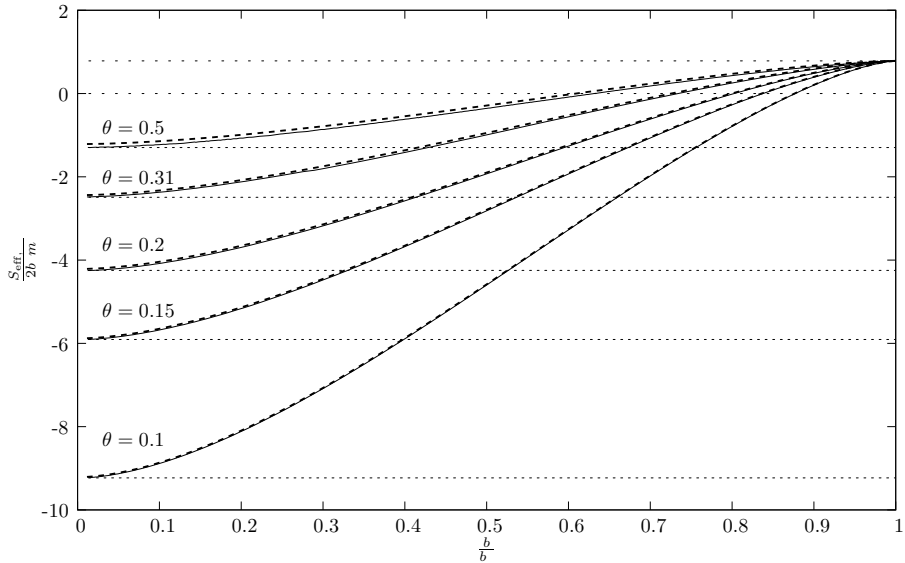} 
  \caption{{\it Effective action as a function of $b$ for various
      values of $\theta$: small $\theta$.} The action is plotted in
    units of $2b_cm=8\pi\alpha_{\rm eff}'m^2$. The key is as in figure 
    Fig.~\ref{fig:2}. The values 0 and $\pi/4$ are plotted for reference
    with a dotted line with sparse dots.}
  \label{fig:4}
\end{figure}
It is now easy to obtain the effective action, after we have computed
one last integral, namely
\begin{equation}
  B(\vphi_0,\tilde\vphi) = 
\Delta \int_0^1 dx \sqrt{1-x^2}[\cosh\tilde\vphi + {\cal O}(\theta)] =
\f{\theta\pi}{8}\cosh\tilde\vphi + {\cal O}(\theta^2)
  \, .
\end{equation}
Substituting $\tvphi$ into the other two terms of
\eqref{eq:eff_noprof} we finally obtain
\begin{equation}
  \label{eq:effact_smallth}
  S_{{\rm eff,\,E}}|_{\lambda\gg 1} =
  \f{b^2}{2\pi\alpha_{\rm eff}'\theta}\acosh\f{b_c}{b} + 
  2\pi^2\alpha_{\rm eff}'m^2 -
  \f{2bm}{\theta}\sqrt{\left(\f{b_c}{b}\right)^2-1}\, , 
\end{equation}
up to order ${\cal O}(\theta^0)$. 

The advantage of this approximation over the other one is that it
extends up to ``large'' values of $b$, i.e., up to $b_c=
4\pi\alpha_{\rm eff}' m$. As we will discuss in detail in the next
Section, the physically interesting region in Minkowski space lies at
large impact--parameter values, and an appropriate extension in $b$
beyond $b_c$ will be required: the expression obtained at large
$\vphi_0$, which is valid only at small $b$, is not reliable for this
purpose. Indeed, we see from Eq.~\eqref{eq:sol_smallth} that in order
to perform this extension we have to pass through the value
$\tvphi=0$, which is clearly inconsistent with the assumption that
$\vphi_0$ is large.

\subsubsection{Numerical results}

\begin{figure}[t]
  \centering
  \includegraphics[width=0.78\textwidth]{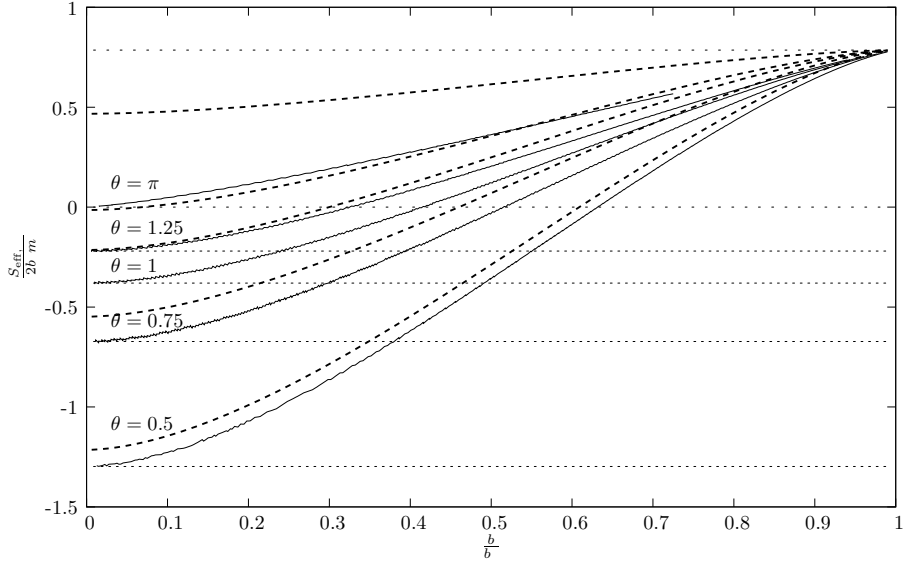}
  \caption{{\it Effective action as a function of $b$ for various
      values of $\theta$: ``large'' $\theta$.} The key is as in figure 
    Fig.~\ref{fig:2}. The values 0 and $\pi/4$ are plotted for reference
    with a dotted line with sparse dots.}
  \label{fig:4bis}
\end{figure}

We discuss now briefly our numerical results. 
It is convenient to perform the numerical calculation by
taking $\vphi_0$ and $\tvphi$ as independent variables, and then
calculate $b$ and $\theta$ as functions of $\vphi_0$ and
$\tvphi$ through Eqs.~\eqref{eq:boundary_1} and
\eqref{eq:boundary_2}. A minor drawback of this approach is that the
space of parameters (i.e., $b$ and $\theta$) 
is not scanned uniformly. In all the figures $b$ is measured in units
of $b_c= 4\pi\alpha_{\rm eff}'m$.

In Fig.~\ref{fig:2} we show the
curves of constant $\vphi_0$ in the $\theta-b$ plane, for various
values of $\vphi_0$, and we compare the numerical results with the
available analytic expressions. It is clear from this figure that the
analytic results cover only a small portion of the region of
$\theta-b$ plane where a real positive solution for $\vphi(s)$
exists. In the bulk of this region none of the two conditions
$\lambda\gg 1$ and $\vphi_0\gg 1$ apply, so that we cannot use the
approximations discussed above. Moreover, it turns out that for small
enough $b$ there are solutions with $\theta>2\pi$, or even larger:
this means that there are several branches for the solution,
corresponding to multiple ``twists'' of the helicoid. It is evident
that the expression obtained for $\vphi_0\gg 1$ is already a good
approximation at $\vphi_0=2$, being practically indistinguishable from
the numerical results. However, large values of $\vphi_0$ correspond
to small values of $b$, and the approximation does not work in the
region $b\sim b_c$, which will turn out to be physically relevant. 
On the other hand, in the region of small $\theta$ the
expression obtained for $\lambda\gg 1$ is a good approximation in the
whole range for $b$, up to $b=b_c$.

In Fig.~\ref{fig:3bis} we show the curves of constant $\tvphi$. Again,
the large--$\vphi_0$ approximation works well only at small $b$, were
it describes rather accurately the numerical results. The
large--$\lambda$ approximation gives a constant value for $\tvphi$ as
a function of $\theta$, which however coincides with the
small--$\theta$ limit of the numerical results in the whole $b$--range.

In Figs.~\ref{fig:4} and \ref{fig:4bis}  we show the results for the
effective action, plotted as a function of $b$ for various values of
$\theta$. The action is plotted in units of $8\pi\alpha_{\rm eff}'m^2
= 2b_cm$. The numerical results show that it is a monotonic function
of $b$ and $\theta$, which moreover is bounded by the value $S_{{\rm
    eff,\,E}} = 2\pi^2\alpha_{\rm eff}'m^2 = \f{\pi}{2}b_cm$. 
It turns out that, at fixed $\theta$, $S_{{\rm eff,\,E}}$ vanishes at some
point $b=\bar{b}(\theta)$, and it is positive for
$b>\bar{b}(\theta)$. This simply means that, while for
$b<\bar{b}(\theta)$ a connected helicoidal surface is
more convenient, for $b>\bar{b}(\theta)$ it is less convenient than a
disconnected configuration. As for the comparison with the analytic
results, the large--$\vphi_0$ approximation, which gives for the
effective action a $b$--independent function, correctly describes the
$b\to 0$ limit. On the other hand, the large--$\lambda$ approximation
is a very good approximation for $S_{{\rm eff,\,E}}$ in the whole
$b$--range for $\theta\lesssim 0.5$.

\section{Analytic continuation into Minkowski space--time}
\label{Mink}

In this Section we want to discuss the physical predictions that can
be obtained after analytic continuation of the Euclidean effective
action. As already remarked, an approximate solution as the ones
discussed in the previous Section is not expected to capture the exact
analytic form of the effective action. Although this is
not rigorous from a mathematical point of view, we can 
nevertheless perform the analytic continuation into Minkowski space--time
of our approximate expressions, and use physical arguments as a guide 
in order to judge the validity of the results obtained for the
Reggeon--exchange amplitude. 
For instance, the expression obtained at large $\vphi_0$ becomes, after
analytic continuation,
\begin{equation}
  \label{eq:an_cont_lphi}
    S_{{\rm eff,\,E}}|_{\vphi_0\gg 1} \to 
    \f{imb_c} 2 \left(\chi \coth^2\f{\chi}{2} 
         -\f{\chi}{\sinh^2\f{\chi}{2}}  -
    2\coth\f{\chi}{2}  - i \pi \right)
\, ,
\end{equation}
which is essentially an energy--dependent phase, so that the resulting
impact--para\-me\-ter amplitude is an oscillating function of
energy. However, as we have explained in the previous Section, this
result corresponds to a region in which the impact parameter $b$ is
very small, namely $b\ll b_c=4\pi\alpha_{\rm eff}'m$, and since
$b_c\to 0$ when $m\to 0$, its contribution to the scattering amplitude
would vanish as $m^2$ at fixed $\chi$, thus not allowing for a
suitable limit of zero quark mass. Moreover, the flat--space  
approximation, described in Section \ref{AdSBH}, which we are using
for the calculation of the minimal surface, is not expected to be
reliable in the region of small $b$. 

As we discuss below, relevant contributions to the physical scattering
amplitude, i.e., after analytic continuation to Minkowski space, 
come from the region of  $b > b_c$, in such a way that the result has
a non--zero limit when $m\to 0$. Hence, a more significant result is
obtained starting from the solution obtained at large $\lambda$, which
as we have explained describes the whole $b \le b_c$ region at small
$\theta$, and which is therefore more suitable for an extension to
larger values of $b$. This is supported by the comparison with
our numerical results for the Euclidean effective action
(Figs.~\ref{fig:4} and \ref{fig:4bis}), which shows a fairly good
agreement with the large--$\lambda$ analytic expression as regards the
dependence on $b$. In particular, the agreement improves as $b$ tends
towards $b_c$. 

\subsection{Subcritical Region $b\le b_c$}

Making then the substitution
$\theta\to -i\chi$ in Eq.~\eqref{eq:effact_smallth}, we obtain
\begin{equation}
\label{eq:mink_smallth}
  S_{{\rm eff,\,E}}\vert_{\lambda\gg 1} \to S_{{\rm eff,\,M}} = \phantom{+}
 i \f{b^2}{2\pi\alpha_{\rm eff}'\chi}\acosh\f{b_c}{b} -
 i\f{2bm}{\chi}\sqrt{\left(\f{b_c}{b}\right)^2-1} 
+   2\pi^2\alpha_{\rm eff}'m^2 \, .
\end{equation}
The real part of expression Eq.~\eqref{eq:mink_smallth} consists
simply of a $b,\chi$--independent term, while the whole
$b,\chi$--dependence is contained in terms which are purely imaginary
for $b\le b_c$, and which moreover are vanishing in the limit
$\chi\to\infty$. As we show in Appendix \ref{app:C}, the phases
$\Phi({\cal C}_{\vec{u}_i})$ in the contribution of the spin factor
remain real in this region after analytic continuation, and they are
independent of energy. Therefore, 
in the region $b\le b_c$, corresponding to the
region where a real solution exists in Euclidean space, the
impact--parameter amplitude is an oscillating
function. However, this region --which we can call the {\it core}
region-- has an energy--independent size, so that  
its contribution at small quark mass $m$ is of the order
of\footnote{For $b\lesssim R_0$ one should also include corrections
  due to the curvature, which cannot be neglected for small impact
  parameter. The corresponding contribution to the amplitude is
  however limited by the unitarity bound on the impact--parameter
  amplitude, and it is subleading with respect to contributions from
  the ``tail'', i.e., from $b>b_c$, discussed below.} 
\begin{equation}
  \label{eq:ampl_largechi}
  |\A^{core}|  \propto \Big\vert\int_0^{b_c} db\, b\, J_0(bq)
  e^{-i{\rm Im} S_{\rm eff,\,M}} e^{-2\pi^2\alpha_{\rm eff}'m^2}
  e^{i\Phi_{\rm spin}} \Big\vert \le 
   \f{b_c^2}{2} =  {\cal O}(m^2)\, ,
\end{equation}
and thus vanishing in the limit $m\to 0$. Therefore, as anticipated,
the relevant contributions to the amplitude come from the region
$b>b_c$: in the remaining part of this Section we discuss how this
region can be reached, and how expression Eq.~\eqref{eq:mink_smallth}
is modified.

\subsection{Analytic continuation towards $b > b_c$}

As we have already pointed out, a real solution of the
saddle--point equation in Euclidean space exists only in a limited
range of impact--parameter values. The limitation to real solutions is
dictated by the fact that the path--integral Eq.~\eqref{eq:sym_PI} is
over real paths ${\cal C}_\pm$ in Euclidean space, leading in turn to
an integral over real $\tau^\pm$. The limitation $b\le b_c$ can be
seen also in the effective action, since $b_c$ is a 
branch point for this quantity, beyond which it acquires an
imaginary component. Nevertheless, since we are mainly
interested in the impact--parameter amplitude in Minkowski space,
which is generally a complex quantity, we can think of extending the
result beyond $b_c$, leaving aside the limitations coming from the
requirement of reality, by making use of an appropriate analytic
continuation. To justify this procedure from a mathematical
point of view we can invoke analyticity in the impact parameter, which
allows us to determine the value of the impact--parameter amplitude
for $b>b_c$ up to fixing the ambiguity in the choice of the Riemann
sheet. 

As we have said above, $b_c$ is a branch point for $S_{{\rm
    eff,\,M}}$, and so we need to specify a prescription in order to
go from $b<b_c$ to $b>b_c$.  To this extent, we choose the usual
``$-i\varepsilon$'' prescription, making the substitution $m^2\to
m^2-i\varepsilon$, or equivalently $b_c\to b_c -i\varepsilon$, in
Eq.~\eqref{eq:mink_smallth}. Defining $y=b_c/b$, 
this prescription amounts to going from $y>1$ to $y<1$ passing in the
lower half of the complex $y$--plane, so that the phase of $y-1$ goes
from $-\varepsilon$ to $-\pi+\varepsilon$. We have then 
\begin{equation}
  \label{eq:an_cont_b}
  \begin{aligned}
    \sqrt{y^2 -i\varepsilon -1} &\mathop \to_{y>1 \to y<1}
    -i\sqrt{1-y^2}        \, , \\
  \acosh y &\mathop \to_{y>1 \to y<1}   -i\acos y\, ,
  \end{aligned}
\end{equation}
and therefore the effective action becomes for $b>b_c$
\begin{equation}
    \label{eq:act_large_b}
   S_{{\rm eff,\,M}} \to  \phantom{+} \f{b^2}{2\pi\alpha_{\rm
       eff}'\chi}\acos\f{b_c}{b}  -
    \f{2bm}{\chi}\sqrt{1-\left(\f{b_c}{b}\right)^2} +\ 
    2\pi^2\alpha_{\rm eff}'m^2  \,  .
\end{equation}
The effective action is then real at $b>b_c$; moreover, 
for very large $b\gg b_c$ the
expression simplifies to
\begin{equation}
  \label{eq:small_m}
  S_{{\rm eff,\,M}} \simeq \f{b^2}{4\alpha_{\rm eff}'\chi} - \f{4bm}{\chi} 
     + 2\pi^2\alpha_{\rm eff}' m^2\, ,
\end{equation}
which yields then a Gaussian--like 
impact--parameter amplitude. The results Eqs.~\eqref{eq:act_large_b}
and \eqref{eq:small_m} represent the main physical
output of our calculation of the Reggeon--exchange amplitude, as
anticipated in the Introduction.\footnote{\label{foot:c2} The term
  neglected in Eq.~\eqref{eq:walls} would give an extra contribution
  $\delta S_{{\rm eff,\,M}} = \delta c (4b/\chi)\sqrt{1-(b_c/b)^2}$
 to Eq.~\eqref{eq:act_large_b}. For large $b$ this contribution is
 approximately $\delta S_{{\rm eff,\,M}} \simeq  \delta c
 (4b/\chi)$. } 

Equations \eqref{eq:act_large_b}
and \eqref{eq:small_m} call for comments. The large--$b$ expansion of
Eq.~\eqref{eq:small_m} can be equivalently seen as a small--$m$
expansion, up to order ${\cal O}(m^2)$.\footnote{More precisely, up to order
${\cal O}(\alpha_{\rm eff}'m^2)$.}  
In particular, taking $m$ to zero we obtain the same result
of~\cite{Jani}, which corresponds to a complex constant Euclidean
profile $\tau(\sigma)\equiv -ib/\theta$, i.e., $\vphi(s)\equiv
-i\pi/2$. As we will show in the next Section, 
taking the Fourier transform with respect to $\vec{b}$ one obtains for
the amplitude a Regge--pole behaviour $\A_{\cal R}\propto
s^{\alpha_{\cal R}(t)}$, with a linear Reggeon trajectory
$\alpha_{\cal R}(t)=\alpha_{\rm eff}' t$ with intercept
$\alpha_0=0$.\footnote{Our expression for the spin factor is enhanced
  by a factor of $s$ with respect to the one found in~\cite{Jani},
  which would apparently raise the intercept by 1. However, an extra
  suppressing factor $s^{-1}$ appears when taking properly into
  account the fact that the quarks and antiquarks are partons inside
  of mesons~\cite{Giordano_wp1}.}  
In~\cite{Jani} also the effect of quadratic fluctuations
of the world--sheet around the classical solution were considered,
which yielded a contribution $\delta\alpha_0 = n_\perp/24$ to the
Reggeon intercept, with $n_\perp$ the number of transverse directions
in which the string could fluctuate. In this paper we do not have
computed quantum 
fluctuations, which require more work due to the non trivial form of
the classical solution.

\begin{figure}
\centering
\includegraphics[width=.5\textwidth]{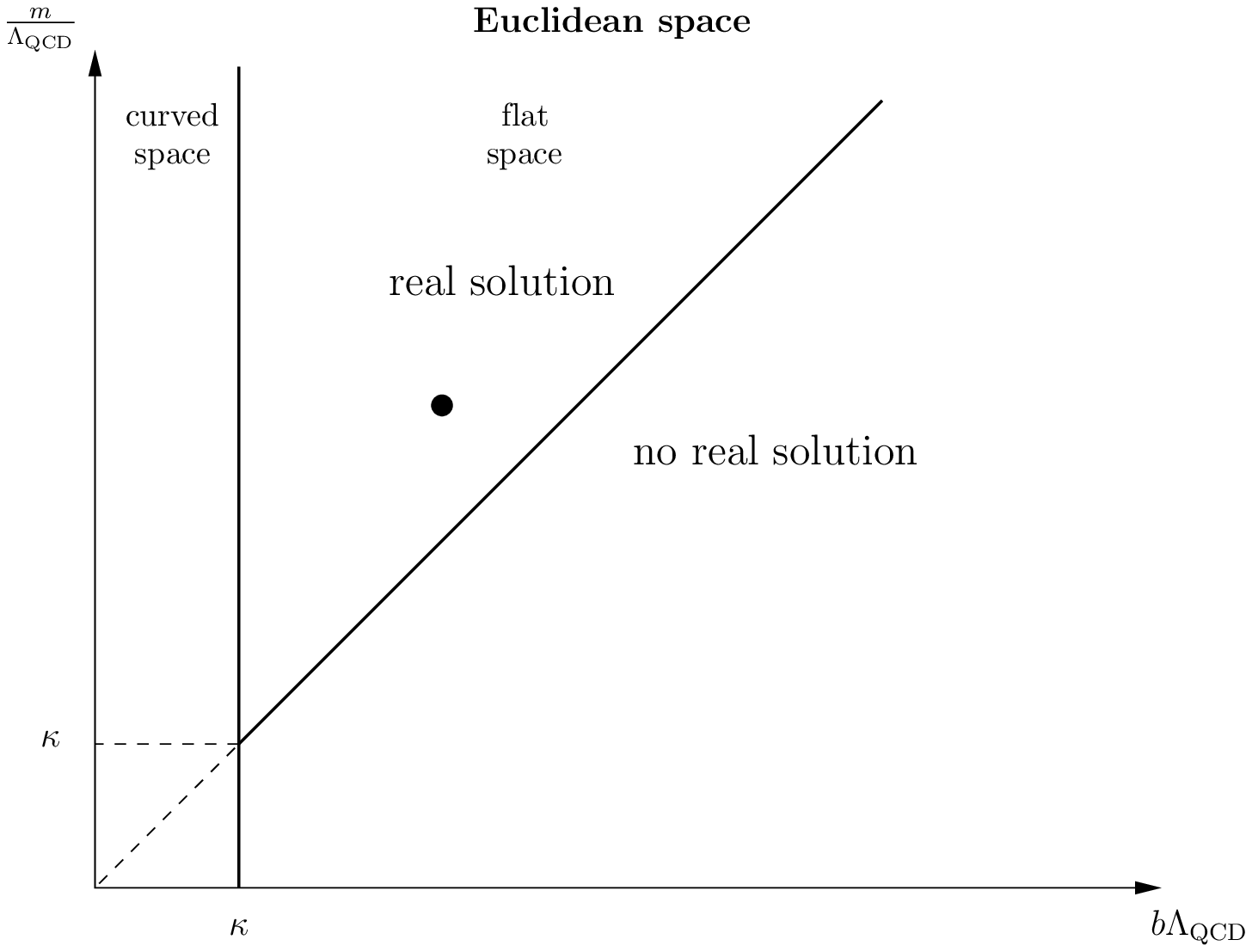}\includegraphics[width=.5\textwidth]{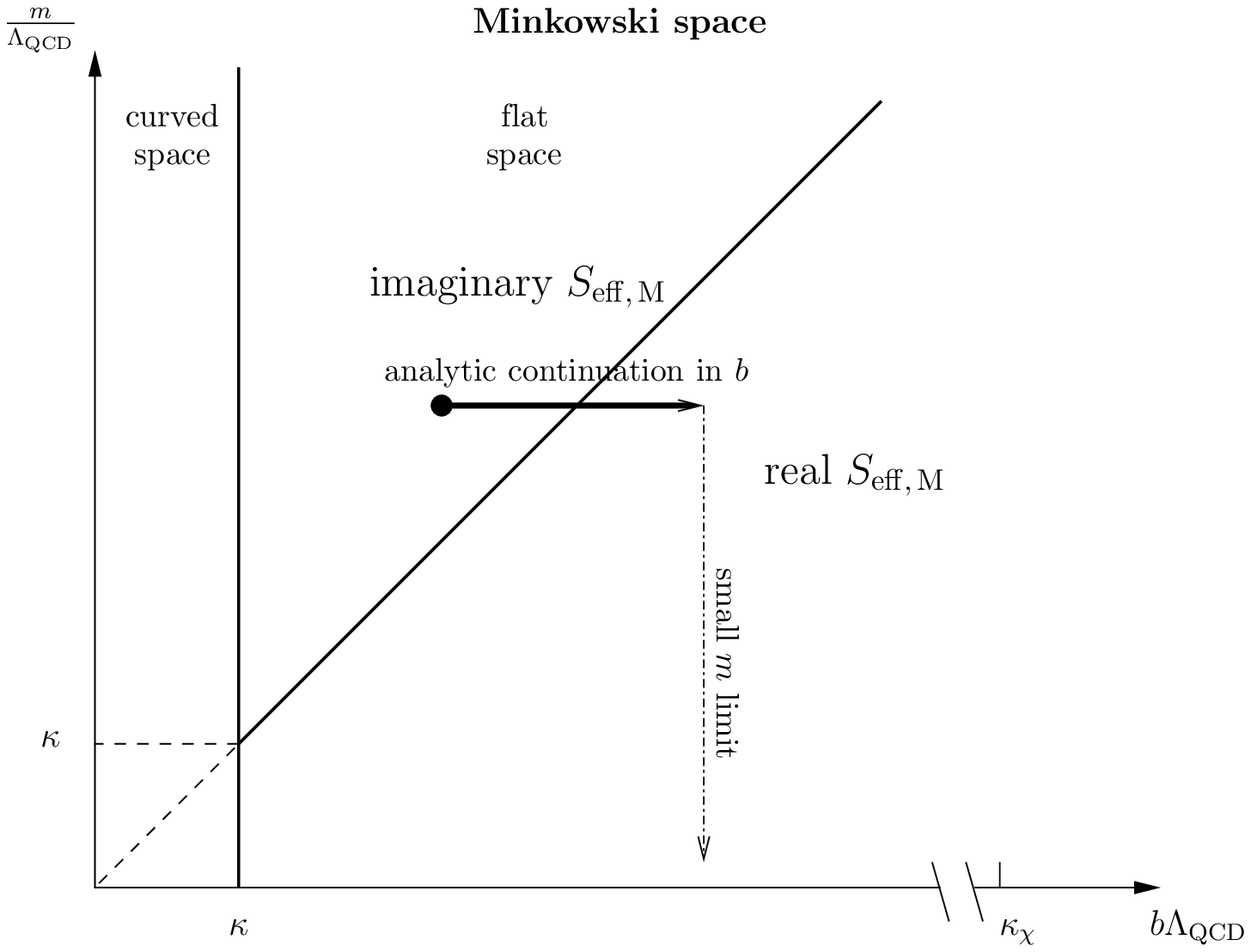} 
\caption{{\it Analytic continuation in the $(b,m)$ plane
    (in dimensionless units)%  from Euclidean (left) to Minkowski 
    % spacetime (right)
    .} 
\newline 
{\it Euclidean space (left).}
 Vertical line at $b= R_0 = \kappa   \Lambda_{\rm QCD}^{-1}$:
separates the region where the relevant geometry is essentially flat
($b\Lambda_{\rm QCD}\ge \kappa$) from the one where the curvature
cannot be neglected.  Tilted straight line
  $b\Lambda_{\rm QCD}   = m \Lambda_{\rm QCD}^{-1}$:  a real solution
to the variational problem exists above this line.  Together 
with the vertical line it defines the ``wedge'' where our approximation
is valid. 
\newline 
  {\it Minkowski spacetime
  (right).} After analytic continuation $\theta\to -i\chi$, which
  connects the black dots, it is possible to perform a further
  analytic continuation in $b$, which allows then to take the 
  small--$m$ limit. The region of $b\Lambda_{\rm QCD}$ relevant to
  Reggeon exchange extends (approximately) up to   the value
  $\kappa_\chi = \sqrt{\chi}$.} 
\label{fig:phase_trans}
\end{figure}
The possibility to take the small--$m$ limit seems to be in
contradiction with our previous remarks on the domain of applicability
of our approximation. We recall that the flat--space approximation is
expected to be valid for $b\ge R_0$, which is the scale at which the
linear potential sets in, and that a real solution in Euclidean
space exists for $b\le b_c=4\pi\alpha'_{\rm eff} m$. Setting
$\Lambda_{\rm QCD}^{-2}=4\pi\alpha'_{\rm eff}$, and $R_0=\kappa
\Lambda_{\rm QCD}^{-1}$, with $\kappa$ of order 1, we have then that
$b$ has to be in the window $\kappa \le b\Lambda_{\rm QCD} \le m
\Lambda_{\rm QCD}^{-1}$, which disappears when $ m
\Lambda_{\rm QCD}^{-1} < \kappa$ (see Fig.~\ref{fig:phase_trans}
left). However, we have shown that, after analytic continuation
$\theta\to -i\chi$ to Minkowski space, it is possible to further
extend the Reggeon-exchange amplitude to larger values of $b$, i.e.,
to $b> m\Lambda_{\rm QCD}^{-1}$, by means of analytic continuation in
$b$ (see Fig.~\ref{fig:phase_trans} right). In this region, which is
connected to the axis $m=0$, the
Minkowskian effective action becomes real, %  while it is (essentially)
% purely imaginary for $b< m\Lambda_{\rm QCD}^{-1}$.
and there is no further obstruction (at least at the given level of 
approximation) to take the limit $m\to 0$.

We have then shown that in order to obtain rigorously a non--zero
result in the singular $m=0$ case, one needs to start from $m\neq 0$
and then perform an analytic continuation in the impact parameter $b$
beyond the branch point $b_c$: indeed, the contribution from the
region $b\le b_c \propto m$ is proportional to $m^2$ at high energy
and in the limit $m\to 0$, and thus the amplitude would
vanish.\footnote{More precisely, it would reduce to the ``curved''
  contribution from the region $b\lesssim R_0$, which is however not
  under control at the present stage, but which is subleading in
  energy with respect to the ``tail'' contribution.} Following 
Eq.~\eqref{eq:small_m}, the impact--parameter region giving the major
contribution to the Regge amplitude extends to $b\to
\sqrt{\alpha_{\rm eff}'\chi}\gg \alpha_{\rm eff}'m$ (cfr. $\kappa_\chi
\gg \kappa$ in Fig.~\ref{fig:phase_trans} right). % Hence the regions of
% suitable analytic continuation from Euclidean to Minkowski spaces are
% well under control. 

The result obtained through analytic continuation in $\theta$ of the
small--$\theta$ solution (which is, strictly speaking, a
large--$\lambda$ solution), and the successive analytic continuation
in $b$ past the branch--point $b_c$, is then sensible from the
physical point of view. % Indeed, , 
% About the result of the continuation itself,
Nevertheless, two important analyticity issues are present. 
\begin{itemize}
\item The terms that we have neglected contain
higher positive powers of $\theta$, and so, although small in
Euclidean space, could give important, and in principle also dominant
contributions to the amplitude after analytic continuation.  
Indeed, terms of order ${\cal O}(\theta^n)$ would give
larger and larger ${\cal O}(\chi^n)$ terms as $n$ increases, which
could possibly lead to violations of the Froissart
bound~\cite{Froissart1,Froissart2,Froissart3}. However, to correctly
perform the analytic  
continuation one should first resum all orders in $\theta$, which
amounts to obtain the exact solution in explicit form, and only after
that take $\theta\to -i\chi$. This could easily lead to modifications
of the Minkowskian effective action which at large
$\chi$ become irrelevant.\footnote{As an illustrative example, one
  could find that the $1/\theta$ factor in
  Eq.~\eqref{eq:effact_smallth} is modified to
  $g(\theta)=1/\theta[1+\theta^2/(2+\cos\theta)]$. While the small--$\theta$
  expansion gives $g(\theta)\simeq 1/\theta + \theta/3+{\cal
    O}(\theta^2)$, which is compatible with our result, the analytic
  continuation $\theta\to -i\chi$ would lead to $g(-i\chi)=i/\chi -
  i\chi/(2+\cosh\chi)$, which reduces to $g(-i\chi)\simeq i/\chi$ for
  $\chi\to\infty$, and thus would not change our result for the Regge
  trajectory.}   
\item In order to fix the ambiguity of the analytic continuation in $b$, we
have chosen the ``$-i\varepsilon$'' prescription, passing from
$b_c/b>1$ to $b_c/b<1$ with a clockwise half--turn in the complex
$b_c/b$--plane. The correctness of this choice is clear from a
physical point of view: indeed, if we had chosen the opposite 
prescription, i.e., if we had passed from $b_c/b>1$ to $b_c/b<1$
moving in the upper half of the complex plane, we would have obtained
an unphysical, divergent impact--parameter amplitude at large
$b$. However, a completely satisfactory explanation from a
mathematical point of view is lacking at the moment. It is possible
that the ``$-i\varepsilon$'' prescription would turn out naturally by
taking into account the exact dependence on $\theta$ in the Euclidean 
effective action.\footnote{In order to have this prescription built in
  the exact expression for the Euclidean effective action, $b_c$
  should appear multiplied by an appropriate function of $\theta$:
  such a function must tend to 1 as $\theta\to 0$, and it should have
  a small negative imaginary component when $\theta\to -i\chi +
  \varepsilon$.} Another interesting possibility, which we consider
in the next Section, is that the whole multi--sheet structure of the
Minkowskian effective action has physical relevance. 
\end{itemize}
These delicate analyticity problems are
currently open, and require further work to be solved. Nevertheless,
although the results cannot be taken too ``literally'', 
it is interesting to investigate the possible physical consequences of
Eq.~\eqref{eq:act_large_b}, in particular the effects of a small
fermion mass on the Reggeon singularity.

\section{The Reggeon amplitude} 
\label{mass_effects}

It is interesting to investigate the effects of a small fermion mass
on the Reggeon singularity, computing the Reggeon--exchange amplitude  
Eq.~\eqref{eq:ampli} by performing the Fourier transform of the 
impact--parameter amplitude, namely
\begin{equation}
  \label{eq:fourier}
  \A_{\cal R}(s,t=-q^2) \equiv
-2is  \int {d^2b}\ e^{i\vec q \cdot \vec b}\
a(\vec{b},\chi)=-4i\pi s\int_0^\infty db\, b\, J_0(qb)\ a({b},\chi)\, 
\, , 
\end{equation}
where in the last passage we have used azimuthal invariance, and with
a small abuse of notation we have denoted $a(\vec{b},\chi)=a({b},\chi)$.

The impact--parameter amplitude is given by the product of several  
factors. The first factor is the contribution $e^{-S_{{\rm eff,\,M}}}$ of the 
saddle point, which up to order ${\cal O}(m)$ reads (see also
Fig.~\ref{fig:impact}) 
\begin{equation}
\label{eq:gaussian_fact}
e^{-S_{{\rm eff,\,M}}} = e^{-\f{b^2}{4\alpha_{\rm
      eff}'\chi}}\left(1+\f{4bm}{\chi}\right)+ {\cal O}(m^2)\,. 
\end{equation}
A second factor is the contribution of the spin factors, evaluated at the 
saddle point and contracted with the bispinors corresponding to the 
interacting quarks and antiquarks.
As we show in Appendix \ref{app:C}, the calculation of this
contribution can be performed exactly, but the result contains an
implicit dependence on $\chi$ and $b$ which we have not been able to
determine explicitly in the general case. We have obtained an explicit
expression in the large--$\lambda$ approximation, as we have done for
the effective action, but a comparison with numerical results 
shows that in this case the extrapolation of the analytic result to
the region $b>b_c$ cannot be trusted. Nevertheless, spin effects are not
expected to affect the behaviour of the Reggeon trajectory. For this
reason, we have preferred not to include the spin factor in our
analysis, delaying a detailed study to a future publication. 

Two other factors should in principle be included, namely the
contributions from the string fluctuations around the minimal surface, i.e., 
the factor ${\cal F}$ in Eq.~\eqref{eq:wloop_conf}, and the contribution of 
quadratic fluctuations of the floating boundary around the
saddle--point. At the present stage these contributions are not known
(except for ${\cal F}$ in the case $m=0$, where it is ${\cal F}|_{m=0}
\propto s^{\f{n_\perp}{24}}$), and they could easily introduce further
dependence on $b$ and $\chi$, thus modifying the form of the
impact--parameter amplitude.

However, an implicit assumption of 
the saddle--point approximation was that these contributions are not of 
exponential type, and so the term $e^{-S_{{\rm eff,\,M}}}$ will not change if
the approximation method works. On the other hand, power--like
factors are not completely under control; the 
same happens for the overall power of $s$, and for logarithmic prefactors 
$\chi\sim \log s$. It is therefore sensible, in a first approximation, to 
consider only the contribution Eq.~\eqref{eq:gaussian_fact} from the 
saddle--point, ignoring all the other factors, and to determine the Reggeon 
trajectory in this case. Clearly, an overall factor $s^{\delta\alpha}$
would simply change the value of the intercept of an amount
$\delta\alpha$. Moreover, the presence of factors $b^{n_b}$ in the
impact--parameter amplitude, or logarithmic $\chi^{n_\chi}$ prefactors
(with $n_b, n_\chi$ positive integers), would modify the nature of the
singularity but not the Reggeon trajectory. We will discuss this issue
in detail in subsection \ref{prefactors}. 
As a final remark, notice that the extra factor of $s$ in front of the
Fourier transform in Eqs.~\eqref{eq:ampli} and \eqref{eq:fourier} is
cancelled by a compensating factor $s^{-1}$, which appears when taking
properly into 
account the fact that the quarks and antiquarks are partons inside of  
mesons~\cite{Giordano_wp1}, as already mentioned in the previous
Section.

\subsection{The Reggeon singularity and small quark--mass effects}

In a first approximation, we therefore consider the following
expression for the Reggeon--exchange amplitude, 
\begin{equation}
  \label{eq:mass_1}
  \begin{aligned}
  \A_{\cal R}(s,t) \approx  &\,\, \f{1}{2\alpha_{\rm
      eff}'\chi}\int_0^\infty  db\, b\,
  e^{-\f{b^2}{4\alpha_{\rm eff}'\chi}}\left(1+\f{4bm}{\chi}\right) J_0(qb) +
  {\cal O}(m^2)  \\   \phantom{\bigg()} =&  \,\,
    {\cal T}_0(\chi,t)  + m{\cal T}_1(\chi,t)  
+ {\cal O}(m^2)\, ,  
  \end{aligned}
\end{equation}
where we are ignoring all numerical prefactors and the dependence on
spin, which are actually irrelevant for the following discussion.
Moreover, in Eq.~\eqref{eq:mass_1} we have introduced the quantities
\begin{equation}
  \label{eq:mass_2}
  \begin{aligned}
  {\cal T}_0(\chi,t) &=  e^{-\alpha_{\rm eff}'\chi q^2}\,,\\
  {\cal T}_1(\chi,t) &=  4\sqrt{\f{\pi\alpha_{\rm eff}'}{\chi}}
  \left\{ 
      \tilde{I}_0\left(\alpha_{\rm eff}'\chi \f{q^2}{2}\right) -
      \alpha_{\rm eff}'\chi 
      q^2\left[\tilde{I}_0\left(\alpha_{\rm eff}'\chi 
        \f{q^2}{2}\right) - \tilde{I}_1\left(\alpha_{\rm eff}'\chi
        \f{q^2}{2}\right)\right]\right\}\\ 
&= 8\sqrt{\pi\alpha_{\rm eff}'}\f{\de}{\de\chi}\left[\sqrt{\chi}
  \tilde{I}_0\left(\alpha_{\rm eff}'\chi \f{q^2}{2}\right)  
\right]\, ,
  \end{aligned}
\end{equation}
where $\tilde{I}_n(z) \equiv e^{-z}{I}_n(z)$, with ${I}_n(z)$ the
modified Bessel functions. The factor $(2\alpha_{\rm 
  eff}'\chi)^{-1}$ has been inserted ``by hand'' in order to remove an
extra logarithmic prefactor, and to fix (arbitrarily) the
normalisation. As explained above, such prefactors are not 
completely under control, but they do not change the Reggeon
trajectory. It is therefore sensible to start from the simpler
``basic'' expression without any extra power of $\chi$; the extension
to the more general case is discussed in the next subsection.
 
\begin{figure}[t]
  \centering
  \includegraphics[width=0.8\textwidth]{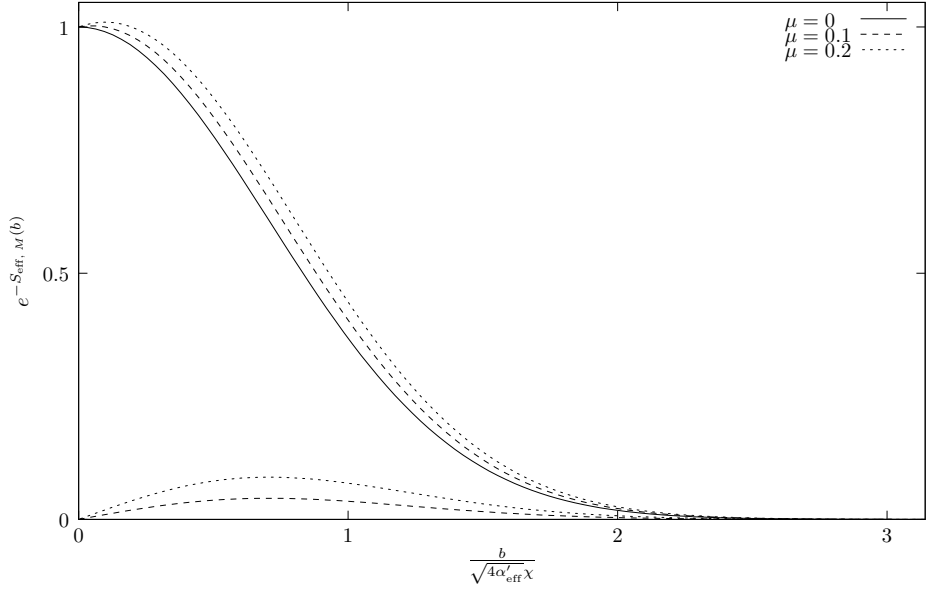}
  \caption{{\it Impact--parameter amplitude at small $m$.} Plot of the full
    saddle--point contribution $e^{-S_{{\rm 
          eff},\,M}}$ in Eq.~\eqref{eq:gaussian_fact} to the
    impact--parameter amplitude (upper curves) and 
    of the mass--dependent term alone (lower curves) for different values of
    $\mu=8m\sqrt{\alpha_{\rm eff}'/\chi}$.} 
  \label{fig:impact}
\end{figure}

As a first remark, notice that the slope of the amplitude at $t=0$,
given by
\begin{equation}
  \label{eq:slope}
  \f{\de\A_{\cal R}}{\de t}(s,t=0) = \alpha_{\rm eff}'\left(\chi  +
  6m\sqrt{\pi\alpha_{\rm eff}'\chi}\right)\,,
\end{equation}
is increased by the effect of the quark mass. Moreover, the dependence
of the slope on energy is stronger when $m\ne 0$. These effects are
related to the effective increase of the width of the
impact--parameter amplitude, which can be seen in Fig.~\ref{fig:impact}.

To uncover the nature of the Reggeon singularity we compute the Mellin
transform of the amplitude. If we write
the amplitude as $\A_{\cal R}(s,t)=\A(\chi,t)$, with $\chi\simeq
\log(s/m_1 m_2)$ at large energy, we can conveniently express
the Mellin transform as an integral over $\chi$, i.e.,
\begin{equation}
  \label{eq:mellin}
  \A^{(M)}(\omega,t) = \int_0^\infty d\chi
  e^{-\omega\chi}\A(\chi,t)\, .
\end{equation}
The Mellin transform is clearly linear, and moreover it has the
following properties,
\begin{equation}
  \label{eq:mellin_2}
  \begin{aligned}
    (\chi f)^{(M)}(\omega) &= -\f{\de}{\de\omega}f^{(M)}(\omega) \, ,\\
    \left(\f{\de}{\de\chi}f\right)^{(M)}\!(\omega) &= \omega f^{(M)}(\omega) \, ,
  \end{aligned}
\end{equation}
which will be useful in the following. We thus write, discarding the subleading
${\cal O}(m^2)$ terms,
\begin{equation}
  \label{eq:mellin_3}
  \A^{(M)}(\omega,t) = 
    {\cal T}_0^{(M)}(\omega,t) + m {\cal T}^{(M)}_1(\omega,t) \, .
\end{equation}
The first term is easily evaluated, and yields
\begin{equation}
  \label{eq:mellin_4}
  {\cal T}^{(M)}_0(\omega,t) = \f{1}{\omega + \alpha_{\rm eff}'q^2} =
  \f{1}{\omega - \alpha_{\rm eff}'t} \, .
\end{equation}
This is the only term when $m=0$, and it clearly corresponds to a
simple pole at $\omega=\alpha_{\rm eff}'t$; the corresponding Reggeon
trajectory is linear, as found in~\cite{Jani}. Partially anticipating
the discussion of the next subsection, we easily determine the effect of
logarithmic prefactors on this term: exploiting the first property
in Eq.~\eqref{eq:mellin_2}, an overall prefactor $\chi^{n}$ would
simply transform the simple pole in Eq.~\eqref{eq:mellin_4} into an
$n$--th order pole, without changing its position. 

In order to evaluate the second term, 
\begin{equation}
  \label{eq:mellin_5}
  {\cal T}^{(M)}_1(\omega,t) =
  8\sqrt{\pi\alpha_{\rm eff}'}\,\omega\int_0^\infty d\chi  e^{-\omega\chi}\sqrt{\chi}
  \tilde{I}_0\left(\alpha_{\rm eff}'\chi \f{q^2}{2}\right)\, ,
\end{equation}
we exploit the integral representation for the modified Bessel
functions, which allows to write for $\tilde{I}_0$
\begin{equation}
  \label{eq:bess}
  \tilde{I}_0(z) = 
  \f{2}{\pi}\int_0^{\f{\pi}{2}} d\phi e^{-2z\sin^2\phi} \, ,
\end{equation}
and thus
\begin{equation}
  \label{eq:mellin_6}
  {\cal T}^{(M)}_1(\omega,t) =
 8\sqrt{\pi\alpha_{\rm eff}'}\omega  \f{2}{\pi} \int_0^{\f{\pi}{2}} d\phi
 (\omega+\alpha_{\rm eff}' q^2  \sin^2\phi)^{-\f{3}{2}}\int_0^\infty d\chi
 \sqrt{\chi} e^{-\chi}
\, .
\end{equation}
The $\chi$ integral is now easily evaluated, and yields $\int_0^\infty
d\chi \sqrt{\chi} e^{-\chi} = \f{\sqrt{\pi}}{2}$, and so
\begin{equation}
  \label{eq:mellin_7}
  \begin{aligned}
  {\cal T}^{(M)}_1(\omega,t) =&
  -16\sqrt{\alpha_{\rm eff}'}\omega\f{\de}{\de\omega}
  \omega^{-\f{1}{2}}\int_0^{\f{\pi}{2}} d\phi \left(1+\f{\alpha_{\rm eff}'
    q^2}{\omega} \sin^2\phi\right)^{-\f{1}{2}}  =\\ &
-16\sqrt{\alpha_{\rm eff}'}\omega\f{\de}{\de\omega} \omega^{-\f{1}{2}}
K\left(-\f{\alpha_{\rm eff}'q^2}{\omega}\right)\, ,  
  \end{aligned}  
\end{equation}
where $K(z)$ is the complete elliptic integral of the first kind
(evaluated at a negative argument; see e.g.~\cite{AS}).
All in all, we have
\begin{equation}
  \label{eq:mellin_8}
  \A^{(M)}(\omega,t) =  
\f{1}{\omega - \alpha_{\rm eff}'t} - 16\sqrt{\alpha_{\rm eff}'}m
\omega\f{\de}{\de\omega} \omega^{-\f{1}{2}} 
K\left(\f{\alpha_{\rm eff}'t}{\omega}\right) 
\, .
\end{equation}
Since $t<0$ in the physical $s$--channel, this function is regular for
all $\omega>0$, and it has a branch--point singularity at
$\omega=0$. Analytically continuing in $t$ to $t>0$, i.e., to the
physical $t$--channel, the Reggeon singularity
moves on the positive real half--axis. As we have already said, 
the first term is a pole at $\alpha_{\rm eff}'t$. Also the
second term is singular at $\alpha_{\rm eff}'t$, the singularity being
governed by the behaviour of the elliptic integral near 1,
\begin{equation}
  \label{eq:elliptic_nearone}
  K(z) \mathop \simeq_{z\to 1} \f{1}{2}\log\f{16}{1-z}\, .
\end{equation}
Explicitly,
\begin{equation}
  \label{eq:mellin_9}
  \begin{aligned}
    \omega\f{\de}{\de\omega} \omega^{-\f{1}{2}} 
    K\left(\f{\alpha_{\rm eff}'t}{\omega}\right)  \mathop \simeq_{\omega \to
      \alpha_{\rm eff}'t}& 
    -\f{1}{2}\omega^{-\f{1}{2}}\left\{\f{1}{2}\log\f{16\omega}{\omega-
        \alpha_{\rm eff} ' t} +\f{\alpha_{\rm eff}'
        t}{\omega-\alpha_{\rm eff}' t}\right\} 
    \\ \mathop \simeq_{\omega \to \alpha_{\rm eff}'t} &
    - \f{1}{2(\alpha_{\rm eff}'
      t)^{\f{1}{2}}}\left\{\f{1}{2}\log\f{16\alpha_{\rm eff}' 
        t}{\omega-\alpha_{\rm eff}' t} +
      \f{\alpha_{\rm eff}' t}{\omega-\alpha_{\rm eff}' t}\right\}\,,
  \end{aligned}
\end{equation}
so that putting everything together we have
\begin{equation}
  \label{eq:singularity}
  \A^{(M)}(\omega,t) \mathop \simeq_{\omega\to \alpha_{\rm eff}' t} 
  \f{1+8\alpha_{\rm
      eff}'m{t}^{\f{1}{2}}}{\omega-\alpha_{\rm eff}' t} +
  4mt^{-\f{1}{2}}
  \log\f{16\alpha_{\rm eff}' t}{\omega-\alpha_{\rm eff}' t}
  \, .
\end{equation}
The leading singularity of $\A^{(M)}$ is then a pole at
$\omega=\alpha_{\rm eff}'t$, with residue (up to numerical factors)
\begin{equation}
  \label{eq:resid}
  {\rm Res} =  1 + 8\alpha_{\rm    eff}'{t}^{\f{1}{2}}m \, .
\end{equation}
Moreover, there is a logarithmic branch--point singularity at
$\omega=\alpha_{\rm eff}'t$ due to the second term of $\A^{(M)}$. 
At $t=0$ this singularity becomes an algebraic one, since in that case
$\A^{(M)}\sim\omega^{-\f{1}{2}}$ near $\omega=0$. 
Nevertheless, although the nature of the singularity
seems more complicated than in the massless case, involving also Regge
cuts, the Reggeon trajectory is still linear after the inclusion of
terms of order ${\cal O}(m)$. 
Of course, this result is based on a certain number of approximations
and assumptions; nevertheless, it shows how a non trivial Regge
singularity can emerge from quark--mass effects. 

\subsection{Prefactors}
\label{prefactors}

Moreover, we show now that the linearity of the trajectory is a quite
robust result, which does not change under the inclusion of 
the factor $2\alpha_{\rm
  eff}'\chi$ removed ``by hand'' in Eq.~\eqref{eq:mass_1}; more
generally,  we show that it does not change under the 
inclusion of possible extra factors of the form
$s^{\delta\alpha}\chi^{n_\chi} b^{n_b}$, which could come from the string
fluctuations and from the quadratic fluctuations around the saddle
point. Clearly, a factor $s^{\delta\alpha}$ simply shifts the
trajectory, changing the intercept of an amount
$\delta\alpha$. According to the properties of the Mellin transform
given in Eq.~\eqref{eq:mellin_2}, factors of the type $\chi^{n_\chi}$
(with $n_\chi>0$) require $n_\chi$ derivatives with
respect to $\omega$, so increasing the order of poles, but
leaving the linearity of the trajectory unaltered. Finally,
factors of the type $b^{n_b}$ (with $n_b>0$) can be of two types, namely
$n_b$ even or $n_b$ odd. Notice that the term ${\cal T}_0$ in
Eq.~\eqref{eq:mass_2} comes from a two--dimensional integral of the type
Gaussian times an even power of $b$, namely $b^0$, while for the term
${\cal T}_1$ the integral is of the type Gaussian times an odd power
of $b$, namely $b^1$. Let us indicate with ${\cal T}_{0,1}^{(n_b)}$ the
modified integrals obtained including an extra $b^{n_b}$ factor in the
integrand. In the case of $n_b$ even,  $n_b=2k$, the type of integral
of ${\cal T}_{0,1}^{(n_b)}$ is the same as that of ${\cal T}_{0,1}$,
and the extra factors are taken into account by performing
$k$ derivatives as follows,  
\begin{equation}
\label{eq:extra_b}
  {\cal T}_{0,1}^{(2k)} = \left[4\chi\alpha_{\rm eff}'\left(1+
      \alpha_{\rm eff}'\f{\de}{\de \alpha_{\rm eff}'}\right)\right]^k
  {\cal T}_{0,1}  \,,
\end{equation}
which are again seen not to change the linear trajectory. For $n_b$ odd,
$n_b=2k+1$, ${\cal T}_{0}^{(n_b)}$ becomes of the type Gaussian times
an odd power of $b$, and similarly ${\cal T}_{1}^{(n_b)}$ becomes of the
type Gaussian times an even power of $b$. More precisely,
\begin{equation}
  \begin{aligned}
    {\cal T}_{0}^{(n_b)}&={\cal T}_{0}^{(2k+1)} = \f{\chi}{4m} {\cal
      T}_{1}^{(2k)}\,,\\
    {\cal T}_{1}^{(n_b)}&={\cal T}_{1}^{(2k+1)}=\f{4m}{\chi}{\cal T}_{0}^{(2k+2)}\,,
  \end{aligned}
\end{equation}
and the result above in Eq.~\eqref{eq:extra_b} for even $n_b$ can be
applied. In conclusion, the linearity of the Regge trajectory is not
affected by the class of modifications considered here; in particular,
the slope of the trajectory does not change.\footnote{\label{foot:c3}
  It can be shown that the term neglected in Eq.~\eqref{eq:walls}
  would not change the Reggeon trajectory. At the order ${\cal O}(m)$
  considered here, its effect could be taken into account by replacing
  $m\to m-\delta c$ in the formulas of this Section.}

\subsection{Multi--sheet structure of the effective action:
  convolution of Regge amplitudes}

The results discussed so far are based on the use of the
``$-i\varepsilon$'' prescription for the analytic continuation of
Eq.~\eqref{eq:mink_smallth} from $b<b_c$ to $b>b_c$, leading to
Eq.~\eqref{eq:act_large_b} for the Minkowskian effective action.
As we have mentioned in the previous Section, it is possible that the
whole multi--sheet structure of the Minkowskian effective action is
physically relevant. A careful analysis shows that in the most general
case the analytic continuation of Eq.~\eqref{eq:mink_smallth} from
$b<b_c$ to $b>b_c$ leads to
\begin{equation}
 \begin{aligned}
    \label{eq:act_large_b_degenerate}
    S_{{\rm eff,\,M}}|_{b<b_c} & \to S^{(\pm, n)}_{{\rm
        eff,\,M}}|_{b>b_c} = \\ &
    \pm \left\{\f{b^2}{2\pi\alpha_{\rm eff}'\chi}
      \acos\f{b_c}{b}-\f{2bm}{\chi}\sqrt{1-\left(\f{b_c}{b}\right)^2}\right\}+ 
    2\pi^2\alpha_{\rm eff}'m^2 
    + \f{n b^2}{\alpha_{\rm eff}'\chi}\, ,
  \end{aligned}
\end{equation}
with $n\in \mathbb{Z}$, depending on the specific prescription chosen
for the analytic continuation, i.e., on the path in the complex plane
along which the analytic continuation is performed. Here $\acos x$
denotes the principal determination of the inverse cosine function,
i.e., $\acos x\in [0,\pi]$. The last term in
Eq.~\eqref{eq:act_large_b_degenerate} comes from the analytic
continuation of this function, $\acos x\to \pm\acos x + 2n\pi i$,
along paths in the complex plane which wind a certain number of times
around $-1$.  

As we have already said, we do not have a precise mathematical
argument which would select a specific prescription, and so one of the
possibilities $S^{(\pm, n)}_{{\rm eff,\,M}}|_{b>b_c}$ for the 
Minkowskian effective action. As a consequence, we have to use
physical arguments in order to discriminate among the various
possibilities. 
A first requirement, related to the unitarity bound
on the impact--parameter amplitude, is that the resulting
amplitude vanishes for $b\to\infty$. The simplest choice satisfying
this requirement is the ``$-i\varepsilon$'' prescription, i.e.,
$S^{(+,0)}_{{\rm eff,\,M}}|_{b>b_c}$, but it is clearly not the only
one. A second reasonable requirement is that the last term in
Eq.~\eqref{eq:act_large_b_degenerate} may be interpreted as a 
correction to a given basic amplitude for Reggeon exchange. Stated
differently, we ask that setting $n=0$ we obtain a physically
acceptable quantity. These two requirements restrict the possibilities
to $S^{(+, n)}_{{\rm eff,\,M}}|_{b>b_c}$ with $n\in {\mathbb N}$.

We will make now the following working 
hypothesis: we will assume that all the physically sensible choices
$S^{(+, n)}_{{\rm eff,\,M}}|_{b>b_c}$, $n\in {\mathbb N}$, contribute
to the Reggeon--exchange amplitude.
The determination of the full contribution of each of the admissible
terms to the scattering amplitude appears to be a difficult task, 
which would require the knowledge of their relative weights in the 
functional integral Eq.~\eqref{eq:sym_PI} (after analytic continuation
to Minkowski space--time). 
However, from their analytical structure and formal properties, the
new contributions can be put into a relation with physical
processes which are expected to take place in meson--meson scattering
at high energy.   

Indeed, one finds that each contribution to the impact--parameter
amplitude is proportional to the following factorised expression,
\begin{equation}
    \label{eq:exp_act_large_b_degenerate}
   \exp\{-S^{(+,n)}_{{\rm eff,\,M}}\} = \exp\{-S_{{\rm eff,\,M}}\}
   \times \left[\exp\left\{-\f{b^2}{\alpha_{\rm eff}'\chi}
     \right\}\right]^{n} \,  ,  
\end{equation}
where $S_{{\rm eff,\,M}}\equiv S^{(+,0)}_{{\rm eff,\,M}}$ is the
effective action given explicitly in Eq.~\eqref{eq:act_large_b},
corresponding to the Reggeon--exchange amplitude discussed in the
previous Section, and where for notational simplicity we have dropped the
subscript $|_{b>b_c}$. Going from impact--parameter to transverse 
momentum space $via$ Fourier transform, and ignoring possible
$b$--dependent prefactors, which can be treated as discussed in the
previous subsection, one obtains for each component
\begin{equation}
  \label{eq:components}
  \begin{aligned}
  \A^{(+,n)}(s,t=-\vec{q}^{\,2}) &\equiv
\int d^2b\, e^{i\vec{q}\cdot \vec{b}}\,
 \exp\{-S^{(+,n)}_{{\rm eff,\,M}}\} 
\\
&=
\int d^2b\, e^{i\vec{q}\cdot \vec{b}}\,
 \exp\{-S_{{\rm eff,\,M}}\}\left[\exp\left\{-\f{b^2}{\alpha_{\rm
         eff}'\chi}\right\}\right]^{n}  \\
&= i^n \A^{(+,0)}(s,t) \otimes \A_{el}^{\otimes n}(s,t) \,, 
  \end{aligned}
\end{equation}
where $\otimes$ is the sign of a {\it convolution}, defined here as
\begin{equation}
  \label{eq:convol}
  f(t)\otimes g(t) \equiv \f{1}{2s}\int \f{d^2k}{(2\pi)^2} \,
  f\big(-(\vec{q}-\vec{k})^2\big)g\big(-\vec{k}^{\,2}\big) \,, \quad t
  = -\vec{q}^{\,2}\,, 
\end{equation}
and where the amplitudes $\A^{(+,0)}$ and $\A_{el}$ are given by
\begin{equation}
  \label{eq:components_el_inel}
  \begin{aligned}
    \A^{(+,0)}(s,t) &= 
    \int d^2b\, e^{i\vec{q}\cdot \vec{b}}\,
    \exp\{-S_{{\rm eff,\,M}}\}\,, \\
    \A_{el}(s,t) &= -i2s 
    \int d^2b\, e^{i\vec{q}\cdot
      \vec{b}}\exp\left\{-\f{b^2}{\alpha_{\rm eff}'\chi}\right\} = 
    -2i\pi s\ \alpha_{\rm eff}'\chi e^{\f{\alpha_{\rm eff}'t}4\chi}\,.
  \end{aligned} 
\end{equation}
The physical interpretation of the resulting convolution
\eqref{eq:components} becomes quite clear when remarking that the
amplitude $\A_{el}(s,t)$ given by Eq.~\eqref{eq:components_el_inel} is
equal (up to prefactors) to the one obtained for elastic
dipole--dipole scattering within the same formalism in
Ref.~\cite{JP2}. 
This means that the various components $\A^{(+,n)}(s,t)$
represent the contribution of multiple elastic
rescattering interaction between the colliding mesons, occuring
together with the $q\barq$--Reggeon exchange previously
discussed, which corresponds to the amplitude $\A^{(+,0)}(s,t)$. We
find that such elastic contributions are independent of the quark
mass, as it is expected, and moreover of Regge--pole type, with  
Regge trajectory $\alpha_{el}(t)= \alpha_{0\,el} + \alpha'_{el}t$.   
As already noticed in \cite{Jani}, the ``Regge slope'' $\alpha'_{el} =
{\alpha_{\rm eff}'}/4$ of the elastic amplitude 
\eqref{eq:components_el_inel} is one--fourth of the one obtained in
the case of $q\!-\!\bar q$ exchange, and the ``Regge intercept'' is
$\alpha_{0\,el}=1$ (up to fluctuations, see \cite{Janik}).

From a phenomenological point of view, such contributions are 
expected to come from the long interaction time allowed by the
softness of the interactions at strong coupling in QCD (although a
complete theoretical derivation is not yet available).
We see here that they may appear in the gauge/gravity framework in
relation with the 
multi--sheet structure of the effective action, if one assumes that
all the sheets which are physically sensible (in the sense discussed
above) contribute to the scattering amplitude. Although a satisfactory
mathematical justification of this assumption is lacking at the
moment, a possible origin of these extra contributions is the
following. When formulated in terms of the variable $\vphi(s)$, the
Euclidean variational problem is invariant under the
reparameterisation $\vphi(s)\to \vphi(s)+2n\pi i$. On the other hand,
the expression Eq.~\eqref{eq:eff_noprof} for the Euclidean effective
action is not: while it is obviously possible to write it in an
explicit reparameterisation--invariant form, in doing so one would
lose analyticity in $\tvphi$. Since an analytic expression is required
in order to go from Euclidean to Minkowski space, one has to impose a
``gauge choice'' (e.g., $\Im \vphi(s)=0$), and use the corresponding
expression for the Euclidean effective action (which in this case
would be Eq.~\eqref{eq:eff_noprof}). As a result, it is possible that
the completely {\it equivalent} choices $\vphi(s)+2n\pi i$ in Euclidean
space are mapped into {\it different} solutions of the corresponding
variational problem in Minkowski space, each one contributing to the
path integral a quantity proportional to expression
Eq.~\eqref{eq:exp_act_large_b_degenerate}. This possibility is
currently under investigation.

\section{Conclusions and outlook}
\label{concl}

In this paper we have investigated the problem of $q\barq$--Reggeon
exchange in {\it soft} high--energy meson--meson scattering, in the
framework of the gauge/gravity duality in a generic 
confining background, along the lines of~\cite{Jani}.
Reggeon exchange is described as quark--antiquark exchange in the $t$
channel between the two mesons, represented as wave packets of
colourless quark--antiquark dipoles. After Wick rotation to Euclidean
space, the corresponding impact--parameter amplitude is represented in
terms of a path--integral of Wilson loops, running along the
trajectories of the constituent partons. While the trajectories of the
``spectator'' quark and antiquark can be dealt with in an eikonal
approximation, the trajectories of the exchanged fermions have to be
integrated over, with a weight depending on their length. In the dual
gravity picture, where a Wilson loop corresponds to a minimal 
surface having the loop contour as boundary, the exchanged--fermion
trajectories become therefore what we have called {\it floating}
boundaries, which should in principle be integrated over. 
In the case of heavy mesons, corresponding to small dipole 
sizes, the {\it floating} boundaries which give the dominant
contribution to the Euclidean path--integral are expected to lie on a
helicoid, determined by the eikonal trajectories of the partons, and
they can be determined in a saddle--point approximation by solving
what we have called a ``minimal surface problem with floating
boundary'', involving both the area of the surface and the length of
the boundary in the minimisation procedure.  
The properties of the Reggeon trajectory are therefore related to the
properties of the solution of this problem, which we have 
investigated in detail. Including the effects of a small but non--zero
{\it constituent} quark mass $m$, 
we have found a real solution to such an equation, in a limited
interval of values of the impact--parameter $b\le b_c \propto
m$. After analytic continuation into Minkowski space--time, and a
subsequent analytic continuation in $b$ to extend the result to the
region $b>b_c$, we have derived an expression for the amplitude in the
case of non--zero quark mass, which reduces to the result for massless 
quarks discussed in~\cite{Jani} in the limit $m\to 0$. 

The advantage of keeping the quark mass different from zero is
twofold: on one side, it regularises the calculation, allowing a
rigorous analysis of the solution of the saddle--point equation in
Euclidean space, and of the physical amplitude obtained after analytic
continuation.  On the other side, it allows to compute
mass--dependent corrections to the amplitude, and to investigate the
modifications of the Reggeon singularity due to the quark mass. To
first order in $m$, it turns out that the Reggeon singularity is more
complicated than a Regge pole, but that nevertheless the Reggeon
trajectory is the same found in the massless case, namely
$\alpha_{\cal R}(t)=\alpha_{\rm eff}' t$, if we neglect string
fluctuations around the minimal surface, and quadratic fluctuations of
the boundary around the saddle--point solution.  We have discussed a
quite large class of possible modifications of the amplitude due to
these effects, and we have shown that while the nature of the
singularity can change, the linearity of the Reggeon trajectory is not
affected; in particular, the slope of the trajectory does not change.  

Let us now discuss the outlook on possible future directions of
investigation. 
As discussed in Section \ref{Mink}, in order to perform correctly the
analytic continuation from Euclidean to Minkowski space--time one should
know the exact dependence on the Euclidean angle $\theta$. Although we
have been able to write down the exact solution to the saddle--point
equation, nevertheless we could not obtain it in a sufficiently
explicit form, displaying the exact dependence on the relevant
variables. Our conclusions regarding the Reggeon trajectory rely on an 
approximate explicit expression, analytically continued to Minkowski
space--time, and require therefore further investigation to be
consolidated. However, the qualitative agreement with the
phenomenology let us hope that more precise calculations of the
$\theta$--dependence would not change too much the result.

In the Appendix we have computed the contribution of the spin factor
in Euclidean space in exact implicit form, and in an approximate
explicit form suitable for the analytic continuation to Minkowski
space--time. However, we are unable for the moment to perform 
reliably the analytic continuation to $b>b_c$, and so we have
preferred not to include the spin factor in our analysis. Although
spin effects are not expected to change the Reggeon trajectory, a
detailed study is needed to clarify this issue. 

The corrections due to string fluctuations have been computed
in~\cite{Jani} in the massless case, where they have been shown to
give a contribution $\delta\alpha_0=n_\perp/24$ to the Reggeon
intercept, but we have not performed the corresponding computation in
the massive case considered in this paper. Moreover, at the present
stage the effect of fluctuations around the saddle--point solution are
not known. This point deserves further investigation. 

Another open issue is that of the origin of the companion contributions,
discussed in Section \ref{mass_effects}. These contributions,
identified with the effect of rescattering interactions between the
colliding mesons, have been obtained from the multi--sheet structure
of the Minkowskian effective action, and it has been suggested that
they are due to the non--uniqueness of the solution of the variational 
problem when formulated in Minkowski space. A detailed investigation
of this problem is needed in order to better substantiate this
suggestion. 

As we have already remarked in Section \ref{saddle}, the basic formula
for the Reggeon--exchange amplitude has been suggested
in~\cite{Jani}, rather than having been directly derived from QCD
first principles. Such a derivation is in progress, and it seems
to confirm essentially the expression used in~\cite{Jani} and in this
paper: a detailed report will be published in a forthcoming
paper~\cite{Giordano_wp1}. 

Another interesting issue is that of corrections 
related to the inclusion of dynami\-cal--fermion effects, which are
subleading in a $1/N_c$ expansion but which could be relevant for the
dependence on energy of the Reggeon--exchange amplitude. Using a
path--integral representation for the fermion--matrix determinant,
such corrections can be computed with the same minimal--surface
formalism employed here (see for example~\cite{Armoni}). This
computation is in progress, and will be discussed in a separate
publication~\cite{Giordano_wp2}. 

In conclusion, we hope that the renewed interest in the study of {\it
  soft} high energy scattering in the modern framework of
gauge/gravity duality will lead to a better understanding of the old
but nevertheless still open problem of Regge amplitudes.

\acknowledgments

We acknowledge useful discussions with R.~Janik and G.~Korchemsky.
M.G. wants to thank the Institut de Physique Th\'eorique, Saclay,
where part of this work has been done, for the kind hospitality. This
work has been partly funded by a grant of the ``Fondazione Angelo
Della Riccia'' (Firenze, Italy). M.G. is supported by MICINN under the
CPAN project CSD2007-00042 from the Consolider-Ingenio2010 programm,
as well as under the grant FPA2009-09638.

\appendix

\section{Evaluation of the spin factor}
\label{app:A}

In this Appendix we critically repeat the calculation of~\cite{Jani}
for the spin factor
$I_D[\dot{x}(\nu)]$~\cite{Polyakov,Korchemsky,Korchemsky2} in $D=4$
Euclidean space, for the special case of a 
path contained in a $3D$ hyperplane. Here $\dot{x}(\nu)$ is the
derivative of the path ${x}(\nu)$ with respect to the natural parameter
$\nu$, so that $(\dot{x}(\nu))^2=1$. The spin factor is defined as 
\begin{equation}
  \label{eq:spin_fact_def}
  \begin{aligned}
  I_D[\dot{x}(\nu)] &= \lim_{N\to\infty} \prod_{k=1}^{N}
   \f{1+\slatwo{\dot{x}}(k\tau)}{2} 
   \\  \tau &= \f{L}{N}\,,
  \end{aligned}
\end{equation}
where $L$ is the length of the path, so that $\nu\in[0,L]$. We have
denoted $\slan=n_\mu\gamma_{E\mu}$ with $\gamma_{E\mu}$ the
Euclidean Dirac matrices, satisfying the Euclidean Clifford algebra
$\{\gamma_{E\mu},\gamma_{E\nu}\}=2\delta_{\mu\nu}$, which in four
dimensions read 
\begin{equation}
  \label{eq:dir_mat}
  \gamma_{E4}=\gamma^0=\left(
    \begin{array}{cc}
      \one_2 & 0 \\ 
      0 & -\one_2
    \end{array}\right)\,, \qquad
  \gamma_{Ej}= -i\gamma^j = \left(\begin{array}{cc} 
      0 & -i\sigma^j \\ 
      i\sigma^j & 0
    \end{array}\right)\,,
\end{equation}
where $\one_D$ is the $D$--dimensional identity matrix, 
$\gamma^\mu$ are the Minkowskian gamma--matrices,  
and $\sigma^j$ are the Pauli matrices
\begin{equation}
  \label{eq:Pauli}
  \sigma^1 = \left(
    \begin{array}{cc}
      0 & 1 \\ 1 & 0
    \end{array}\right)\,, \quad
  \sigma^2 = \left(
    \begin{array}{cc}
      0 & -i \\ i & 0
    \end{array}\right)\,, \quad
  \sigma^3 = \left(
    \begin{array}{cc}
      1 & 0 \\ 0 & 1
    \end{array}\right)\, .
\end{equation}
It is immediate to see that for $n^2=1$ the quantity
\begin{equation}
  \label{eq:proj}
  P(n) = \f{1+\slan}{2}
\end{equation}
is a projector, i.e., $[P(n)]^2=[P(n)]^\dag = P(n)$.
Let us consider now the case of interest, namely $D=4$
\begin{equation}
  \label{eq:path_spin}
  \dot{x}(\nu) = (\dot{x}_4,\dot{x}_1,\dot{x}_2,\dot{x}_3)\,, \quad
  \dot{x}_4^2+\dot{x}_1^2+\dot{x}_2^2+\dot{x}_3^2 =1\,, 
\end{equation}
and a path contained in the $3D$ hyperplane $x_3=const.$, i.e.,
$\dot{x}_3=0$. The projector \eqref{eq:proj} has therefore the form
\begin{equation}
  \label{eq:proj_4D}
  P(\dot{x}(\nu)) = \f{1}{2}\left(
    \begin{array}{cccc}
      1 + \dot{x}_4 & 0 & 0 & -i\dot{x}_1-\dot{x}_2 \\
      0 & 1+\dot{x}_4 & -i\dot{x}_1+\dot{x}_2 & 0 \\
      0 & i\dot{x}_1+\dot{x}_2 & 1-\dot{x}_4 & 0 \\
      i\dot{x}_1-\dot{x}_2 & 0 & 0 & 1-\dot{x}_4
    \end{array}\right)\, ,
\end{equation}
which is easily recognised as the direct sum of two two--dimensional
projectors. To see this explicitly, the matrix $P(\dot{x})$ can be
brought to block--diagonal form,
\begin{equation}
  \label{eq:block_diag}
  \begin{aligned}
     P(\dot{x}(\nu)) &= M^T \bar{P}(\dot{x}(\nu)) M\,, &&&
    \bar{P}(\dot{x}(\nu)) &= \left(
      \begin{array}{c|c}
        \bar{P}_1 & 0 \\ \hline 
        0 & \bar{P}_2
      \end{array}\right)
     \,, \\ 
     \bar{P}_1(\dot{x}(\nu))  &= \frac{1}{2}\left(\begin{array}{cc}
      1 + \dot{x}_4 & -i\dot{x}_1-\dot{x}_2 \\
       i\dot{x}_1-\dot{x}_2  & 1-\dot{x}_4 
    \end{array}\right) \,, &&&
\bar{P}_2(\dot{x}(\nu)) 
    &= \frac{1}{2}\left(
    \begin{array}{cc}
       1+\dot{x}_4 & -i\dot{x}_1+\dot{x}_2 \\
       i\dot{x}_1+\dot{x}_2 & 1-\dot{x}_4
    \end{array}\right) \,, 
\end{aligned}
\end{equation}
where the matrix $M$ is given by
\begin{equation}
  \begin{aligned}
    &M_{ij} = \left\{
      \begin{aligned}
        1& && {\rm for}\,\,(i,j)=(1,1)\,,(2,4)\,,(3,2)\,,(4,3)\,, \\
        0& && {\rm otherwise}
      \end{aligned}\right.\,,
    &&  M^T M = \one_4
  \end{aligned}
\end{equation}
and one can easily verify that
$\bar{P}_{1,2}^2=\bar{P}_{1,2}^\dag=\bar{P}_{1,2}$. Moreover,\footnote{Here 
  and in the rest of the Appendices, we denote with $\vec{v}$ a
  three--dimensional vector, while two--dimensional vectors are
  denoted as $\vec{v}_\perp$.}
\begin{equation}
  \label{eq:two_dim_proj}
  \begin{aligned}
    \bar{P}_1 &= \f{1+\vec{u}_1\cdot \vec{\sigma}}{2} =
    \bar{P}_1(\vec{u}_1(\nu)) \,,&&& \vec{u}_1(\nu) &= 
    (-\dot{x}_2,\dot{x}_1,\dot{x}_4)\,, &&& \vec{u}_1^2&=1\,, \\
    \bar{P}_2 &= \f{1+\vec{u}_2\cdot \vec{\sigma}}{2} =
    \bar{P}_2(\vec{u}_2(\nu))\,, &&& \vec{u}_2(\nu) &= 
    (\dot{x}_2,\dot{x}_1,\dot{x}_4)\,, &&& \vec{u}_2^2&=1\,,
  \end{aligned}
\end{equation}
and since in three dimensions the gamma--matrices are equal to the
Pauli matrices, $\bar{P}_{1,2}$ are exactly the projectors entering
the definition of the three--dimensional spin factor. One can thus
write
\begin{equation}
  \label{eq:four_three}
I_4[\dot{x}] =  M^T\left(
  \begin{array}{c|c}
    I_3[\vec{u}_1] & 0 \\ \hline 0 & I_3[\vec{u}_2]
  \end{array}\right)M \,,
\end{equation}
and exploit the explicit expression for the three--dimensional
spin factor~\cite{Korchemskaya},
\begin{equation}
  \label{eq:three_spin}
  \begin{aligned}
    I_3[\vec{u}_j] &=  \f{1+\vec{u}_j(L)\cdot\vec{\sigma}}{2}\,
    e^{-\f{i}{2}\Phi({\cal C}_{\vec{u}_j})}\,
  \f{1+\vec{u}_j(0)\cdot\vec{\sigma}}{2}\, \left(\f{1 + \vec{u}_j(L)\cdot
      \vec{u}_j(0)}{2}\right)^{-\f{1}{2}}
  \,, 
\end{aligned}
\end{equation}
where $\Phi({\cal C}_{\vec{u}_j})$ is the area of the portion of
sphere delimited by the closed path ${\cal C}_{\vec{u}_j}$ 
made up of the path $\vec{u}_j(\nu)$
and by the segment of great circle connecting the points
$\vec{u}_j(L)$ and $\vec{u}_j(0)$ (see
Fig.~\ref{fig:sphere}). Explicitly,
\begin{equation}
  \label{eq:phase}
  \begin{aligned}
    \Phi({\cal C}_{\vec{u}_j}) &= \oint_{{\cal C}_{\vec{u}_j}} dt \dot{\phi}
    (1-\cos\omega)\,, \\ 
    \vec{u}_j(\nu) &= (\sin\omega\cos\phi,\sin\omega\sin\phi,\cos\omega)\,. 
  \end{aligned}
\end{equation}
It is immediate to see that this quantity changes sign under inversion
of the orientation of the path; moreover, it is invariant under
rotations of the path, and it changes sign under
parity.\footnote{Possible extra contributions coming from a
  non--trivial winding of the path around the sphere are proportional
to $4\pi$, and thus irrelevant in the phase factor.}  
As a consequence, since $\vec{u}_2={\cal P}{\cal R} \vec{u}_1$, where
${\cal P}$ is parity and ${\cal R}$ an appropriate rotation, and
$\vec{u}_1 = {\cal R}'\vec{u}$ with  
\begin{equation}
  \label{eq:uvect}
  \vec{u} = (\dot{x}_4,\dot{x}_1,\dot{x}_2)
\end{equation}
for an appropriately chosen rotation ${\cal R}'$, we can write
\begin{equation}
  \label{eq:rotation}
  \Phi({\cal C}_{\vec{u}_1}) = \Phi({\cal C}_{\vec{u}}) \, ,\qquad
  \Phi({\cal C}_{\vec{u}_2}) = -\Phi({\cal C}_{\vec{u}_1}) = -\Phi({\cal
    C}_{\vec{u}})\, . 
\end{equation}
This form will be useful when applying the general expression to our
specific case in the Reggeon--exchange calculation. Finally, noting
that $\vec{u}_1(L)\cdot \vec{u}_1(0) = \vec{u}_2(L)\cdot \vec{u}_2(0) =
\dot{x}(L)\cdot \dot{x}(0)$, and setting
\begin{equation}
  \label{eq:u_mat_til}
  \bar{\cal U} = \left(
  \begin{array}{c|c}
     e^{-\f{i}{2}\Phi({\cal C}_{\vec{u}})} & 0 \\ \hline 0 &
     e^{\f{i}{2}\Phi({\cal C}_{\vec{u}})}  
  \end{array}\right)\,,\qquad  {\cal N} = \left(\f{1 + \dot{x}(L)\cdot
    \dot{x}(0)}{2}\right)^{-\f{1}{2}} \, ,
\end{equation}
we can write the final expression
\begin{equation}
  \label{eq:four_three_final}
  I_4[\dot{x}] =  {\cal N} P(\dot{x}(L))\, {\cal U}\,  P(\dot{x}(0))\,, 
\end{equation}
where
\begin{equation}
  \label{eq:u_mat}
  {\cal U} = M^T \bar{\cal U} M = {\rm diag}\left(e^{-\f{i}{2}\Phi({\cal
      C}_{\vec{u}})}, e^{\f{i}{2}\Phi({\cal
      C}_{\vec{u}})},e^{\f{i}{2}\Phi({\cal
      C}_{\vec{u}})},e^{-\f{i}{2}\Phi({\cal C}_{\vec{u}})}\right)
\end{equation}
is a diagonal matrix which commutes with the four--dimensional projectors
$P(\dot{x}(L))$ and $P(\dot{x}(0))$, and which is easily seen to
induce opposite rotations on the two two--spinor components of a Dirac
four--spinor. Defining
\begin{equation}
\Sigma^3\equiv \left(\begin{array}{c|c}
    \sigma^3 & 0\\ \hline
    0 & \sigma^3
  \end{array}\right)
\end{equation}
we can write ${\cal U}$ as
\begin{equation}
   {\cal U} = \cos\left(\f{\Phi({\cal
      C}_{\vec{u}})}{2}\right) \one_4 - i\sin
\left(\f{\Phi({\cal C}_{\vec{u}})}{2}\right)\gamma^0\,\Sigma^3
\end{equation}
where for future utility we have made use of the Minkowskian
gamma--matrix $\gamma^0$. 

\section{Application to the Reggeon--exchange amplitude}
\label{app:B}

We apply now the results of Appendix \ref{app:A} to the case of the
Reggeon--exchange amplitude. We begin with the contraction of the spin
factor, after analytic continuation to Minkowski space, 
with the bispinors associated to the scattering quarks and
antiquarks. In the following Subsection we will evaluate the phase
factors corresponding to the relevant saddle--point solution.

\subsection{Contraction with the bispinors}

We are interested in the two following quantities,
\begin{equation}
  \begin{aligned}
    {\cal Q}_+ &=
    \bar{u}^{(s_q')}(p_q')I_4[\dot{X}_+]\Big|_{\theta\to-i\chi}
    v^{(t_{\barq}')}(p_\barq')  \,,\\
    {\cal Q}_- &=
    \bar{v}^{(t_\barq)}(p_\barq)I_4[\dot{X}_-]\Big|_{\theta\to-i\chi}
    u^{(s_q)}(p_q) \,,  
  \end{aligned}
\end{equation}
where $X_\pm$ are given in Eq.~\eqref{eq:path_curved}, and, taking
into account the softness of the process,\footnote{In
a more rigorous treatment of meson--meson scattering, the mass of the
quark $m$ in Eq.~\eqref{eq:vector_app} should be substituted with the
meson--mass fraction carried by the constituent
quarks~\cite{Giordano_wp1}.}
\begin{equation}
  \label{eq:vector_app}
  \begin{aligned}
    p_\barq'&\simeq p_\barq \simeq (E,p,\vec{0}_\perp)= m
    \tilde{u}_1\,, 
 &&&
     p_q' &\simeq p_q =(E,-p,\vec{0}_\perp)= m \tilde{u}_2\,,\\ 
 \tilde{u}_1&=\left(\cosh\textstyle\f{\chi}{2},
   \sinh\textstyle\f{\chi}{2},\vec{0}_\perp\right)\,, 
 &&&
    \tilde{u}_2&=\left(\cosh\textstyle\f{\chi}{2},
      -\sinh\textstyle\f{\chi}{2},\vec{0}_\perp\right)\,,
  \end{aligned}
\end{equation}
with $m$ the mass of the light quarks. 
Since we expect that the relevant paths deviate from the eikonal
trajectory only near the interaction region, we have that
\begin{equation}
  \begin{aligned}
    &\dot{X}_+(0)=-u_1\,,&&&
    &\dot{X}_+(L)=u_2\,,\\ 
    &\dot{X}_-(0)=u_2\,,&&&
    &\dot{X}_-(L)=-u_1\,,\\ 
    &u_1=\left(\cos\textstyle\f{\theta}{2},
      \sin\textstyle\f{\theta}{2},\vec{0}_\perp\right)\,, &&& 
    &u_2=\left(\cos\textstyle\f{\theta}{2},
      -\sin\textstyle\f{\theta}{2},\vec{0}_\perp\right)\,,  
  \end{aligned}
\end{equation}
and moreover
\begin{equation}
{\cal N} =  \left(\f{1 - u_1\cdot
      u_2}{2}\right)^{-\f{1}{2}} = \left(\f{1 -
      \cos\theta}{2}\right)^{-\f{1}{2}}\, .
\end{equation}
Since the straight--line parts of the paths do not contribute to the
phase factors, as we will show below, the spin factor should be
independent of $T$, and therefore only the analytic continuation in the
angular variable has to be performed. This is actually the case for
the solution of the saddle--point equation.
Performing now the analytic continuation, we obtain
\begin{equation}
  \begin{aligned}
    \sla{\dot{X}}_+(0) &= -\slatwo{u}_1 \to
   -\slatwo{\tilde{u}}_1 = 
     -\displaystyle\f{\slap_\barq\/'}{m}\,,
   &&& \sla{\dot{X}}_+(L) &= \slatwo{u}_2 \to \slatwo{\tilde{u}}_2 =
  \displaystyle\f{\slap_q\/'}{m}\,, \\
     \sla{\dot{X}}_-(0) &= \slatwo{u}_2 \to
   \slatwo{\tilde{u}}_2 = 
    \displaystyle\f{\slap_q}{m} \,,
   &&& \sla{\dot{X}}_-(L) &= -\slatwo{u}_1 \to -\slatwo{\tilde{u}}_1 =
    -\displaystyle\f{\slap_\barq}{m}\,,
  \end{aligned}
\end{equation}
where it is understood that $\slatwo{\tilde{u}}_j=
{\tilde{u}}_{j\mu}\gamma^\mu$. In the high--energy, low momentum
transfer limit we are 
interested in, the bispinors can be approximated as
\begin{equation}
  \begin{aligned}
    u^{(s)}(p) &= \sqrt{E+m}\left(
      \begin{array}{c}
        \phi^{(s)}\\
        \f{\vec{p}\cdot\vec{\sigma}}{E+m}\phi^{(s)}
      \end{array}\right)
    \to \sqrt{E+m}\left(
      \begin{array}{c}
        \phi^{(s)}\\
        \f{p\sigma^1}{E+m}\phi^{(s)}
      \end{array}\right)\,,\\
    v^{(t)}(p) &= \sqrt{E+m}\left(
      \begin{array}{c}
        \f{\vec{p}\cdot\vec{\sigma}}{E+m}\tilde{\phi}^{(t)}\\
        \tilde{\phi}^{(t)}        
      \end{array}\right)
    \to \sqrt{E+m}\left(
      \begin{array}{c}
        \f{p\sigma^1}{E+m}\tilde{\phi}^{(t)}\\
        \tilde{\phi}^{(t)}        
      \end{array}\right)\,,
  \end{aligned}
\end{equation}
where $\phi^{(s)}$ and $\tilde{\phi}^{(t)}$ are two--component
spinors. As a consequence, the bispinors are eigenstates of the projectors
acting on them, so that
\begin{equation}
  \begin{aligned}
    {\cal Q}_+ &= \tilde{\cal N}\bar{u}^{(s_q')}(p_q')\tilde{\cal
      U}_+v^{(t_{\barq}')}(p_\barq')\,,\\ 
    {\cal Q}_- &= \tilde{\cal N}\bar{v}^{(t_\barq)}(p_\barq)\tilde{\cal
      U}_-u^{(s_q)}(p_q)     \,,
  \end{aligned}
\end{equation}
where
\begin{equation}
  \tilde{\cal N} = {\cal N}\Big|_{\theta\to-i\chi} = \left(\f{1 -
      \cosh\chi}{2}\right)^{-\f{1}{2}} 
  \mathop\simeq_{\chi\to\infty} 
  \left(1-\f{s}{4m_1m_2}\right)^{-\f{1}{2}}\, ,
\end{equation}
and 
\begin{equation}
  \begin{aligned}
  &\tilde{\cal U}_\pm   = {\rm
    diag}\left(e^{-\f{i}{2}\Phi^{(M)}({\cal C}_{\vec{u}_\pm})},
    e^{\f{i}{2}\Phi^{(M)}({\cal C}_{\vec{u}_\pm})},e^{\f{i}{2}\Phi^{(M)}({\cal 
      C}_{\vec{u}_\pm})},e^{-\f{i}{2}\Phi^{(M)}({\cal
      C}_{\vec{u}_\pm})}\right)\, ,\\
&\Phi^{(M)}({\cal C}_{\vec{u}_\pm}) = \Phi({\cal C}_{\vec{u}_\pm})|_{\theta\to-i\chi}
\,.  
  \end{aligned}
\end{equation}
It is now straightforward to evaluate these quantities, obtaining
\begin{equation}
\label{eq:spin_fact_temp}
  \begin{aligned}
    {\cal Q}_+ &= \tilde{\cal
      N}2p\phi^{(s_q')\dag}\left(
e^{\f{i}{2}\Phi^{(M)}({\cal C}_{\vec{u}_+})}\sigma_-
+e^{-\f{i}{2}\Phi^{(M)}({\cal C}_{\vec{u}_+})}\sigma_+
\right)\tilde{\phi}^{(t_\barq')}\,,\\
{\cal Q}_- &= \tilde{\cal
      N}2p\tilde{\phi}^{(t_\barq)\dag}\left(e^{\f{i}{2}\Phi^{(M)}({\cal
          C}_{\vec{u}_-})}\sigma_+
+e^{-\f{i}{2}\Phi^{(M)}({\cal C}_{\vec{u}_-})}\sigma_-
\right){\phi}^{(s_q)}\,,
  \end{aligned}
\end{equation}
where $\sigma_\pm$ are the usual raising and lowering operators,
\begin{equation}
  \sigma_\pm = \f{\sigma^1 \pm i\sigma^2}{2}\, .
\end{equation}
The spinor base which gives the simplest representation is that of the
eigenvectors of $\sigma^3$,
\begin{equation}
  \begin{aligned}
 & \sigma^3{\phi}^{(s)} = s{\phi}^{(s)}\,,\,s=\pm 1\,,&& && \tilde{\phi}^{(t)} =
  {\phi}^{(-t)}\,, \,t=\pm 1\,,  \\
    &{\phi}^{(1)} = \left(
      \begin{array}{c}
        1 \\ 0
      \end{array}\right)\,, && &&
    {\phi}^{(-1)} = \left(
      \begin{array}{c}
        0 \\1
      \end{array}\right)\,,
  \end{aligned}
\end{equation}
and for this choice
\begin{equation}
  \begin{aligned}
    &\phi^{(s_q')\dag}\sigma_+
    \tilde{\phi}^{(t_\barq')} =
    \delta_{s_q',1}\delta_{t_\barq',1}\,,
    &&\phi^{(s_q')\dag}\sigma_-
    \tilde{\phi}^{(t_\barq')} =
    \delta_{s_q',-1}\delta_{t_\barq',-1}\,,\\
    & \tilde{\phi}^{(t_\barq)\dag}\sigma_-
    {\phi}^{(s_q)} =
    \delta_{s_q,1}\delta_{t_\barq,1}\,,
    &&\tilde{\phi}^{(t_\barq)\dag}\sigma_+
    {\phi}^{(s_q)} =
    \delta_{s_q,-1}\delta_{t_\barq,-1} \,.
  \end{aligned}
\end{equation}
The evaluation of the other two terms contributing to the complete
spin factor is trivial: since in that case $\dot{x}=u_{1,2}$ is
constant along the trajectory, the path ${\cal C}_{\vec{u}}$ contracts
to a point and the corresponding phase vanishes. Since the
bispinors are again eigenstates of the projectors, one
obtains simply ($p_Q'\simeq p_Q$, $p_{\barQ'}'\simeq p_{\barQ'}$)
\begin{equation}
  \begin{aligned}
    {\cal Q}_1 &= \bar{u}^{(s_Q')}(p_Q')I_4[\dot{X}_1]u^{(s_Q)}(p_Q) =
    \bar{u}^{(s_Q')}(p_Q)u^{(s_Q)}(p_Q) =
    2m_Q\delta_{s_Q',s_Q}\,,\\
    {\cal Q}_2&=-\bar{v}^{(t_{\barQ'})}(p_{\barQ'})I_4[\dot{X}_2]
    v^{(t_{\barQ'}')}(p_{\barQ'}')   
    = -\bar{v}^{(t_{\barQ'})}(p_{\barQ'})v^{(t_{\barQ'}')}(p_{\barQ'}) 
    =  2m_{\barQ'}\delta_{t_{\barQ'},t_{\barQ'}'}\,,
\end{aligned}
\end{equation}
where $m_Q$ and $m_{\barQ'}$ are the masses of the heavy quark and
antiquark, respectively.\footnote{The minus sign in the contribution of
  the heavy antiquark compensates for an extra minus sign included in
  the eikonal approximation for the antiquark propagator.} 

\subsection{Evaluation of the phase factor on the solution of the
  saddle--point equation}

We evaluate now the phase factor 
\begin{equation}
\label{eq:pathspsol}
  \begin{aligned}
&\Phi({\cal C}_{\vec{u}})= \oint_{{\cal C}_{\vec{u}}} d\nu \dot{\phi}
    (1-\cos\omega)    \,,\\
 & \vec{u} = (\dot{x}_4,\dot{x}_1,\dot{x}_2) =
(\sin\omega\cos\phi,\sin\omega\sin\phi,\cos\omega) \, ,
  \end{aligned}
\end{equation}
for the solution of the saddle--point equation.
The paths which we are interested in are given by  
\begin{equation}
   x_4^{(+)} = \tau(-\sigma) \cos\left(-\f{\theta\sigma}{b}\right)\,, \quad
   x_1^{(+)} = \tau(-\sigma) \sin\left(-\f{\theta\sigma}{b}\right)\,, \quad
   x_2^{(+)} = -\sigma\,, 
\end{equation}
and
\begin{equation}
   x_4^{(-)} = -\tau(\sigma) \cos\left(\f{\theta\sigma}{b}\right)\,, \quad
   x_1^{(-)} = -\tau(\sigma) \sin\left(\f{\theta\sigma}{b}\right)\,, \quad
   x_2^{(-)} = \sigma\,, 
\end{equation}
with $\sigma\in [-b/2,b/2]$ in both cases, and $\tau(\sigma)$ being
the solution of the saddle--point equation
Eq.~\eqref{eq:variational}. The minus signs are due to the orientation
of the path, which is in principle relevant in this calculation, while
it was not in the saddle--point equation. Although $\sigma$ is not 
the natural parameter, so that in the expression above it stands for
$\sigma=\sigma(\nu)$, we find convenient for notational simplicity to
not show explicitly its dependence on $\nu$. Notice that the 
symmetry of the path implies that $\sigma(L-\nu)=-\sigma(\nu)$, i.e.,
reversing the orientation is equivalent to flip the sign of
$\sigma$. Moreover, since $\tau(\sigma)=\tau(-\sigma)$ we can write 
\begin{equation}
   x_4^{(+)} = \tau(\sigma) \cos\left(\f{\theta\sigma}{b}\right)\,, \quad
   x_1^{(+)} = -\tau(\sigma) \sin\left(\f{\theta\sigma}{b}\right)\,, \quad
   x_2^{(+)} = -\sigma\,, 
\end{equation}
and thus both paths are seen to be connected by a rotation to the path
\begin{equation}
\label{eq:pathspsol2}
   x_4 = \tau(\sigma) \cos\left(\f{\theta\sigma}{b}\right)\,, \quad
   x_1 = \tau(\sigma) \sin\left(\f{\theta\sigma}{b}\right)\,, \quad
   x_2 = \sigma\,, 
\end{equation}
and so are their derivatives with respect to the natural
parameter. As a consequence the phases $\Phi({\cal C}_{\vec{u}_\pm})$
are both equal to the phase $\Phi({\cal
  C}_{\vec{u}})$ for the path $x$,\footnote{This can be seen even more
  directly for $X_+$. By definition the path $X_+$ is the path $x$
  with its orientation reversed, $X_+(\nu)=x(L-\nu)$, so that 
  $\dot{X}_+(\nu)=-\dot{x}(L-\nu)={\cal P}\dot{x}(L-\nu)$. Since the
  phase $\Phi({\cal C}_{\vec{u}})$ changes sign under parity and when
  reversing the path, the desired equality follows.} so that the spin
factor ${\cal Q} \equiv {\cal Q}_+{\cal Q}_-{\cal Q}_1{\cal Q}_2$ reduces to 
\begin{equation}
  \label{eq:spin_total}
  \begin{aligned}
&      {\cal Q} =       {\cal K}   \delta_{s_Q',s_Q}
  \delta_{t_{\barQ'},t_{\barQ'}'}\delta_{s_q t_\barq} 
\\ &  \phantom{diacciacciaccio} \times 
  \bigg[ 
\delta_{s_q s_q'}\delta_{t_\barq t_\barq'}\bigg(
e^{i\Phi^{(M)}({\cal C}_{\vec{u}})}\delta_{s_q,-1}
+ e^{-i\Phi^{(M)}({\cal C}_{\vec{u}})} \delta_{s_q,1}\bigg) 
+
 \delta_{s_q,-s_q'}\delta_{t_\barq,-t_\barq'}
\bigg]\, ,\\
 &     {\cal K} = 4(2p\tilde{\cal
    N})^2 m_Q m_{\barQ'} \mathop \simeq_{s\to\infty} -16 m_1 m_2 m_Q
  m_{\barQ'} \,.
\end{aligned}
\end{equation}
\begin{figure}[t]
  \centering
  \includegraphics[width=0.5\textwidth]{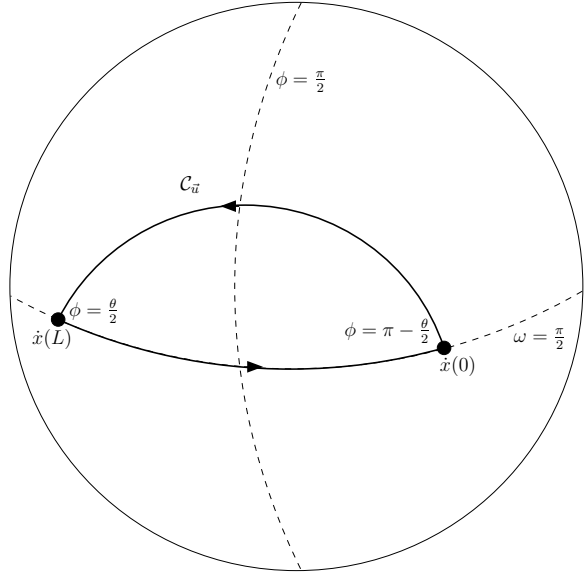}
  \caption{{\it Phase factor for the saddle--point solution.}
    Schematic representation of the path $\vec{u}(\nu)$, see 
    Eqs.~\eqref{eq:pathspsol} and \eqref{eq:pathspsol2}. The  
    phase $\Phi({\cal C}_{\vec{u}})$ is given by the area enclosed by
    the path.}
  \label{fig:sphere}
\end{figure}
We turn now to the computation of $\Phi({\cal C}_{\vec{u}})$. 
Using the dimensionless variables $t(s)=p\tau(\sigma)$, $s=p\sigma$,
with $p=\theta/b$, the derivatives $\dot{x}_\mu = d x_\mu /d\nu $ with 
respect to the natural parameter $\nu$ are written as
\begin{equation}
  \left\{\begin{aligned}
    \dot{x}_4 &= \f{1}{\sqrt{1+t^2+t'^2}}(t'\cos s - t \sin s)\,,\\
    \dot{x}_1 &= \f{1}{\sqrt{1+t^2+t'^2}}(t'\sin s + t \cos s)\,,\\
    \dot{x}_2 &= \f{1}{\sqrt{1+t^2+t'^2}}\,,
  \end{aligned}\right.
\end{equation}
where $s\in[-\theta/2,\theta/2]$ and the prime denotes derivative with
respect to $s$, and 
\begin{equation}
\label{eq:nat_par_hel}
  \f{ds}{d\nu} = \f{\theta}{b}\f{1}{\sqrt{1+t^2+t'^2}}=
  \f{\theta}{b}\,\dot{x}_2 \,.
\end{equation}
Changing variables to $t=\sinh\vphi$ we have
\begin{equation}
  \left\{\begin{aligned}
    \dot{x}_4 &=
    \f{1}{\cosh\vphi\sqrt{1+\vphi'^2}}(\vphi'\cosh\vphi\cos s -
    \sinh\vphi  \sin s) = \dot{x}_2f_4\,,\\
    \dot{x}_1 &=
    \f{1}{\cosh\vphi\sqrt{1+\vphi'^2}}(\vphi'\cosh\vphi\sin s + \sinh\vphi \cos
    s)=\dot{x}_2f_1\,,\\ 
    \dot{x}_2 &= \f{1}{\cosh\vphi\sqrt{1+\vphi'^2}}\,.
  \end{aligned}\right.
\end{equation}
Recalling now the properties of the solution $\vphi(s)$,
\begin{equation}
  \lim_{s\to\pm \f{\theta}{2}} \vphi'(s) = \pm\infty\,, \quad
  \vphi(-s) = \vphi(s) \ge 0\,, \quad \vphi'(s)\ge 0\,, \quad \vphi'(0)=0\,,
\end{equation}
one finds\footnote{In Eqs.~\eqref{eq:points_sphere_1} and
  \eqref{eq:points_sphere_2} the arguments of the various functions
  refer to the variable $s$.} 
\begin{equation}
\label{eq:points_sphere_1}
  \begin{aligned}
  &\textstyle \dot{x}_4(\pm \f{\theta}{2}) = \pm\cos\f{\theta}{2}\,, && && 
   \textstyle \dot{x}_1(\pm \f{\theta}{2}) = \sin\f{\theta}{2}\,, && &&
   \textstyle \dot{x}_2(\pm \f{\theta}{2}) = 0\,, \\
  & \dot{x}_4(0) = 0\,, && &&
    \dot{x}_1(0) = \tanh\vphi_0\,, && &&
    \dot{x}_2(0) = \f{1}{\cosh\vphi_0}\,, 
\end{aligned}
\end{equation}
so that
\begin{equation}
\label{eq:points_sphere_2}
  \begin{aligned}
  &\textstyle\omega\left(-\f{\theta}{2}\right)
  =\displaystyle\f{\pi}{2}\,, &&  && 
  \omega(0) = \acos\f{1}{\cosh\vphi_0}\,, && &&  
  \textstyle\omega\left(\f{\theta}{2}\right) =\displaystyle\f{\pi}{2}\,, \\
  &\textstyle\phi\left(-\f{\theta}{2}\right) = \displaystyle
  \pi-\f{\theta}{2}\,,  
  && && \phi(0)=
  \f{\pi}{2}\,, &&  &&
  \textstyle\phi\left(\f{\theta}{2}\right) = \displaystyle \f{\theta}{2}  \,.
  \end{aligned}
\end{equation}
One can prove that $\phi\ne 0,\pi$, $\omega\ne 0,\pi$, so that the
path lies in the sector $0<\phi<\pi$, $0<\omega<\pi/2$ of the sphere
and so does not wind around it; also, the sign of $\dot{\omega}$ is
the same as the sign of $\vphi'$. 
A schematic representation of the path is given in Fig.~\ref{fig:sphere}. 
Note that the segment of great circle closing the path lies at
$\omega=\pi/2$. We have therefore
\begin{equation}
\label{eq:phase_spin}
   \Phi({\cal C}_{\vec{u}})= \oint_{{\cal C}_{\vec{u}}} d\nu\,
   \dot{\phi}(1-\cos\omega) =
   -\int_{0}^{L}d\nu\,
   \dot{\phi}\cos\omega\,. 
\end{equation}
Using now $\tan\phi=\dot{x}_1/\dot{x}_4$ and
$\cos\phi=\dot{x}_4/\sqrt{1-\dot{x}_2^2}$, we can show that
\begin{equation}
  \dot{\phi} = \f{ds}{d\nu}\f{\dot{x}_2^2}{1-\dot{x}_2^2}(f_1'f_4-f_1f_4')\,,
\end{equation}
and a straightforward calculation gives
\begin{equation}
  \begin{aligned}
  f_1'f_4-f_1f_4' &= 2\vphi'^2 + (\sinh\vphi)^2(1+\vphi'^2) -
  \sinh\vphi\cosh\vphi\vphi''\\ &= 2\vphi'^2 -\lambda
   \sinh\vphi(\cosh\vphi)^2(1+\vphi'^2)^{\f{3}{2}}\,,  
  \end{aligned}
\end{equation}
where in the last passage we have substituted the equations of motion
in $\vphi''$. All in all, we get
\begin{equation}
  \dot{\phi} =
  \f{ds}{d\nu}\f{2\vphi'^2
    -\lambda
    \sinh\vphi(\cosh\vphi)^2(1+\vphi'^2)^{\f{3}{2}}}{(\cosh\vphi)^2(1+\vphi'^2)-1}
  \,. 
\end{equation}
Plugging this in Eq.~\eqref{eq:phase_spin} and exploiting the
symmetries of $\vphi(s)$ we obtain
 \begin{equation}
  \begin{aligned}
  \Phi({\cal C}_{\vec{u}}) 
  &= -2\int_{\vphi_0}^{\tvphi}d\vphi \f{1}{\vphi'}
  \f{2\vphi'^2   
    -\lambda
    \sinh\vphi(\cosh\vphi)^2(1+\vphi'^2)^{\f{3}{2}}}{[\cosh\vphi\sqrt{1+
      \vphi'^2}][(\cosh\vphi)^2(1+\vphi'^2)-1]}\,,
 \end{aligned} 
\end{equation}
and substituting the exact solution Eq.~\eqref{eq:exact_v}
\begin{equation}
  \sqrt{1+\vphi'^2}=v(\vphi) =
  \f{\cosh\vphi}{\f{\lambda}{2}(f(\tvphi)-f(\vphi))} \equiv
  \f{\cosh\vphi}{r(\vphi)} 
\end{equation}
we finally obtain
 \begin{equation}
\label{eq:phase_saddle_exact}
  \Phi({\cal C}_{\vec{u}}) = -2\int_{\vphi_0}^{\tvphi}d\vphi
  \f{r(\vphi)\left[2r(\vphi)((\cosh\vphi)^2-
      (r(\vphi))^2)-
      \lambda\sinh\vphi(\cosh\vphi)^5\right]}{(\cosh\vphi)^2[(\cosh\vphi)^4-
    (r(\vphi))^2] \sqrt{(\cosh\vphi)^2- (r(\vphi))^2}  } \,.
 \end{equation}

\section{Spin--factor contribution in the large--$\lambda$  case}
\label{app:C}

The calculations of the previous Appendix are exact, but in order to
obtain an analytic expression for the spin--factor contribution we
have to make some approximation. We consider therefore the case of
large--$\lambda$, corresponding to small $\Delta=\tvphi-\vphi_0$, in
which case the calculation of the phase
Eq.~\eqref{eq:phase_saddle_exact} can be explicitly performed. Up to
order ${\cal O}(\Delta^2)$ we have
\begin{equation}
  \Delta \simeq \f{\theta}{2}\,, \quad \lambda\Delta\cosh\tvphi\simeq
  \f{\lambda\theta}{2}\cosh\tvphi\simeq 1\,,
\end{equation}
so that setting $ x = (\tvphi-\vphi)/\Delta$, and expanding 
\begin{equation}
  \begin{aligned}
    \cosh\vphi &=  \cosh\tvphi - \Delta x \sinh\tvphi + {\cal
      O}(\Delta^2)\,, \\ 
    r(\vphi) &= x(\cosh\tvphi - \Delta x \sinh\tvphi)
    +  {\cal O}(\Delta^2)\,,
  \end{aligned}
\end{equation}
we obtain to leading order
\begin{equation}
  \Phi({\cal C}_{\vec{u}}) = 2\int_0^1 dx\,
  \f{x}{\sqrt{1-x^2}}\f{\sinh\tvphi}{(\cosh\tvphi)^2-x^2} + {\cal
    O}(\Delta) = \pi - 2\atan\sinh\tvphi + {\cal O}(\theta)\,.
\end{equation}
Consistently with what we have done in the calculation of the
effective action, we neglect higher--order terms when performing the
analytic continuation to Minkowski space-time, and since $\tvphi$ does
not depend on $\theta$ in the given approximation,
$\sinh\tvphi=\sqrt{(b_c/b)^2-1}$, we have
\begin{equation}
\label{eq:phase_smalltheta}
    \Phi^{(M)}({\cal C}_{\vec{u}}) = \Phi({\cal
      C}_{\vec{u}})\Big|_{\theta\to -i\chi}  
= \pi -
  2\atan\left(\sqrt{\left(\f{b_c}{b}\right)^2-1}\right)\,.
\end{equation}
Therefore, the quantity $\Phi^{(M)}({\cal C}_{\vec{u}})$ remains real
in the region $b\le b_c$ after the analytic continuation to Minkowski
space. 

At this point, in order to go to $b>b_c$ we have to perform a second
analytic continuation, which according to the ``$-i\varepsilon$''
prescription leads to 
\begin{equation}
\label{eq:phase_smalltheta_mink}
  \Phi^{(M)}({\cal C}_{\vec{u}}) \mathop \to_{b>b_c} = \pi -
  \f{2}{i}\log\left[\f{b}{b_c}\left(1+\sqrt{1-
        \left(\f{b_c}{b}\right)^2}\right)\right]\,   ,
\end{equation}
where we have used the representation
\begin{equation}
  \atan\, x = \f{1}{2i}\log\f{1+ix}{1-ix}
\end{equation}
for the arctangent. Exponentiating this expression we obtain 
\begin{equation}
  e^{\pm {i}\Phi^{(M)}({\cal C}_{\vec{u}})} \mathop \to_{b>b_c}
    = -\left(\f{b}{b_c}\right)^2\left(1\mp\sqrt{1- 
        \left(\f{b_c}{b}\right)^2}\right)^2
\,.
\end{equation}
However, a comparison of the analytic result
Eq.~\eqref{eq:phase_smalltheta} with numerical calculations for the
exact expression Eq.~\eqref{eq:phase_saddle_exact} shows that, although
there is good agreement for small values of $b/b_c$, the distance
between analytic and numerical results increases as $b$ tends to 
$b_c$. In contrast, the same comparison for the Euclidean effective
action shows that the distance between analytic and numerical results
decreases as $b$ tends to $b_c$. 
For this reason, the extrapolation
Eq.~\eqref{eq:phase_smalltheta_mink} of $\Phi^{(M)}({\cal
  C}_{\vec{u}})$ to $b>b_c$ cannot be used,    
and have thus preferred not to include this result in the main
analysis of the present work, delaying a more careful study to a
future publication.


\begin{thebibliography}{999}


\bibitem{Polyakov1}
  A.~M.~Polyakov,
  %``String theory and quark confinement,''
  Nucl.\ Phys.\ Proc.\ Suppl.\  {\bf 68} (1998) 1
  [arXiv:hep-th/9711002].
  %%CITATION = NUPHZ,68,1;%%


\bibitem{adscft1} {J.~M.~Maldacena}, {Adv.\ Theor.\ Math.\ Phys.\ } {\bf 2}
  (1998) 231 [arXiv:hep-th/9711200].

\bibitem{adscft2} {S.~S.~Gubser, I.~R.~Klebanov and
A.~M.~Polyakov}, {Phys.\ Lett.\ } B {\bf 428}  (1998) 105
[arXiv:hep-th/9802109]. 

\bibitem{adscft3} {E.~Witten}, {Adv.\ Theor.\ Math.\ Phys.\ } {\bf 2}
  (1998) 253 [arXiv:hep-th/9802150].

\bibitem{QGP1}
  G.~Policastro, D.~T.~Son and A.~O.~Starinets,
 % ``The shear viscosity of strongly coupled N = 4 supersymmetric Yang-Mills
 % plasma,''
  Phys.\ Rev.\ Lett.\  {\bf 87}, 081601 (2001) [arXiv:hep-th/0104066].

\bibitem{QGP2}
  R.~A.~Janik and R.~B.~Peschanski,
 % ``Asymptotic perfect fluid dynamics as a consequence of AdS/CFT,''
  Phys.\ Rev.\  D {\bf 73}, 045013 (2006) [arXiv:hep-th/0512162].

\bibitem{QGP3}
  A.~Bernamonti and R.~Peschanski,
  {\it Time--dependent AdS/CFT correspondence and the Quark--Gluon
    plasma}, arXiv:1102.0725 [hep-th]. 
  
\bibitem{AdSBH} E.~Witten, Adv.\ Theor.\ Math.\ Phys.\ {\bf 2} (1998) 505
  [arXiv:hep-th/9803131].

\bibitem{Polchinski} J.~Polchinski and M.J.~Strassler,
  %``Hard scattering and gauge / string duality,''
  Phys.\ Rev.\ Lett.\  {\bf 88} (2002) 031601
  [arXiv:hep-th/0109174].

\bibitem{Karch}
  A.~Karch, E.~Katz, D.T.~Son and M.A.~Stephanov,
  %``Linear Confinement and AdS/QCD,''
  Phys.\ Rev.\  D {\bf 74} (2006) 015005
  [arXiv:hep-ph/0602229].

\bibitem{JP1}
R.~A.~Janik and R.~Peschanski, Nucl.\ Phys.\  B {\bf 565} (2000) 193
[arXiv:hep-th/9907177].

\bibitem{JP2}
R.~A.~Janik and R.~Peschanski, Nucl.\ Phys.\ B {\bf 586} (2000) 163
[arXiv:hep-th/0003059].

\bibitem{Janik}
R.~A.~Janik, Phys.\ Lett.\  B {\bf 500} (2001) 118 [arXiv:hep-th/0010069].

\bibitem{Jani} {R.~A.~Janik and R.~B.~Peschanski}, 
  {Nucl.\ Phys.\ } B {\bf 625} (2002) 279 [arXiv:hep-th/0110024].

\bibitem{GP} M.~Giordano and R.~Peschanski, {JHEP} {\bf 1005} (2010)
  037 [arXiv:1003.2309 [hep-ph]].


\bibitem{Collins} P.~D.~B.~Collins, {\it An Introduction to Regge Theory
    and High-Energy Physics}, (Cambridge University Press, Cambridge,
  1977).

\bibitem{Nacht} {O.~Nachtmann}, {Ann.\ Phys.\ } {\bf 209} (1991) 436.

\bibitem{DFK} {H.~G.~Dosch, E.~Ferreira and A.~Kr{\"a}mer},
  {Phys.\ Rev.\ D} {\bf 50} (1994) 1992 [arXiv:hep-ph/9405237]. 

\bibitem{Meggiolaro96}
  E.~Meggiolaro,
  %``A Remark on the high-energy quark quark scattering and the eikonal
  %approximation,''
  Phys.\ Rev.\  D {\bf 53} (1996) 3835 
  [arXiv:hep-th/9506043].


\bibitem{Nachtr} {O.~Nachtmann}, {\it High
Energy Collisions and Nonperturbative QCD\/}, 
 in {\it Perturbative and nonperturbative aspects of 
quantum field theory\/}, proceedings of the {\it 35th International
University School Of Nuclear And Particle Physics}, 2-9 Mar 1996,
Schladming, Austria, edited by H.~Latal and W.~Schweiger 
(Springer-Verlag, Berlin, Heidelberg, 1997), 49; in {\it Lectures on
  QCD: Applications}, edited by H.~W.~Grie\ss hammer, F.~Lenz and
D.~Stoll (Springer-Verlag, Berlin, 
Heidelberg, 1997), 1 [arXiv:hep-ph/9609365].


\bibitem{BN} 
{E.~R.~Berger and O.~Nachtmann}, {Eur.\ Phys.\ J.\ C} {\bf 7}
   (1999) 459 [arXiv:hep-ph/9808320].

\bibitem{Dosch}
H.~G.~Dosch, in {\it At the frontier of Particle Physics -- Handbook of QCD
(Boris Iof\mbox{}fe Festschrift)}, edited by M.~Shifman (World
Scientif\mbox{}ic, Singapore, 2001), vol. 2, 1195--1236.


\bibitem{Meggiolaro01}
  E.~Meggiolaro,
  %``Eikonal propagators and high-energy parton parton scattering in gauge
  %theories,''
  Nucl.\ Phys.\   B {\bf 602} (2001) 261 [arXiv:hep-ph/0009261].


\bibitem{LLCM1} 
{A.~I.~Shoshi, F.~D.~Stef\mbox{}fen and H.~J.~Pirner}, 
  {Nucl.\ Phys.\ A} {\bf 709} (2002) 131 [arXiv:hep-ph/0202012]. 

\bibitem{LLCM2} {A.~I.~Shoshi, F.~D.~Steffen, H.~G.~Dosch and
    H.~J.~Pirner}, {Phys.\ Rev.\ D} {\bf 68} (2003) 074004
  [arXiv:hep-ph/0211287]. 

\bibitem{ILM} {E.~Shuryak and I.~Zahed}, {Phys.\ Rev.\ D} {\bf 62} 
    (2000) 085014 [arXiv:hep-ph/0005152]. 

\bibitem{Latt}
  {M.~Giordano and E.~Meggiolaro}, 
  %``High-energy hadron-hadron (dipole-dipole) scattering from lattice QCD,''
  {Phys.\ Rev.\ D} {\bf 78} (2008) 074510 [arXiv:0808.1022 [hep-lat]].

\bibitem{Latt2}  M.~Giordano and E.~Meggiolaro, {Phys.\ Rev.\ D} {\bf 81}
  (2010) 074022 [arXiv:0910.4505 [hep-ph]].

\bibitem{Brow0}
  {R.~C.~Brower, J.~Polchinski, M.~J.~Strassler and C.~I.~Tan}, 
  %``The Pomeron and Gauge/String Duality,''
  {JHEP} {\bf 0712} (2007) 005 [arXiv:hep-th/0603115].
  %%CITATION = JHEPA,0712,005;%%
%\cite{Brower:2007xg}

\bibitem{Brow1}
  {R.~C.~Brower, M.~J.~Strassler and C.~I.~Tan},
  %``On the Eikonal Approximation in AdS Space,''
  {JHEP} {\bf 0903} (2009) 050 [arXiv:0707.2408 [hep-th]]. 
  %%CITATION = JHEPA,0903,050;%%%\cite{Levin:2009kk}
\bibitem{Brow2}
  {R.~C.~Brower, M.~J.~Strassler and C.~I.~Tan},
  %``On The Pomeron at Large 't Hooft Coupling,''
  {JHEP} {\bf 0903}(2009) 092 [arXiv:0710.4378 [hep-th]]. 
  %%CITATION = JHEPA,0903,092;%%
%\cite{Brower:2008cy}
\bibitem{Brow3}
  {R.~C.~Brower, M.~Djuric and C.~I.~Tan},
  %``Odderon In Gauge/String Duality,''
  {JHEP} {\bf 0907} (2009) 063 [arXiv:0812.0354 [hep-th]].
%\cite{Brower:2009bh}
\bibitem{Brow4}
  {R.~Brower, M.~Djuric and C.~I.~Tan},
  {\it Elastic and Diffractive Scattering after AdS/CFT}, %{}
  {arXiv:0911.3463 [hep-ph].}
  %%CITATION = ARXIV:0911.3463;%%

\bibitem{Corn1}
  {L.~Cornalba, M.~S.~Costa, J.~Penedones and R.~Schiappa},
  %``Eikonal approximation in AdS/CFT: From shock waves to four-point
  %functions,''
  {JHEP} {\bf 0708} (2007) 019 [arXiv:hep-th/0611122].
  %%CITATION = JHEPA,0708,019;%%%\cite{Cornalba:2006xm} 

\bibitem{Corn2}
  {L.~Cornalba, M.~S.~Costa, J.~Penedones and R.~Schiappa}, 
  %``Eikonal approximation in AdS/CFT: Conformal partial waves and finite N
  %four-point functions,''
  {Nucl.\ Phys.\ } B {\bf 767} (2007) 327 [arXiv:hep-th/0611123].
  %%CITATION = NUPHA,B767,327;%%%\cite{Cornalba:2007zb}

\bibitem{Corn3}
  {L.~Cornalba, M.~S.~Costa and J.~Penedones}, 
  %``Eikonal Approximation in AdS/CFT: Resumming the Gravitational Loop
  %Expansion,''
  {JHEP} {\bf 0709} (2007) 037 [arXiv:0707.0120 [hep-th]].
  %%CITATION = JHEPA,0709,037;%%%\cite{Cornalba:2007fs}

\bibitem{Corn4}
  {L.~Cornalba}, 
  {\it Eikonal Methods in AdS/CFT: Regge Theory and Multi-Reggeon
    Exchange}, 
  {arXiv:0710.5480 [hep-th]}.
  %%CITATION = ARXIV:0710.5480;%%
%\cite{Cornalba:2008qf}

\bibitem{Corn5}
  {L.~Cornalba, M.~S.~Costa and J.~Penedones}, 
  %``Eikonal Methods in AdS/CFT: BFKL Pomeron at Weak Coupling,''
  {JHEP} {\bf 0806} (2008) 048 [arXiv:0801.3002 [hep-th]]. %\cite{Brower:2}


\bibitem{Tali1}
  {J.~L.~Albacete, Y.~V.~Kovchegov and A.~Taliotis}, 
  %``DIS on a Large Nucleus in AdS/CFT,''
  {JHEP} {\bf 0807} (2008) 074 [arXiv:0806.1484 [hep-th]].
%\cite{Taliotis:2009ne}

\bibitem{Tali2}
  {A.~Taliotis},
  %``DIS from the AdS/CFT correspondence,''
  {Nucl.\ Phys.\ } A {\bf 830} (2009) 299C [arXiv:0907.4204 [hep-th]]. 
  %%CITATION = NUPHA,A830,299C;%%%\cite{Mueller:2008bt}

\bibitem{Muel}
  {A.~H.~Mueller, A.~I.~Shoshi and B.~W.~Xiao},
  %``Deep inelastic and dipole scattering on finite length hot $\mathcal{N}=4$
  %SYM matter,''
  {Nucl.\ Phys.\ } A {\bf 822} (2009) 20 [arXiv:0812.2897 [hep-th]].

\bibitem{LP}
  {E.~Levin and I.~Potashnikova}, 
  %``Soft interaction at high energy and N=4 SYM,''
  {JHEP} {\bf 0906} (2009) 031 [arXiv:0902.3122 [hep-ph]].
  %%CITATION = JHEPA,0906,031;%%

\bibitem{Khar}
  {D.~E.~Kharzeev and E.~M.~Levin}, 
  %{D-instantons and multiparticle production in N=4 SYM,}
  {JHEP} {\bf 1001} (2010) 046 [arXiv:0910.3355 [hep-ph]].

\bibitem{Yoshi} E.~Avsar, Y.~Hatta, and T.~Matsuo, JHEP {\bf 1003}
  % {\bf 1003} 
  (2010) 037 {[arXiv:0912.3806 [hep-th]]}.

\bibitem{Makeenko} Y.~Makeenko, Phys.\ Rev.\ D {\bf 83} (2011) 026007
  [arXiv:1012.0708 [hep-th]]. 

\bibitem{Brandt} {R.~A.~Brandt, F.~Neri, and D.~Zwanziger},
  {Phys.\ Rev.\ D} {\bf 19} (1979) 1153. 
\bibitem{Brandt2} {R.~A.~Brandt, A.~Gocksch,
    M.~Sato, and F.~Neri}, {Phys.\ Rev.\ D} {\bf 26} (1982) 3611.

\bibitem{Polyakov} A.~M.~Polyakov, {Mod.\ Phys.\ Lett.\ A} {\bf 3}
  (1988) 325.
\bibitem{Korchemsky} G.~P.~Korchemsky, {Phys.\ Lett.\ } B {\bf 232} (1989)
  334.
\bibitem{Korchemsky2} G.~P.~Korchemsky, {Int.\ J.\ Mod.\ Phys.\ A} {\bf 7}
  (1992) 339.

\bibitem{Meggiolaro97} {E.~Meggiolaro} {Z.\ Phys.\ C} {\bf 76} (1997)
  523 [arXiv:hep-th/9602104].
\bibitem{Meggiolaro98} {E.~Meggiolaro}, {Eur.\ Phys.\ J.\ C} {\bf 4}
  (1998) 101 [arXiv:hep-th/9702186].
\bibitem{Meggiolaro02} {E.~Meggiolaro}, {Nucl.\ Phys.\ } B {\bf 625}
  (2002) 312 [arXiv:hep-ph/0110069].
\bibitem{Meggiolaro05}{E.~Meggiolaro}, {Nucl.\ Phys.\ } B {\bf 707}
  (2005) 199 [arXiv:hep-ph/0407084].


\bibitem{crossing}
{M.~Giordano and E.~Meggiolaro}, {Phys.\ Rev.\ D} {\bf 74} (2006)
016003 [arXiv:hep-ph/0602143].
\bibitem{Meggiolaro07} {E.~Meggiolaro}, {Phys.\ Lett.\ } B {\bf 651}
  (2007) 177 [arXiv:hep-ph/0612307].

\bibitem{EMduality} {M.~Giordano and E.~Meggiolaro}, {Phys.\ Lett.\ }
  B {\bf 675}  (2009) 123 [arXiv:0902.4145 [hep-ph]].


\bibitem{Alday} L.~F.~Alday, J.~M.~Maldacena, JHEP {\bf 0706} (2007)
  064 [arXiv:0705.0303 [hep-th]].

%\cite{Naculich:2007ub}
\bibitem{Naculich:2007ub}
  S.~G.~Naculich and H.~J.~Schnitzer,
  %``Regge behaior of gluon scattering amplitudes in N=4 SYM theory,''
  Nucl.\ Phys.\  B {\bf 794}  (2008) 189
  [arXiv:0708.3069 [hep-th]].
  %%CITATION = NUPHA,B794,189;%%
%\cite{Naculich:2007ub}

\bibitem{us} M.~Giordano, R.~Peschanski, and S.~Seki, {\it ${\cal N} = 4$
  SYM Regge Amplitudes and Minimal Surfaces in AdS/CFT
  Correspondence}, arXiv:1110.3680 [hep-th]. 

\bibitem{Wilson} {J.~M.~Maldacena}, {Phys.\ Rev.\ Lett.\ } {\bf 80} (1998)
  {4859}. [arXiv:hep-th/{9803002}]. 

\bibitem{Wilson2} {S.-J.~Rey and J.~Yee}, {Eur.\ Phys.\ Jour.\ C} {\bf
    22} (2001) {379} [arXiv:hep-th/{9803001}].

\bibitem{GO} D.~J.~Gross and H.~Ooguri,
  %``Aspects of large N gauge theory dynamics as seen by string theory,''
  Phys.\ Rev.\  D {\bf 58} (1998) 106002
  [arXiv:hep-th/9805129].

\bibitem{DGO} N.~Drukker, D.~J.~Gross and H.~Ooguri,
  %``Wilson loops and minimal surfaces,''
  Phys.\ Rev.\  D {\bf 60} (1999) 125006
  [arXiv:hep-th/9904191].

\bibitem{Sonn0} A.~Brandhuber, N.~Itzhaki, J.~Sonnenschein and
S.~Yankielowicz, 
  %``Wilson loops, confinement, and phase transitions in large N
%  gauge  theories 
  %from supergravity,''
  JHEP {\bf 9806} (1998) 001
  [arXiv:hep-th/9803263].

\bibitem{Sonn1} Y.~Kinar, E.~Schreiber and J.~Sonnenschein,
  %``Q anti-Q potential from strings in curved spacetime: Classical results,''
  Nucl.\ Phys.\  B {\bf 566} (2000) 103
  [arXiv:hep-th/9811192].

\bibitem{Sonn2}  Y.~Kinar, E.~Schreiber, J.~Sonnenschein and N.~Weiss,
  %``Quantum fluctuations of Wilson loops from string models,''
  Nucl.\ Phys.\  B {\bf 583} (2000) 76
  [arXiv:hep-th/9911123].

\bibitem{Frampton} P.~A.~Frampton, {\it Dual resonance models and superstrings}
  (World Scientific, Singapore, 1986).


\bibitem{Giordano_wp1} M.~Giordano, to appear.

\bibitem{Armoni}
  A.~Armoni,
  %``Beyond The Quenched (or Probe Brane) Approximation in Lattice (or
  %Holographic) QCD,''
  Phys.\ Rev.\ D {\bf 78} (2008) 065017 [arXiv:0805.1339 [hep-th]].

\bibitem{Giordano_wp2} M.~Giordano and R.~Peschanski, to appear.

\bibitem{min_surf} U.~Dierkes, S.~Hildebrandt, A.~K\"uster, and
  O.~Wohlrab, {\it Minimal Surfaces I: Boundary Value Problems}
  (Springer-Verlag, Berlin, 1992).
  

\bibitem{Froissart1} {M.~Froissart}, {Phys.\ Rev.\ } {\bf 123} (1961)
    1053.

\bibitem{Froissart2}
    {A.~Martin}, {Il Nuovo Cimento A} {\bf 42} (1966) 930.

\bibitem{Froissart3}
    {L.~\L ukaszuk and A.~Martin}, {Il Nuovo Cimento A} {\bf 52} (1967)
    122. 

\bibitem{AS} M.~Abramowitz and I.~A.~Stegun, {\it Handbook of
    Mathematical Functions with Formulas, Graphs, and Mathematical
    Tables} (Dover, New York, 1964).

\bibitem{Korchemskaya} I.~A.~Korchemskaya and G.~P.~Korchemsky,
  J.\ Phys.\ A: Math.\ Gen.\ {\bf 24} (1991) 4511.

\end{thebibliography}
\end{document}